\documentclass[aps,prb,twocolumn,superscriptaddress,floatfix,10pt]{revtex4-1}
\usepackage[utf8]{inputenc}
\usepackage{natbib}
\usepackage{graphicx}
\usepackage{latexsym}
\usepackage{amssymb}
\usepackage{amsmath}
\usepackage{amsfonts}
\usepackage{mathtools}
\usepackage{bm}
\usepackage{enumitem}
\usepackage{hyperref}
\usepackage{lipsum}
\usepackage{braket}
\usepackage{tikz}
\usepackage{comment}
\usepackage{subcaption}
\usepackage{ifthen}
\usepackage{etoolbox}
\usepackage{caption}
\usepackage{bbm}
\usepackage{ragged2e}
\usepackage{multirow}
\usetikzlibrary{arrows,decorations.pathreplacing,decorations.markings,arrows.meta,patterns,3d,shapes.geometric}
\usepackage[normalem]{ulem} 
\graphicspath{{Figures/}} 

\newcommand{\octagon}[3]{
    \pgfmathsetmacro{\angle}{45}
    \pgfmathsetmacro{\startangle}{-22.5}
    \pgfmathsetmacro{\radius}{#1}
    \begin{scope}[decoration={markings, mark=at position #2 with {\arrow{Latex}}}]
        \foreach \i in {0,1,2,3,4,5,6,7} {
            \pgfmathsetmacro{\ii}{\i+1}
            \pgfmathsetmacro{\x}{\radius*sin(\startangle + \angle*\i)}
            \pgfmathsetmacro{\xx}{\radius*sin(\startangle + \angle*\ii)}
            \pgfmathsetmacro{\y}{\radius*cos(\startangle + \angle*\i)}
            \pgfmathsetmacro{\yy}{\radius*cos(\startangle + \angle*\ii)}
            
            \draw[line width = #3 pt, postaction={decorate}] (\xx , \yy) -- (\x , \y);
            
            \pgfmathsetmacro{\tmpxA}{0.5*\radius*sin(45+90*(\i-1)/2)}
            \pgfmathsetmacro{\tmpxB}{0.5*\radius*cos(45+90*(\i-1)/2)}
            \pgfmathsetmacro{\distfactor}{1.5}
            
            \ifodd\i {
            
                \draw[line width = #3 pt, postaction={decorate}] (\xx + \tmpxA  , \yy + \tmpxB)--(\xx , \yy);
                \draw[line width = #3 pt, postaction={decorate}] (\x + \tmpxA  , \y + \tmpxB)--(\x , \y);
                
                \ifnum \i = 1 {
                    \node [] at (\xx + \tmpxA*\distfactor , \yy + \tmpxB*\distfactor) {$\bar{\bar{\ell}}_3$};
                    \node [] at (\x + \tmpxA*\distfactor , \y + \tmpxB*\distfactor) {$\bar{\ell}_3$};
                }\fi
                \ifnum \i = 3 {
                    \node [] at (\xx + \tmpxA*\distfactor , \yy + \tmpxB*\distfactor) {$\ell_4$};
                    \node [] at (\x + \tmpxA*\distfactor , \y + \tmpxB*\distfactor) {$\bar{\ell}_4$};
                }\fi
                \ifnum \i = 5 {
                    \node [] at (\xx + \tmpxA*\distfactor , \yy + \tmpxB*\distfactor) {$\bar{\ell}_1$};
                    \node [] at (\x + \tmpxA*\distfactor , \y + \tmpxB*\distfactor) {$\ell_1$};
                }\fi
                \ifnum \i = 7 {
                    \node [] at (\xx + \tmpxA*\distfactor , \yy + \tmpxB*\distfactor) {$\bar{\bar{\ell}}_2$};
                    \node [] at (\x + \tmpxA*\distfactor , \y + \tmpxB*\distfactor) {$\bar{\ell}_2$};
                }\fi
                     
            }\fi
        }
    \end{scope}
}

\newcommand{\octagonControlledZero}[3]{
    \pgfmathsetmacro{\angle}{45}
    \pgfmathsetmacro{\startangle}{-22.5}
    \pgfmathsetmacro{\radius}{#1}
    \begin{scope}[decoration={markings, mark=at position #2 with {\arrow{Latex}}}]
        \foreach \i in {0,1,2,3,4,5,6,7} {
            \pgfmathsetmacro{\ii}{\i+1}
            \pgfmathsetmacro{\x}{\radius*sin(\startangle + \angle*\i)}
            \pgfmathsetmacro{\xx}{\radius*sin(\startangle + \angle*\ii)}
            \pgfmathsetmacro{\y}{\radius*cos(\startangle + \angle*\i)}
            \pgfmathsetmacro{\yy}{\radius*cos(\startangle + \angle*\ii)}
            
            \ifthenelse{\i = 4}
            {\draw[line width = #3 pt, dashed] (\xx , \yy) -- (\x , \y);}
            {\draw[line width = #3 pt, postaction={decorate}] (\xx , \yy) -- (\x , \y);}

            \pgfmathsetmacro{\tmpxA}{0.5*\radius*sin(45+90*(\i-1)/2)}
            \pgfmathsetmacro{\tmpxB}{0.5*\radius*cos(45+90*(\i-1)/2)}
            \pgfmathsetmacro{\distfactor}{1.5}
            
            \ifodd\i {
            
                \draw[line width = #3 pt, postaction={decorate}] (\xx + \tmpxA  , \yy + \tmpxB)--(\xx , \yy);
                \draw[line width = #3 pt, postaction={decorate}] (\x + \tmpxA  , \y + \tmpxB)--(\x , \y);
                
                \ifnum \i = 1 {
                    \node [] at (\xx + \tmpxA*\distfactor , \yy + \tmpxB*\distfactor) {$\bar{\bar{\ell}}_3$};
                    \node [] at (\x + \tmpxA*\distfactor , \y + \tmpxB*\distfactor) {$\bar{\ell}_3$};
                }\fi
                \ifnum \i = 3 {
                    \node [] at (\xx + \tmpxA*\distfactor , \yy + \tmpxB*\distfactor) {$\ell_4$};
                    \node [] at (\x + \tmpxA*\distfactor , \y + \tmpxB*\distfactor) {$\bar{\ell}_4$};
                }\fi
                \ifnum \i = 5 {
                    \node [] at (\xx + \tmpxA*\distfactor , \yy + \tmpxB*\distfactor) {$\bar{\ell}_1$};
                    \node [] at (\x + \tmpxA*\distfactor , \y + \tmpxB*\distfactor) {$\ell_1$};
                }\fi
                \ifnum \i = 7 {
                    \node [] at (\xx + \tmpxA*\distfactor , \yy + \tmpxB*\distfactor) {$\bar{\bar{\ell}}_2$};
                    \node [] at (\x + \tmpxA*\distfactor , \y + \tmpxB*\distfactor) {$\bar{\ell}_2$};
                }\fi
                    
            }\fi
        }
    \end{scope}
}

\newcommand{\octagonControlled}[3]{
    \pgfmathsetmacro{\angle}{45}
    \pgfmathsetmacro{\startangle}{-22.5}
    \pgfmathsetmacro{\radius}{#1}
    \begin{scope}[decoration={markings, mark=at position #2 with {\arrow{Latex}}}]
        \foreach \i in {0,1,2,3,4,5,6,7} {
            \pgfmathsetmacro{\ii}{\i+1}
            \pgfmathsetmacro{\x}{\radius*sin(\startangle + \angle*\i)}
            \pgfmathsetmacro{\xx}{\radius*sin(\startangle + \angle*\ii)}
            \pgfmathsetmacro{\y}{\radius*cos(\startangle + \angle*\i)}
            \pgfmathsetmacro{\yy}{\radius*cos(\startangle + \angle*\ii)}

            \ifthenelse{\i = 4}
            {
            \draw[line width = 1 pt, postaction={decorate},draw=red] (\xx , \yy) -- (\x , \y);
            }
            {\draw[line width = #3 pt, postaction={decorate}] (\xx , \yy) -- (\x , \y);}

            \pgfmathsetmacro{\tmpxA}{0.5*\radius*sin(45+90*(\i-1)/2)}
            \pgfmathsetmacro{\tmpxB}{0.5*\radius*cos(45+90*(\i-1)/2)}
            \pgfmathsetmacro{\distfactor}{1.5}
            
            \ifodd\i {
            
                \draw[line width = #3 pt, postaction={decorate}] (\xx + \tmpxA  , \yy + \tmpxB)--(\xx , \yy);
                \draw[line width = #3 pt, postaction={decorate}] (\x + \tmpxA  , \y + \tmpxB)--(\x , \y);
                
                \ifnum \i = 1 {
                    \node [] at (\xx + \tmpxA*\distfactor , \yy + \tmpxB*\distfactor) {$\bar{\bar{\ell}}_3$};
                    \node [] at (\x + \tmpxA*\distfactor , \y + \tmpxB*\distfactor) {$\bar{\ell}_3$};
                }\fi
                \ifnum \i = 3 {
                    \node [] at (\xx + \tmpxA*\distfactor , \yy + \tmpxB*\distfactor) {$\ell_4$};
                    \node [] at (\x + \tmpxA*\distfactor , \y + \tmpxB*\distfactor) {$\bar{\ell}_4$};
                }\fi
                \ifnum \i = 5 {
                    \node [] at (\xx + \tmpxA*\distfactor , \yy + \tmpxB*\distfactor) {$\bar{\ell}_1$};
                    \node [] at (\x + \tmpxA*\distfactor , \y + \tmpxB*\distfactor) {$\ell_1$};
                }\fi
                \ifnum \i = 7 {
                    \node [] at (\xx + \tmpxA*\distfactor , \yy + \tmpxB*\distfactor) {$\bar{\bar{\ell}}_2$};
                    \node [] at (\x + \tmpxA*\distfactor , \y + \tmpxB*\distfactor) {$\bar{\ell}_2$};
                }\fi
                    
            }\fi
        }
    \end{scope}
}

\newcommand{\octagonControlledCN}[2]{
    \pgfmathsetmacro{\angle}{45}
    \pgfmathsetmacro{\startangle}{-22.5}
    \pgfmathsetmacro{\radius}{#1}
    \begin{scope}
        \foreach \i in {0,1,2,3,4,5,6,7} {
            \pgfmathsetmacro{\ii}{\i+1}
            \pgfmathsetmacro{\x}{\radius*sin(\startangle + \angle*\i)}
            \pgfmathsetmacro{\xx}{\radius*sin(\startangle + \angle*\ii)}
            \pgfmathsetmacro{\y}{\radius*cos(\startangle + \angle*\i)}
            \pgfmathsetmacro{\yy}{\radius*cos(\startangle + \angle*\ii)}

            \ifthenelse{\i = 4}
            {
            \draw[line width = 1 pt, draw=red] (\xx , \yy) -- (\x , \y);
            }
            {\draw[line width = #2 pt] (\xx , \yy) -- (\x , \y);}
            
            \ifodd \ii {
                \pgfmathsetmacro{\shiftxx}{\radius*sin(\angle*\i)}
                \pgfmathsetmacro{\shiftyy}{\radius*cos(\angle*\i)}
                \pgfmathsetmacro{\shiftfactor}{0.08}
                
                \ifthenelse{\i = 4}{
                    \draw[draw = green!70!black, line width = 1 pt] (\xx + \shiftfactor*\shiftxx , \yy + \shiftfactor*\shiftyy) -- (\x + \shiftfactor*\shiftxx , \y + \shiftfactor*\shiftyy);
                }{
                    \draw[line width = #2 pt] (\xx + \shiftfactor*\shiftxx , \yy + \shiftfactor*\shiftyy) -- (\x + \shiftfactor*\shiftxx , \y + \shiftfactor*\shiftyy);
                }
            }\fi

            \pgfmathsetmacro{\tmpxA}{0.5*\radius*sin(45+90*(\i-1)/2)}
            \pgfmathsetmacro{\tmpxB}{0.5*\radius*cos(45+90*(\i-1)/2)}
            \pgfmathsetmacro{\distfactor}{1.5}
            
            \ifodd\i {
            
                \draw[line width = #2 pt] (\xx + \tmpxA  , \yy + \tmpxB)--(\xx , \yy);
                \draw[line width = #2 pt] (\x + \tmpxA  , \y + \tmpxB)--(\x , \y);
                
                \ifnum \i = 1 {
                    \node [] at (\xx + \tmpxA*\distfactor , \yy + \tmpxB*\distfactor) {$\bar{\bar{\ell}}_3$};
                    \node [] at (\x + \tmpxA*\distfactor , \y + \tmpxB*\distfactor) {$\bar{\ell}_3$};
                }\fi
                \ifnum \i = 3 {
                    \node [] at (\xx + \tmpxA*\distfactor , \yy + \tmpxB*\distfactor) {$\ell_4$};
                    \node [] at (\x + \tmpxA*\distfactor , \y + \tmpxB*\distfactor) {$\bar{\ell}_4$};
                }\fi
                \ifnum \i = 5 {
                    \node [] at (\xx + \tmpxA*\distfactor , \yy + \tmpxB*\distfactor) {$\bar{\ell}_1$};
                    \node [] at (\x + \tmpxA*\distfactor , \y + \tmpxB*\distfactor) {$\ell_1$};
                }\fi
                \ifnum \i = 7 {
                    \node [] at (\xx + \tmpxA*\distfactor , \yy + \tmpxB*\distfactor) {$\bar{\bar{\ell}}_2$};
                    \node [] at (\x + \tmpxA*\distfactor , \y + \tmpxB*\distfactor) {$\bar{\ell}_2$};
                }\fi
                    
            }\fi
        }
    \end{scope}
}

\newcommand{\octagonControlledCNZero}[2]{
    \pgfmathsetmacro{\angle}{45}
    \pgfmathsetmacro{\startangle}{-22.5}
    \pgfmathsetmacro{\radius}{#1}
    \begin{scope}
        \foreach \i in {0,1,2,3,4,5,6,7} {
            \pgfmathsetmacro{\ii}{\i+1}
            \pgfmathsetmacro{\x}{\radius*sin(\startangle + \angle*\i)}
            \pgfmathsetmacro{\xx}{\radius*sin(\startangle + \angle*\ii)}
            \pgfmathsetmacro{\y}{\radius*cos(\startangle + \angle*\i)}
            \pgfmathsetmacro{\yy}{\radius*cos(\startangle + \angle*\ii)}

            \ifthenelse{\i = 4}
            {
            \draw[line width = #2 pt, dashed] (\xx , \yy) -- (\x , \y);
            }
            {\draw[line width = #2 pt] (\xx , \yy) -- (\x , \y);}
            
            \ifodd \ii {
                \pgfmathsetmacro{\shiftxx}{\radius*sin(\angle*\i)}
                \pgfmathsetmacro{\shiftyy}{\radius*cos(\angle*\i)}
                \pgfmathsetmacro{\shiftfactor}{0.08}
                
                \ifthenelse{\i = 4}{
                    \draw[draw = green!70!black, line width = 1 pt] (\xx + \shiftfactor*\shiftxx , \yy + \shiftfactor*\shiftyy) -- (\x + \shiftfactor*\shiftxx , \y + \shiftfactor*\shiftyy);
                }{
                    \draw[line width = #2 pt] (\xx + \shiftfactor*\shiftxx , \yy + \shiftfactor*\shiftyy) -- (\x + \shiftfactor*\shiftxx , \y + \shiftfactor*\shiftyy);
                }
            }\fi

            \pgfmathsetmacro{\tmpxA}{0.5*\radius*sin(45+90*(\i-1)/2)}
            \pgfmathsetmacro{\tmpxB}{0.5*\radius*cos(45+90*(\i-1)/2)}
            \pgfmathsetmacro{\distfactor}{1.5}
            
            \ifodd\i {
            
                \draw[line width = #2 pt] (\xx + \tmpxA  , \yy + \tmpxB)--(\xx , \yy);
                \draw[line width = #2 pt] (\x + \tmpxA  , \y + \tmpxB)--(\x , \y);
                
                \ifnum \i = 1 {
                    \node [] at (\xx + \tmpxA*\distfactor , \yy + \tmpxB*\distfactor) {$\bar{\bar{\ell}}_3$};
                    \node [] at (\x + \tmpxA*\distfactor , \y + \tmpxB*\distfactor) {$\bar{\ell}_3$};
                }\fi
                \ifnum \i = 3 {
                    \node [] at (\xx + \tmpxA*\distfactor , \yy + \tmpxB*\distfactor) {$\ell_4$};
                    \node [] at (\x + \tmpxA*\distfactor , \y + \tmpxB*\distfactor) {$\bar{\ell}_4$};
                }\fi
                \ifnum \i = 5 {
                    \node [] at (\xx + \tmpxA*\distfactor , \yy + \tmpxB*\distfactor) {$\bar{\ell}_1$};
                    \node [] at (\x + \tmpxA*\distfactor , \y + \tmpxB*\distfactor) {$\ell_1$};
                }\fi
                \ifnum \i = 7 {
                    \node [] at (\xx + \tmpxA*\distfactor , \yy + \tmpxB*\distfactor) {$\bar{\bar{\ell}}_2$};
                    \node [] at (\x + \tmpxA*\distfactor , \y + \tmpxB*\distfactor) {$\bar{\ell}_2$};
                }\fi
                    
            }\fi
        }
    \end{scope}
}

\DeclareMathAlphabet\mathbfcal{OMS}{cmsy}{b}{n}

\newcommand{\sfigref}[2]{Fig.\,\hyperref[#1]{\ref{#1}(#2)}}

\hypersetup{
  colorlinks = true,
  urlcolor = blue,
  pdfauthor = {Zongyuan Wang, Xiuqi Ma, Mike Hermele, David Stephen, Xie Chen},
  pdftitle = {Wang et al. - 2022 - Renormalization of Ising cage-net model and generalized foliation}
}

\begin{document}

\title{Renormalization of Ising cage-net model and generalized foliation}
\date{\today}

\author{Zongyuan Wang}
\author{Xiuqi Ma}
\affiliation{Department of Physics and Institute for Quantum Information and Matter, \mbox{California Institute of Technology, Pasadena, California 91125, USA}}
\author{David T. Stephen}
\affiliation{Department of Physics and Center for Theory of Quantum Matter, University of Colorado, Boulder, CO 80309, USA}
\affiliation{Department of Physics and Institute for Quantum Information and Matter, \mbox{California Institute of Technology, Pasadena, California 91125, USA}}
\author{Michael Hermele}
\affiliation{Department of Physics and Center for Theory of Quantum Matter, University of Colorado, Boulder, CO 80309, USA}
\author{Xie Chen}
\affiliation{Department of Physics and Institute for Quantum Information and Matter, \mbox{California Institute of Technology, Pasadena, California 91125, USA}}

\begin{abstract} 
A large class of type-I fracton models, including the X-cube model, have been found to be fixed points of the foliated renormalization group (RG). The system size of such \emph{foliated} models can be changed by adding or removing decoupled layers of $2$D topological states and continuous deformation of the Hamiltonian.  In this paper, we study a closely related model -- the Ising cage-net model -- and find that this model is not foliated in the same sense.  In fact, we point out certain unnatural restrictions in the foliated RG, and find that removing these restrictions leads to a generalized foliated RG under which the Ising cage-net model is a fixed point, and which includes the original foliated RG as a special case.  The Ising cage-net model thus gives a prototypical example of the generalized foliated RG, and its system size can be changed either by condensing / uncondensing bosonic planon excitations near a 2D plane or through a linear depth quantum circuit in the same plane. We show that these two apparently different RG procedures are closely related, as they lead to the same gapped boundary when implemented in part of a plane. Finally, we briefly discuss the implications for foliated fracton phases, whose universal properties will need to be reexamined in light of the generalized foliated RG.
\end{abstract}

\maketitle

\section{Introduction}
\label{sec:intro}

The renormalization group (RG) plays a fundamental role in the characterization and classification of quantum phases of matter.\cite{RevModPhys.46.597, VidalRG, Xie10} It is a piece of conventional wisdom that each phase -- defined as a deformation class of quantum systems -- is characterized by a unique RG fixed point, which encodes the universal long-distance and low-energy properties of the phase.  Moreover, the existence of such a fixed point underlies the key role played by continuum quantum field theory as a tool to describe universal properties of phases (and phase transitions) while discarding extraneous non-universal information.

Fracton models in three spatial dimensions (3D)\cite{Nandkishore_2019,Pretko_2020} provide exceptions to this conventional wisdom, and accordingly challenge our understanding of the relationships among quantum phases of matter, the renormalization group, and quantum field theory. This is nicely illustrated in the X-cube model,\cite{Sagar16} perhaps the simplest fracton model. The defining characteristic of a fracton model is the presence of excitations of restricted mobility, and the X-cube model supports point-like excitations mobile in planes (planons), along lines (lineons), and for which an isolated excitation is fully immobile (fractons). The model is exactly solvable and has zero correlation length, so we might expect it to be a fixed point of the RG, as is the case for toric code and string-net models. \cite{PhysRevLett.100.070404,PhysRevB.79.085118}

However, the X-cube model on a lattice of linear size $L$ is equivalent (under the application of a finite-depth circuit) to an X-cube model on a smaller lattice stacked with 2D toric code layers.\cite{Shirley_2018} Therefore, when trying to coarse-grain the X-cube model, non-trivial 2D layers are left behind. These layers cannot be integrated out or otherwise removed, thus preventing the model from being a fixed point of any conventional RG procedure. This behavior is closely related to the striking system-size dependence of certain properties, such as the ground state degeneracy (GSD) and the number of types of fractional excitations, both of which grow exponentially in the linear system size.\cite{Shirley_2018,Shirley_2019_excitation} Similar phenomena occur in other fracton models, including Haah's cubic code\cite{HaahCode}.

It is interesting to ask whether some fracton models are fixed points of a suitably generalized RG. While there are many schemes and procedures for carrying out RG in different settings, it is important to emphasize that simply finding a new RG scheme is not enough. Instead, a more radical generalization of what we usually mean by RG is needed, because, for instance, any RG procedure that can have the fracton models as fixed points must allow for the increase / decrease in GSD and the addition / removal of fractional excitations in the process.

Along these lines, it was found the X-cube model is a fixed point of a \emph{foliated RG} procedure.\cite{Shirley_2018,Wang_2019,shirley2019twisted,Shirley_2019_checkerboard} It is helpful to recall the conventional RG procedure for gapped phases\cite{VidalRG,Xie10}, which allows, in each RG step, for continuous deformations of the Hamiltonian that keep the gap open, and for the addition/removal of trivial gapped systems (those whose ground state is a product state). In the foliated RG, one also allows addition or removal of decoupled, gapped $2$D systems. Such $2$D systems can be topologically ordered and thus carry non-trivial GSD and fractional excitation types, hence allowing for these properties to change under RG. In the case of the X-cube model, we can remove 2D toric code layers under the foliated RG, thus making the model into a fixed point. More generally, a large class of type-I fracton models\cite{Sagar16} -- those where some of the fractional excitations are mobile -- are fixed points of the foliated RG.

The foliated RG leads to the closely related notion of \emph{foliated fracton phases.}\cite{Shirley_2019_excitation,Shirley_2019} Foliated fracton phases, which we define in Appendix~\ref{app:foliated-phases}, are a coarser equivalence relation on ground states than ordinary phases, and each foliated fracton phase contains a fixed point of the foliated RG. This fixed point captures certain universal properties that are the same everywhere in the foliated phase, and these properties are referred to as \emph{foliated fracton order}. When a model belongs to a foliated fracton phase, it is a convenient shorthand terminology to refer to the model as being foliated. 

An interesting type-I fracton model that has not been investigated from this perspective is the Ising cage-net model.\cite{Prem_2019} The Ising cage-net model is very similar to the X-cube model in many ways. Both are exactly solvable models that can be obtained from a coupled layer construction, based on toric code layers in the X-cube case,\cite{MaLayers,SagarLayers} and doubled-Ising string-net layers in the cage-net case.\cite{Prem_2019} Both have fracton excitations that are created at the corners of a rectangular membrane operator. Both have lineon excitations (abelian in the X-cube model and non-abelian in the cage-net model) that move in the $x$, $y$ and $z$ directions. Both have other planon excitations that move in $xy$, $yz$ or $zx$ planes.

Despite these similarities, it has not been clear whether the Ising cage-net model is foliated in the sense defined above. It is important to emphasize that, while both involve a layer structure, the coupled-layer constructions of X-cube and cage-net models are very different from foliated RG and from the notion of foliated fracton phases. In particular, there is no obvious relationship between whether a model can be obtained by a coupled-layer construction and whether it is foliated. By analogy with the X-cube model, it is natural to guess that the Ising cage-net model is a foliated RG fixed point upon adding/removing doubled-Ising string-net layers. However, this cannot be the case, because the doubled-Ising string-net model contains non-abelian excitations with quantum dimension $\sqrt{2}$, while the cage-net model has excitations with integer quantum dimension only.\cite{Prem_2019} While this argument does not rule out the possibility of a foliated RG fixed point with other 2D topological states as resources, in fact the Ising cage-net model is not foliated. This can be seen by studying the model's GSD, which has been computed by some of the authors in a separate paper.\cite{GSD_IsCN} It is found that the GSD does not grow by integer multiples when the system size grows by unity in the $x$, $y$ or $z$ directions.

The question is then open again: can we think of the Ising cage-net model as a fixed point of a suitably generalized RG? More specifically, can the foliated RG be generalized somehow to include the Ising cage-net model? In fact, we argue in this paper that the foliated RG \emph{should} be extended, independent of the Ising cage-net example. We do this by re-examining foliated RG from two complementary perspectives, one based on planon condensation, and the other based on quantum circuits, and point out that in both these pictures, the foliated RG has unnatural restrictions. These observations lead us to a generalized foliated RG under which, remarkably, the Ising cage-net model is a fixed point.

\begin{figure}[h]
    \centering
    \includegraphics[scale=1]{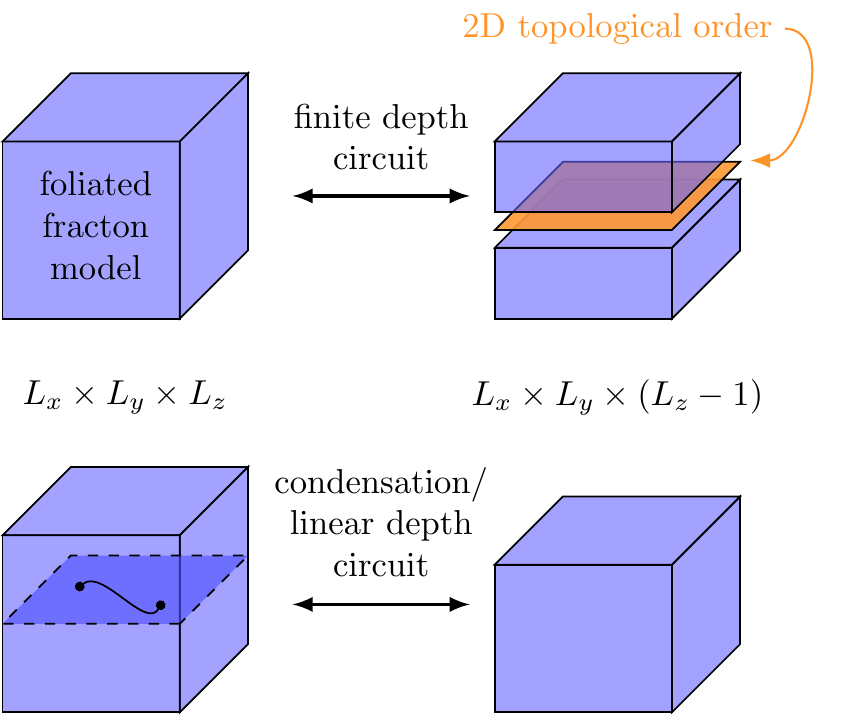}
    \captionsetup{justification=Justified}
    \caption{Top: the foliated RG scheme, where a layer of topologically ordered state (shown in orange) can be added into or removed from a foliated fracton model via a finite depth circuit. Bottom: generalized foliated RG scheme realized by condensation of bosonic planons or a sequential linear depth circuit around the plane.}
    \label{fig:foliation}
\end{figure}

The generalized foliated RG can be carried out either by condensing or uncondensing bosonic planon excitations supported near a 2D plane, or by acting with a quantum circuit, supported near a 2D plane, whose depth scales with the linear size of the plane. We show that either of these operations can be used to decrease or increase the system size of the Ising cage-net model, which is thus a generalized foliated RG fixed point. The two apparently different ways of carrying out the generalized foliated RG are closely related, through a connection that we explain between anyon condensation and a class of linear depth circuits that we refer to as \emph{sequential circuits}. 

We note that the original foliated RG arises as a special case of the generalized procedure introduced here. In particular, for the X-cube model, instead of decoupling a toric code layer and removing it to decrease system size, we can condense the bosonic planon that effectively comes from the toric code layer (either $e$ or $m$), which has the same effect as removing the layer. Alternatively, we can act with a certain linear-depth circuit (more specifically, a sequential circuit) whose effect is to condense the same bosonic planon. Therefore, we can use generalized foliation to study the X-cube model, the Ising cage-net model and many other type-I fracton models within a single framework. Just as foliated RG comes with the notion of foliated fracton phases and foliated fracton order, we expect that the generalized foliated RG comes with corresponding notions of generalized foliated fracton phases and generalized foliated fracton order. It will be interesting to study these notions in future work.

The paper is structured as follows: In Sec.~\ref{sec:foliation_Xc}, we review the original foliated RG by focusing on the X-cube model. In Sec.~\ref{sec:ICN}, we review the Ising cage-net model, which is not foliated according to the original scheme.  Section~\ref{sec:gfoliation} then briefly points out some unnatural restrictions within the original foliated RG, and proposes a generalized foliated RG where these restrictions are removed.  In Sec.~\ref{sec:RG_cond}, we show that the Ising cage-net model is foliated in terms of a generalized foliated RG defined by planon condensation. Then, in Sec.~\ref{sec:RG_circ}, we demonstrate that the generalized foliated RG  can also be implemented by a planar linear depth circuit. The linear depth circuit has a special structure, and we dub it a sequential circuit; in Sec.~\ref{sec:condvcirc} we show how the sequential circuit we use is closely related to the condensation of planons via gapped boundaries. Finally, in Sec.~\ref{sec:summary}, we conclude with a brief discussion on the implications of and outlook for the generalized foliated RG.

\section{Foliation in X-cube}
\label{sec:foliation_Xc}

Before our discussion of the `generalized foliation', it is instructive to review the original notion of foliation and see how the corresponding RG procedure is carried out for the X-cube. The X-cube model has a foliated structure, where layers of the toric code can be added to or removed from the X-cube via a finite depth circuit $\mathcal{S}$.\cite{Shirley_2018} Given an X-cube ground state $\ket{\Psi_\text{X.C.}}$ of the system size $L_x \times L_y \times L_z$ and a toric code ground state $\ket{\Psi_\text{T.C.}}$, $\mathcal{S}$ yields a $\ket{\Psi_\text{X.C.}}$ of the size $L_x \times L_y \times (L_z+1)$. In rest of this section, we review the finite depth circuit $\mathcal{S}$ on the three-torus. 

\begin{figure}[h]
    \centering
    \begin{subfigure}[b]{0.33\textwidth}
        \centering
        \includegraphics[scale = 0.12]{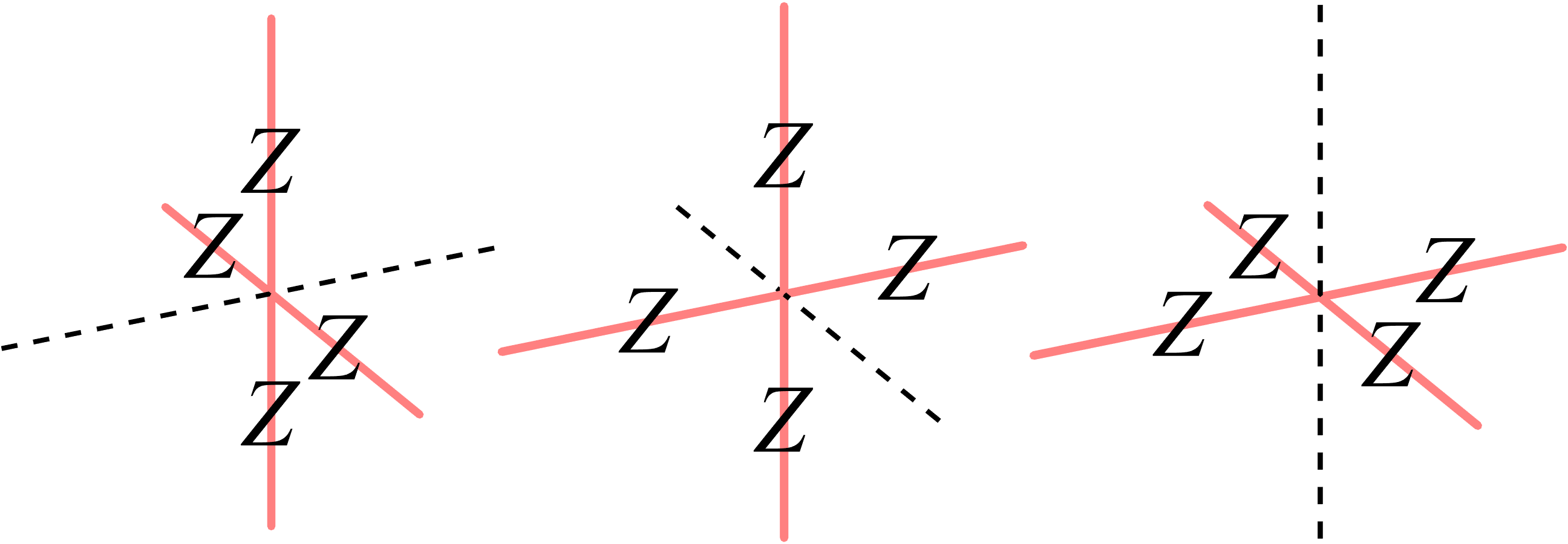}
        \caption{}
        \label{fig:X_cube_vertex_terms}
    \end{subfigure}
    \begin{subfigure}[b]{0.13\textwidth}
        \centering
        \includegraphics[scale = 0.11]{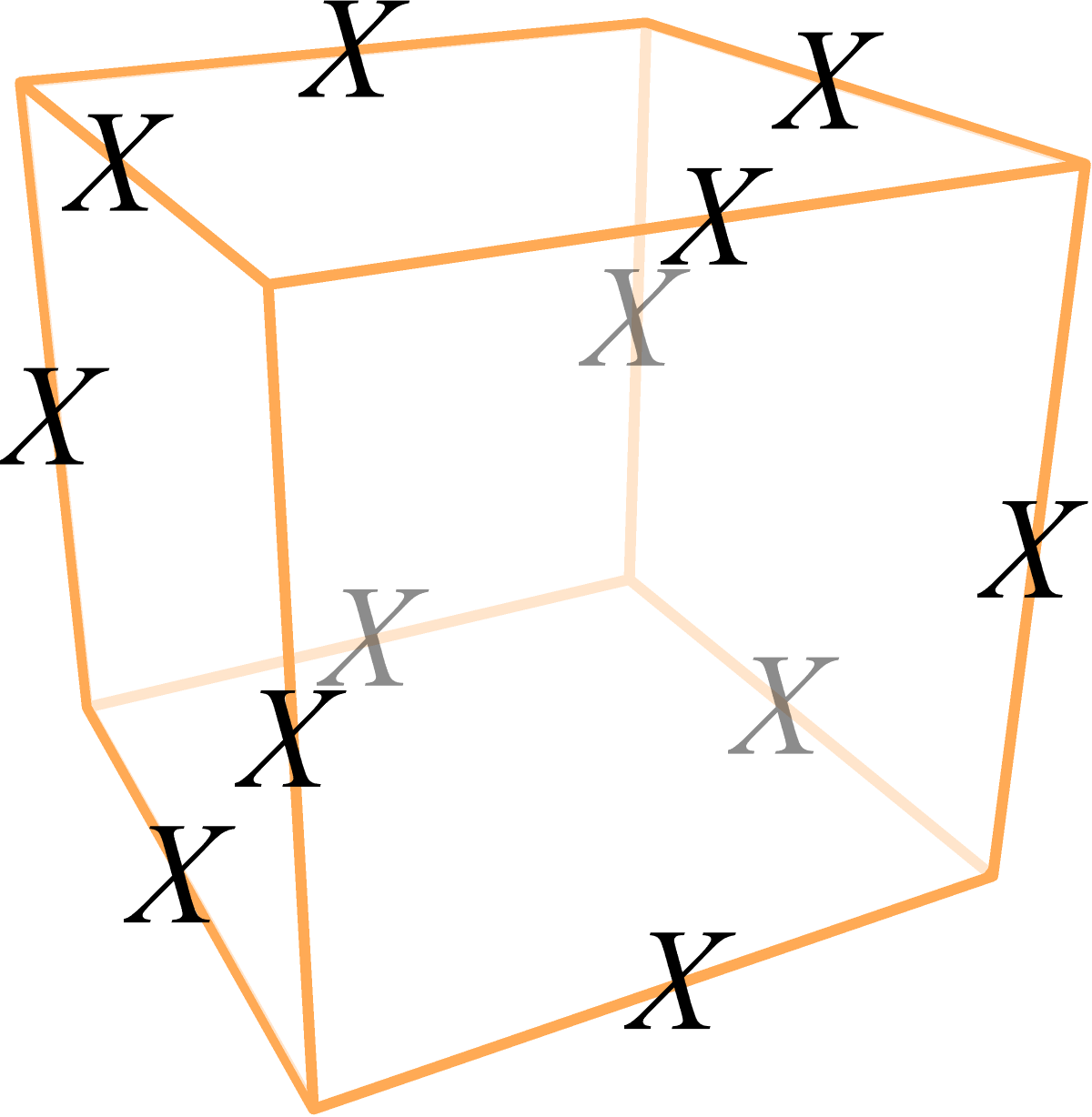}
        \caption{}
        \label{fig:X_cube_cube_terms}
    \end{subfigure}
    \captionsetup{justification=Justified}
    \caption{(a) The three types of vertex terms in the X-cube Hamiltonian $A^x_v$, $A^y_v$, and $A^z_v$, which are tensor products of Pauli-$Z$ operators. (b) The cube term $B_c$.}
    \label{fig:X_cube_Hamiltonian}
\end{figure}

\begin{figure}[h]
    \centering
    \begin{subfigure}[b]{0.23\textwidth}
        \centering
        \includegraphics[scale = 0.12]{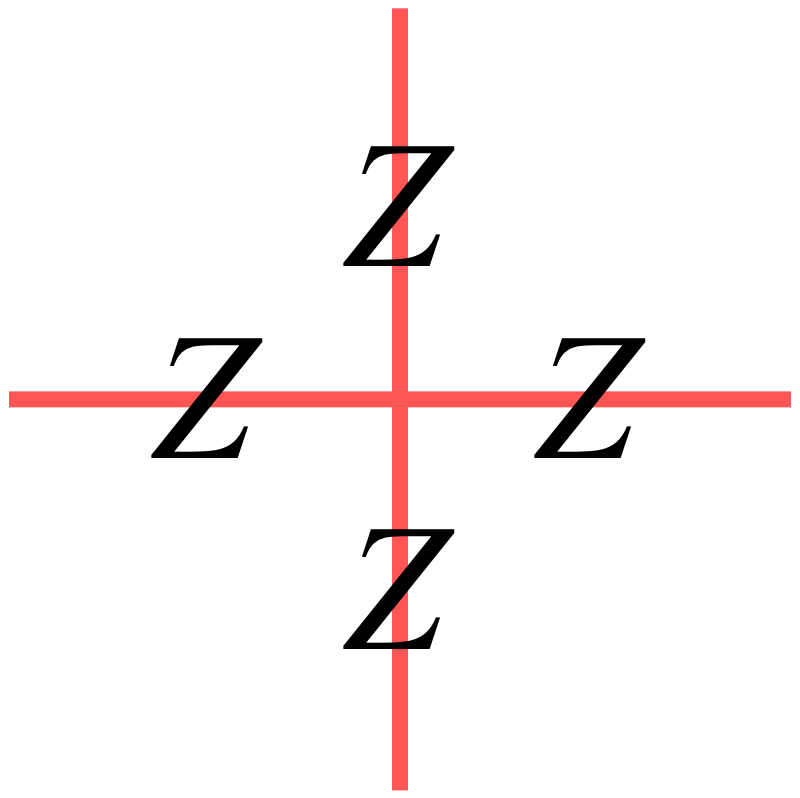}
        \caption{}
        \label{fig:Toric_code_vertex_term}
    \end{subfigure}
    \begin{subfigure}[b]{0.23\textwidth}
        \centering
        \includegraphics[scale = 0.12]{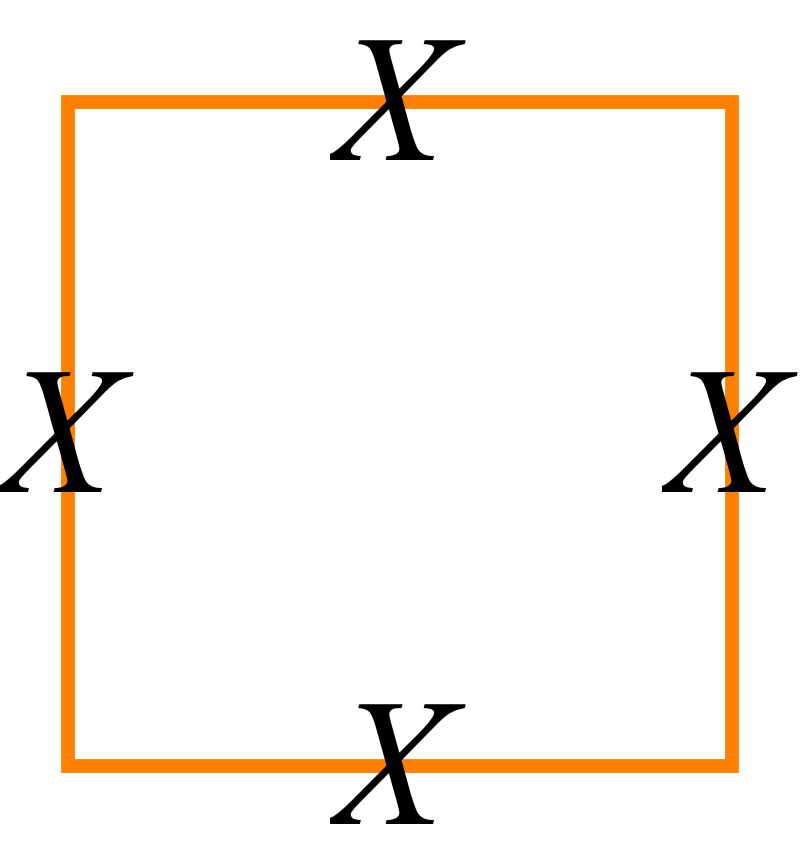}
        \caption{}
        \label{fig:Toric_code_plaquette_term}
    \end{subfigure}
    \captionsetup{justification=Justified}
    \caption{(a) The vertex term $Q_v$ in the toric code Hamiltonian. (b) The plaquette term $B_p$. }
    \label{fig:Toric_code_Hamiltonian}
\end{figure}

Let us consider the X-cube Hamiltonian defined on a cubic lattice on the three-torus; and one copy of the toric code Hamiltonian defined on a square lattice on the two-torus. For both models, the local qubit DOFs are placed on the edges. The X-cube Hamiltonian\cite{Sagar16} 
\begin{equation}
    H_\text{X.C.} = - \sum_v \left(A^x_v + A^y_v + A^z_v \right) - \sum_c B_c
\end{equation}
contains three types of vertex terms $A^x_v$, $A^y_v$, and $A^z_v$; and one type of cube term $B_c$, as shown in Fig.~\ref{fig:X_cube_Hamiltonian}. The toric code Hamiltonian\cite{KitaevToricCode}
\begin{equation}
    H_\text{T.C.} = - \sum_v Q_v - \sum_p B_p
    \label{eq:Toric_code_Hamiltonian}
\end{equation}
is a sum of local terms as shown in Fig.~\ref{fig:Toric_code_Hamiltonian}.

\begin{figure}[ht]
    \centering
    \includegraphics[scale = 0.085]{ 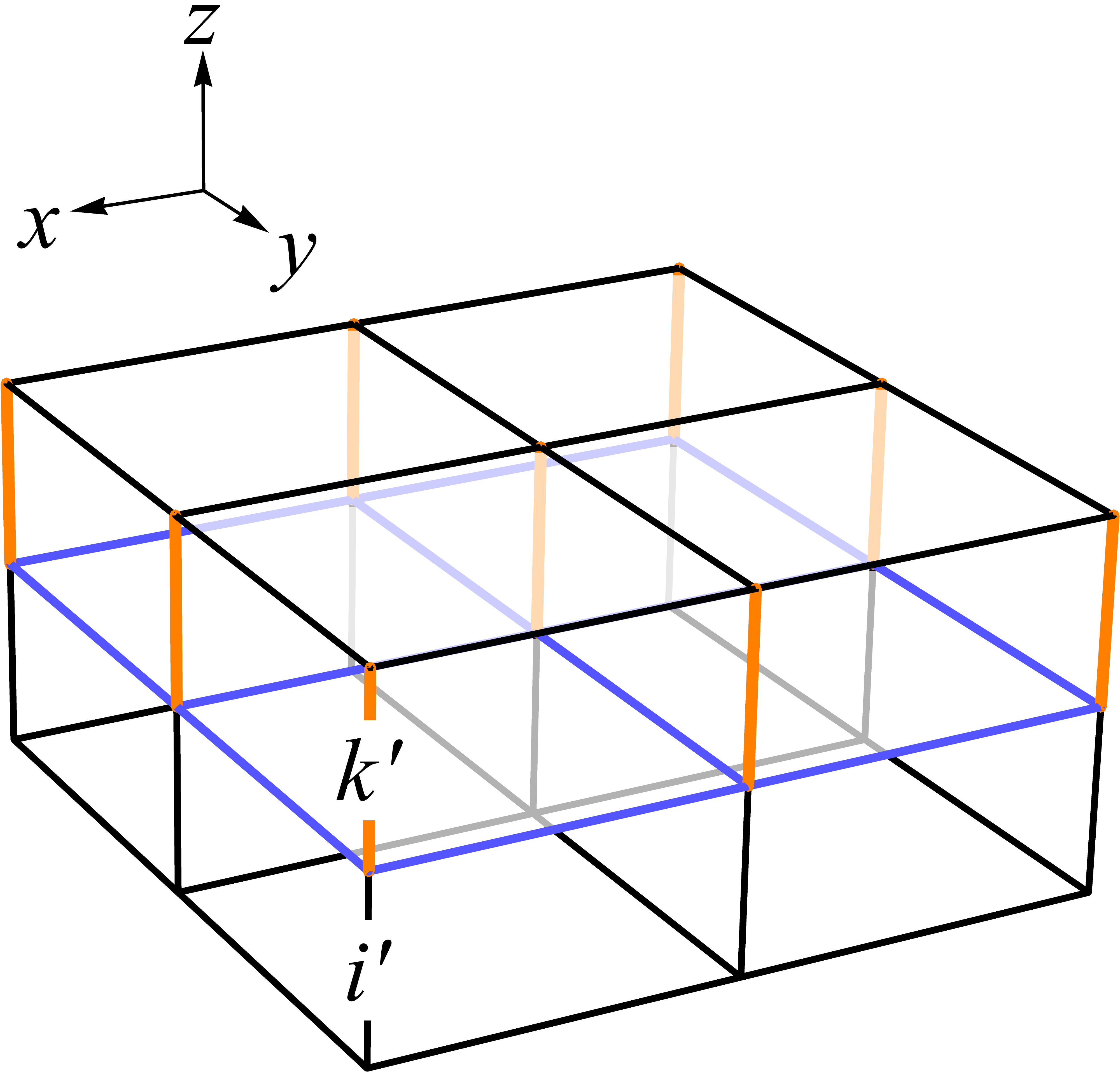}
    \captionsetup{justification=Justified}
    \caption{The insertion of a layer of toric code living on an $xy$-plane (blue colored square lattice) into a cubic lattice, which hosts the X-cube. The inserted layer bisects an edge $i$ near the inserted plane into edges labeled by $i'$ and $k'$. For every bisected edge, the X-cube Hamiltonian is modified by replacing $Z_i \to Z_{i'}$ and $X_i \to X_{i'}$. The new edges $k'$ are product states with the Hamiltonian of $H_0 = -\sum_{\{k'\}}Z_{k'}$. }
    \label{fig:Cubic_lattice_added_plane}
\end{figure}

To construct the circuit, we first insert a decoupled toric code into the X-cube. As depicted in Fig.~\ref{fig:Cubic_lattice_added_plane}, when the inserted toric code lies in the $xy$-plane, it bisects the $z$-direction edges in the X-cube model, thus creating new qubit edges $k'$ colored in orange. These new $k'$ edges are added to the system as product states whose Hamiltonian is chosen to be $H_0 = - \sum_{\{k'\}}Z_{k'}$. For each bisected edge $i$ in the X-cube Hamiltonian, we substitute $Z_i \to Z_{i'}$ and $X_i \to X_{i'}$.

\begin{figure}[ht]
    \centering
    \begin{subfigure}[b]{0.23\textwidth}
        \centering
        \includegraphics[scale = 0.037]{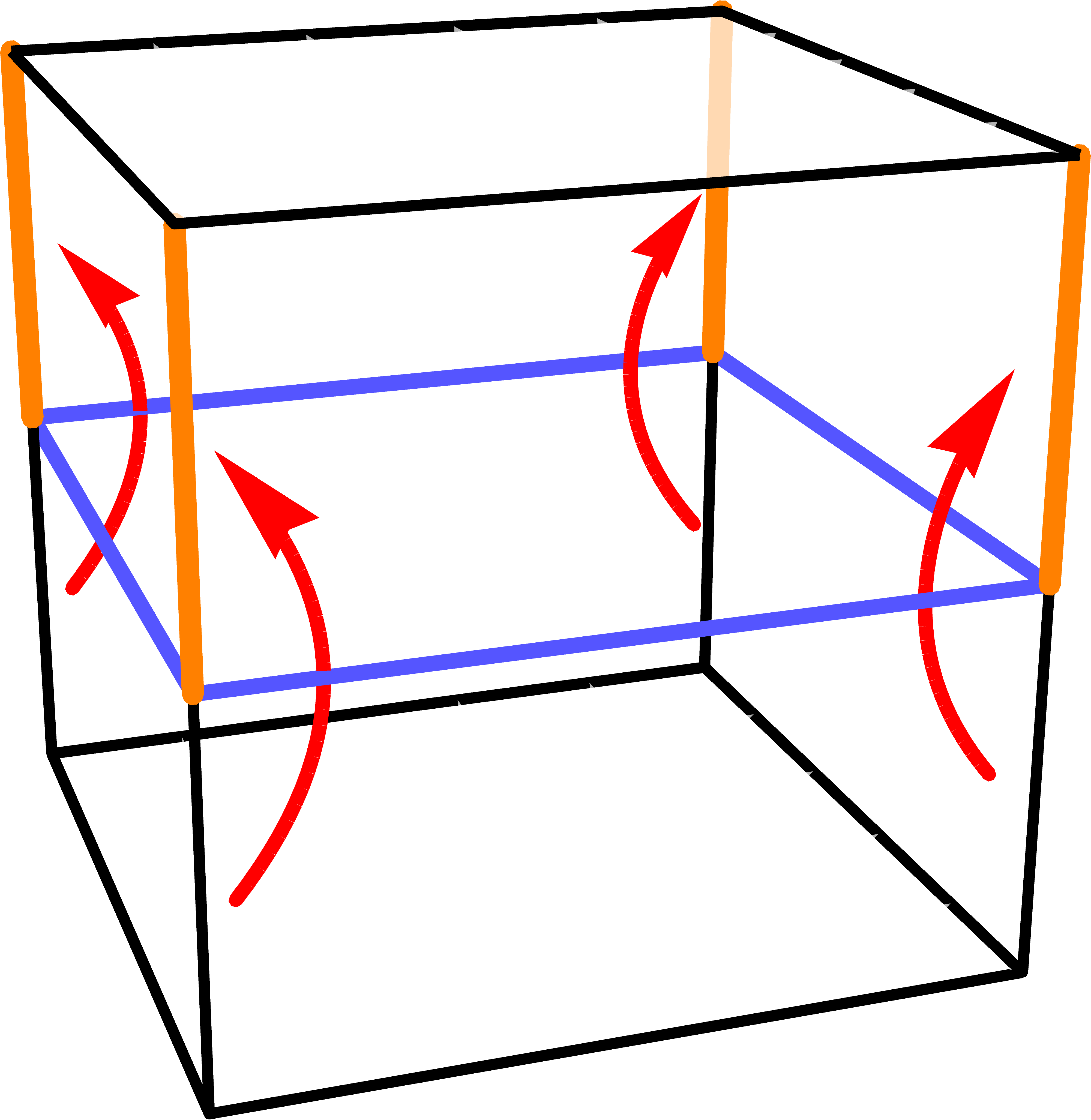}
        \caption{}
        \label{fig:ERG_X_cube_sub_A}
    \end{subfigure}
    \begin{subfigure}[b]{0.23\textwidth}
        \centering
        \includegraphics[scale = 0.036]{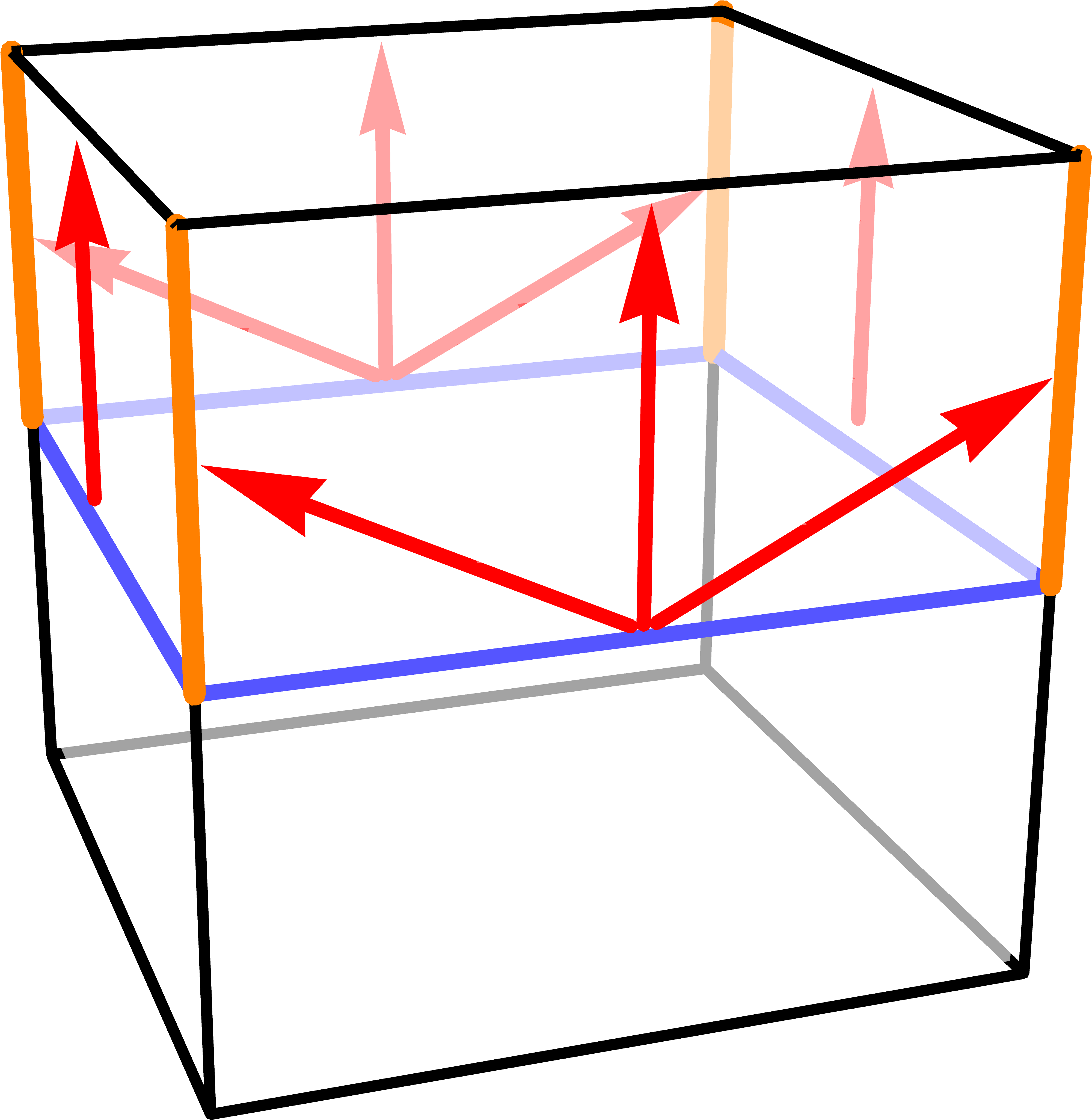}
        \caption{}
        \label{fig:ERG_X_cube_sub_B}
    \end{subfigure}
    \captionsetup{justification=Justified}
    \caption{An illustration of the finite depth circuit $\mathcal{S}=\mathcal{S}_2\mathcal{S}_1$. (a) The action of the circuit $\mathcal{S}_1$ when focus on an elementary cube of the original cubic lattice. The arrows, representing the CNOT gates, point from the controlled qubits to the targets. (b) $\mathcal{S}_2$'s action viewed at a cube.}
    \label{fig:ERG_X_cube}
\end{figure}

The circuit $\mathcal{S}$ is a product of two finite depth circuits $\mathcal{S}_2$ and $\mathcal{S}_1$, $\mathcal{S}=\mathcal{S}_2\mathcal{S}_1$. Each is a product of the controlled-NOT (CNOT) gates. The circuit $\mathcal{S}_1$ acts on the edges of the modified X-cube Hamiltonian, as shown in Fig.~\ref{fig:ERG_X_cube_sub_A}. Every CNOT gate in $\mathcal{S}_1$ has an $i'$ edge serving as the controlled qubit and the corresponding $k'$ edge as the target. On the other hand, $\mathcal{S}_2$ acts on both edges of the X-cube and those of the toric code. Every edge of the toric code serves as the controlled qubit for the CNOT gates whose targets are edges in the modified X-cube. An illustration of $\mathcal{S}_2$ is given in Fig.~\ref{fig:ERG_X_cube_sub_B}. The CNOT gate, acting by conjugation, has the actions of
\begin{equation}
    \begin{aligned}
        &ZI \mapsto ZI, & IZ \leftrightarrow ZZ, \\
        &XI \leftrightarrow XX, & IX \mapsto IX,
    \end{aligned}
\end{equation}
where the first qubit is the control and the second is the target. All the CNOT gates in $\mathcal{S}_1$ or $\mathcal{S}_2$ commute with each other. Therefore, $\mathcal{S}$ is a finite depth circuit. By direct computation, we see that
\begin{equation}
    \mathcal{S}\left(\tilde{H}_\text{X.C.}^{(L_x,L_y,L_z)}+H_\text{T.C.} + H_0\right)\mathcal{S}^\dagger \cong H_\text{X.C.}^{(L_x,L_y,L_z+1)},
\end{equation}
where $\tilde{H}_\text{X.C.}$ is the modified X-cube Hamiltonian, and the symbol $\cong$ denotes that the L.H.S. and the R.H.S. share the same ground space.

\section{Ising cage-net}
\label{sec:ICN}

In this section, we review the basic definition and properties of the Ising cage-net model. 

The Ising cage-net is an exactly solvable model obtained from the coupled layer construction\cite{Prem_2019}, in which decoupled layers of the doubled-Ising string-net\cite{Stringnet,ExtendedStringnet_Hu,Burnell_doubleIsing,SN_thorough,TianLan_stable,ExtendedStringnet_Hu} are coupled together through the particle-loop (p-loop) condensation.

\begin{figure}[ht]
    \centering
    \includegraphics[scale = 1]{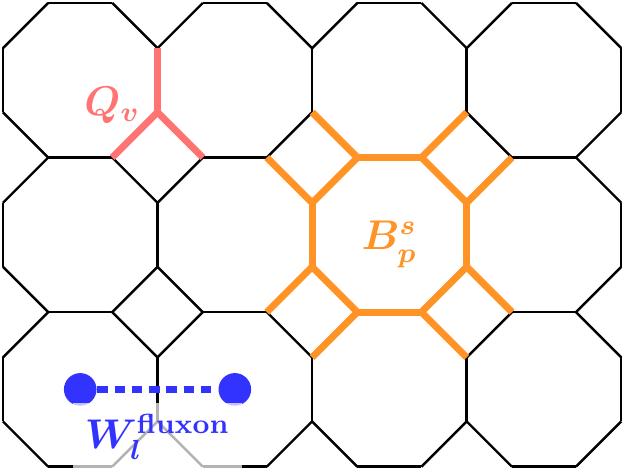}
    \captionsetup{justification=Justified}
    \caption{A square-octagon lattice, where the doubled-Ising string-net is defined. Each edge has a local Hilbert space with a basis $\{\ket{0},\ket{1},\ket{2}\}$. $Q_v$ is defined for every trivalent vertex, and $B_p = \sum_{s=0}^2 (d_s/ D) B^s_p$ is defined for each square and octagonal plaquette. The string-operator for a fluxon excitation $W^\text{fluxon}_l$ violates the two $B_p$ terms containing the edge $l$ and no $Q_v$ term anywhere.}
    \label{fig:Square_octagon_lattice}
\end{figure}

\begin{figure}[h]
    \centering
    \includegraphics[scale = 0.12]{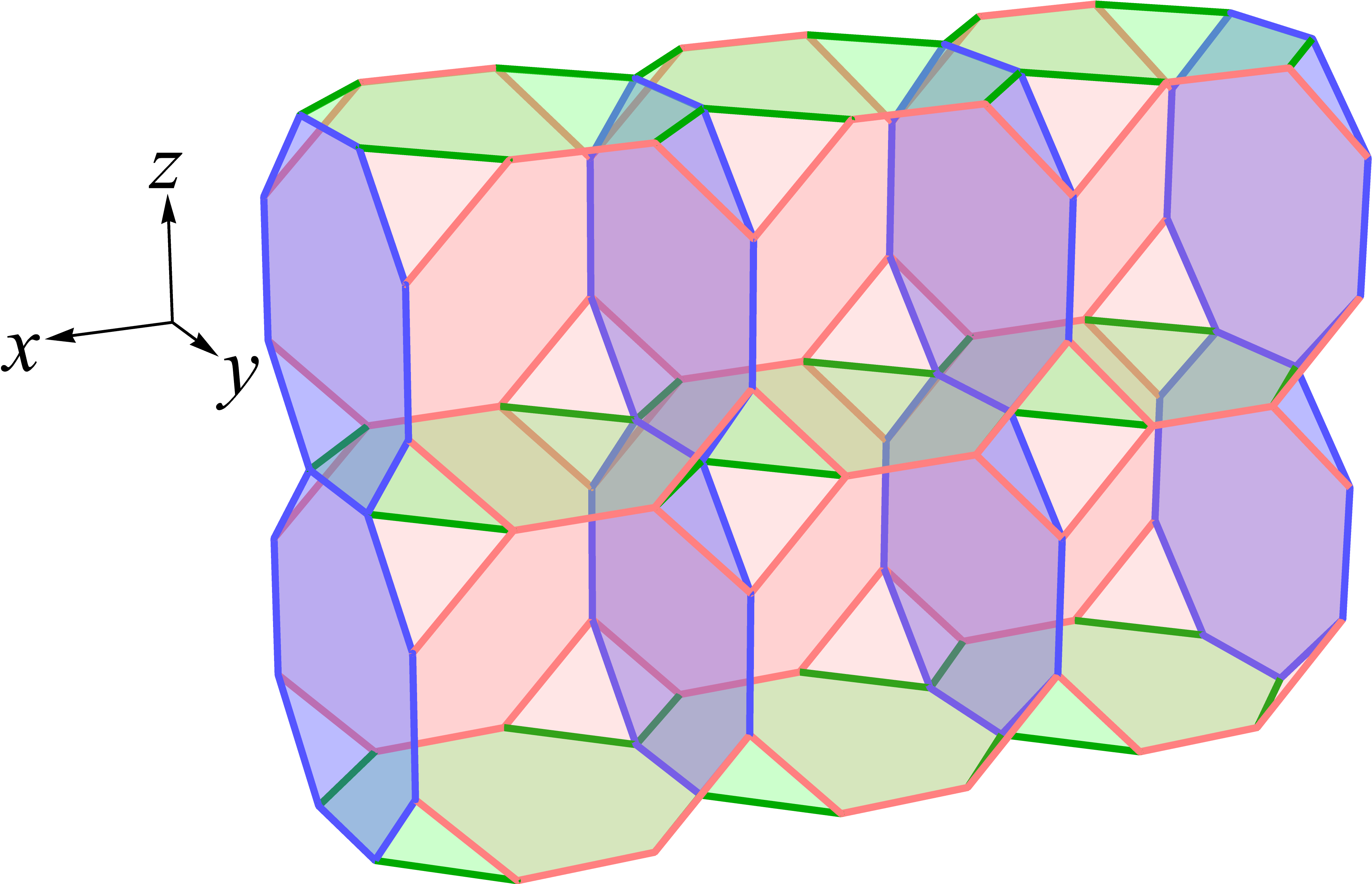}
    \captionsetup{justification=Justified}
    \caption{A truncated cubic lattice. It is formed by intersecting layers of the square-octagon lattice. Every cube has six octagonal faces. At the corners of each cube are octahedrons (see Fig.~\ref{fig:Lineons_turn_corner}). The edges $l$, parallel to $\mu$ direction for $\mu \in \{x,y,z\}$, are called the $\mu$-principal edges, which are denoted by $l_\mu$. For the system of decoupled layers, a $\mu$-principal edge has a nine-dimensional local space given by the tensor product of $( \,\text{span}_\mathbb{C}\{\ket{0},\ket{1},\ket{2}\} \,)^{\otimes 2} $.}
    \label{fig:Truncated_cubic_lattice2}
\end{figure}

Specifically, we take three stacks of the doubled-Ising string-net defined on a square-octagon lattice (see Fig.~\ref{fig:Square_octagon_lattice}), and stack them together to form a truncated cubic lattice, as shown in Fig.~\ref{fig:Truncated_cubic_lattice2}. Each of the six faces of a cube is an octagonal plaquette. We call an edge $l$, parallel to the $\mu$-direction for $\mu \in \{x,y,z\}$, a $\mu$-principal edge, and denote it by $l_\mu$.

As a 2D lattice model, the doubled-Ising string-net is built from the Ising unitary modular tensor category\cite{KitaevAnyons,rowell_classification_2009}, which consists of an index set $\{0,1,2\}$ and a set of symbols $(\delta_{ijk}, d_s, F^{ijm}_{kln}, R^{ij}_{k})$. The model has a three-dimensional local Hilbert space of $\text{span}_\mathbb{C}\{\ket{0},\ket{1},\ket{2}\}$ for each edge of the square-octagon lattice. The states $\ket{0}$, $\ket{1}$, $\ket{2}$ are dubbed as $0$-string, $1$-string, and $2$-string respectively. The commuting projector Hamiltonian
\begin{equation}
    H_\text{D.I.} = - \sum_v Q_v - \sum_p B_p
    \label{eq:Doubled_Ising_Hamiltonian}
\end{equation}
consists of the vertex projector $Q_v$ and the plaquette projector $B_p = \sum_{s=0}^2 (d_s/ D) B^s_p$ (see Fig.~\ref{fig:Square_octagon_lattice}). The symbol $d_s$ takes values in $d_0 = d_2 = 1$, and $d_1 = \sqrt{2}$. $D = \sum_s (d_s)^2$ is \textit{the total quantum dimension} of the model. $Q_v$'s action is defined by
\begin{equation}
    \vcenter{\hbox{\includegraphics[scale=1]{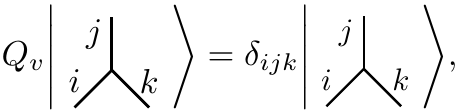}}}
    \label{eq:DI_vertex_rule}
\end{equation}
where the symbol $\delta_{ijk}$ is symmetric under permutation of its indices. The non-zero elements are $\delta_{000} = \delta_{011} = \delta_{211} = \delta_{022} = 1$, up to permutations. The subspace where all the vertex terms $Q_v$ are satisfied is called \textit{the stable vertex subspace} $\mathcal{H}_{Q_v}^\text{D.I.}$.\cite{TianLan_stable} The plaquette operator $B^s_p$'s action are evaluated by the graphical rules, which are defined via the $d$- and $F$-symbols (Appendix \ref{sec:Review_SN}). $B^s_p$ acts on a plaquette by fusing a loop of $s$ into the edges as, for example,
\begin{equation}
    \vcenter{\hbox{\includegraphics[scale = 1]{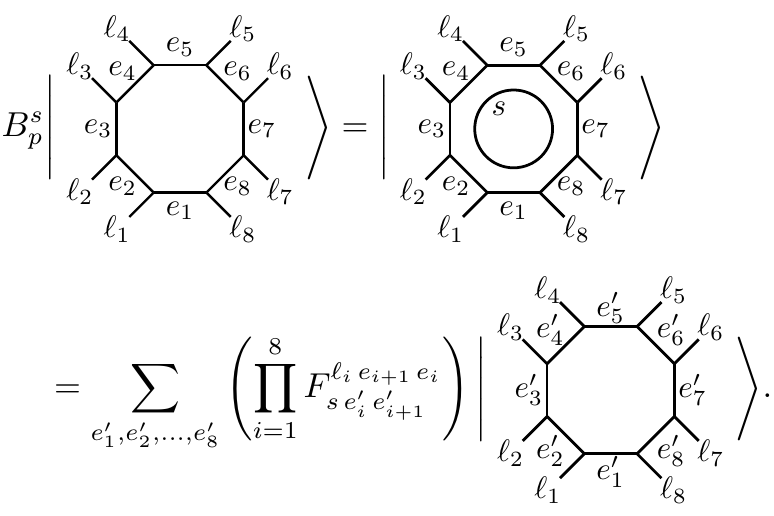}}}
    \label{eq:Double_Ising_Bsp_def}
\end{equation}
For every ground state $\ket{\Psi_\text{D.I.}}$, which is a superposition of different configurations of closed loops satisfying $Q_v$ at each vertex, $B^s_p$ acts as
\begin{equation}
    B^s_p \ket{\Psi_\text{D.I.}} = d_s \ket{\Psi_\text{D.I.}}.
\end{equation}
Moreover, the $B^s_p$ operators form a commutative fusion algebra of
\begin{equation}
    B^i_p B^j_p = \sum_{k=0}^2 \delta_{ijk}B^k_p.
    \label{eq:DoulbeIsing_PlaquetteAlgebra}
\end{equation}

The doubled-Ising string-net has nine topological excitations $\{\mathbbm{1}, \psi, \bar{\psi}, \sigma, \bar{\sigma}, \sigma \bar{\psi}, \psi \bar{\sigma}, \sigma \bar{\sigma}, \psi \bar{\psi}\}$. In terms of the theory of anyons, these excitations come from a copy of the chiral Ising anyon $\{\mathbbm{1}, \sigma, \psi\}$, and an anti-chiral copy $\{\mathbbm{1},\bar{\sigma},\bar{\psi}\}$. The fusion rules for the chiral Ising anyon are 
\begin{equation}
    \begin{tabular}{l|ccc} 
      $\times$ & $\mathbbm{1}$ & $\sigma$ & $\psi$ \\
      \hline
      $\mathbbm{1}$ & $\mathbbm{1}$ & $\sigma$ & $\psi$\\
      $\sigma$ & $\sigma$ & $\mathbbm{1}+\psi$ & $\sigma$\\
      $\psi$ & $\psi$ & $\sigma$ & $\mathbbm{1}$
    \end{tabular}
    \label{eq:Ising_category_fusion_rule}
\end{equation}
The anti-chiral Ising anyon obeys the same fusion rules; we simply replace the anyon labels above with the barred version. Among the nine excitations, the non-abelian $\sigma \bar{\sigma}$ and the abelian $\psi\bar{\psi}$ are bosons. They are also the only non-trivial pure fluxon excitations. A \textit{fluxon} excitation violates exactly one $B_p$ term and none of the $Q_v$ terms. A fluxon string-operator $W^\text{fluxon}_l$ creates the fluxon and its anti-particle on the two adjacent plaquettes sharing the edge $l$ (see Fig.~\ref{fig:Square_octagon_lattice}). In particular, the $\psi\bar{\psi}$ has a string-operator
\begin{equation}
    W^{\psi\bar{\psi}}_{l} = (-1)^{n_1(l)},
\end{equation}
where $n_1(l) = 1$ if the edge $l$ is in the state $\ket{1}$, and $n_1(l) = 0$ otherwise.

\begin{figure}[h]
    \centering
    \includegraphics[scale = 0.07]{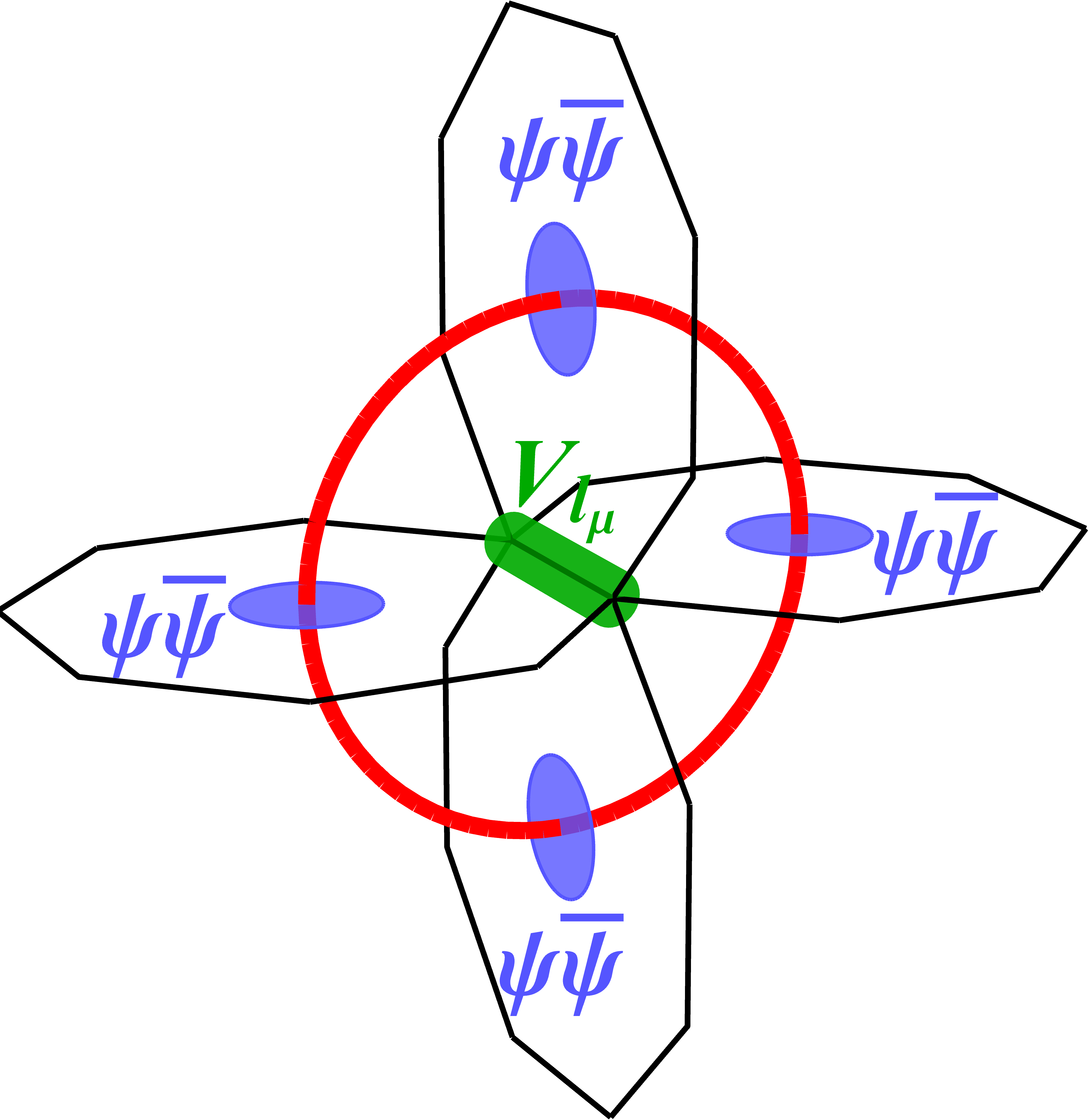}
    \captionsetup{justification=Justified}
    \caption{An elementary $\psi\bar{\psi}$ particle-loop (p-loop), the red loop, created by the coupling operator $V_{l_\mu}$ shown by the green tube. We represent a flux by a line segment normal to the hosting plaquette. Joining the segments together, we have the red loop.}
    \label{fig:p-loop}
\end{figure}

To couple the stacks of the doubled-Ising string-net layers together, we condense the $\psi \bar{\psi}$ p-loop. The p-loop condensation is a direct generalization of anyon condensation in 2D. We review some simple examples of anyon condensation in Appendix~\ref{sec:Anyon_condensation_review}. Illustrated in Fig.~\ref{fig:p-loop} is the smallest $\psi\bar{\psi}$ p-loop created by the coupling operator
\begin{equation}
    V_{l_\mu} = W^{(\psi\bar{\psi})_{\mu\nu}}_{l_\mu}W^{(\psi\bar{\psi})_{\mu\rho}}_{l_\mu},
    \label{eq:V_lmu_coupling_operator}
\end{equation}
which is a product of $\psi\bar{\psi}$ string-operators, from the $\mu\nu$- and $\mu\rho$-planes, acting on the edge $l_\mu$. We add $-V_{l_\mu}$ for every principal edge to the Hamiltonian of the decoupled layers. $-V_{l_\mu}$ penalizes the presence of the states $\ket{01}$, $\ket{10}$, $\ket{21}$, and $\ket{12}$ on $l_\mu$. Using the Brillouin-Wigner degenerate perturbation theory and treating doubled-Ising string-nets as perturbations, we arrive at the Ising cage-net. Hence, on a principal edge, the Ising cage-net has a five-dimensional local Hilbert space of $\text{span}_\mathbb{C}\{\ket{00},\ket{11},\ket{02},\ket{20},\ket{22}\}$. Other edges are unchanged.

The Ising cage-net has a commuting Hamiltonian of
\begin{equation}
        H_\text{I.C.} = - \sum_{{\mu\nu},v} A^{\mu\nu}_v - \sum_{p_\text{s}}B_{p_\text{s}} - \sum_{p_\text{o}} \frac{1}{2}\left(B^0_{p_\text{o}}+B^2_{p_\text{o}}\right) - \sum_\text{c} B_\text{c},
        \label{eq:Ising_CN_Hamiltonian}
\end{equation}
where $A^{\mu\nu}_v$ is the vertex projector in a $\mu\nu$-plane; $B_{p_\text{s}}$ is the doubled-Ising string-net plaquette projector for a square plaquette; $\frac{1}{2}\left(B^0_{p_\text{o}}+B^2_{p_\text{o}}\right)$ is a plaquette term associated with each octagonal plaquette $p_\text{o}$; and 
\begin{equation}
    B_\text{c} = \prod_{{p_\text{o}} \in \text{c}}\frac{\sqrt{2}}{2} B^1_{p_\text{o}}
\end{equation}
is the cube term. The vertex term acts as
\begin{equation}
    A^{\mu\nu}_v
    \Bigg \vert
    \begin{tikzpicture}[baseline={([yshift=-.5ex]current bounding box.center)}, every node/.style={scale=1}]
        \begin{scope}
            \pgfmathsetmacro{\lineW}{0.8}
            \pgfmathsetmacro{\radius}{0.7}
            \pgfmathsetmacro{\SquareL}{2*\radius*cos(22.5)}
            \pgfmathsetmacro{\Lcenterx}{0}
            \pgfmathsetmacro{\Lcentery}{2*\SquareL}
            \pgfmathsetmacro{\Rcenterx}{\SquareL}
            \pgfmathsetmacro{\Rcentery}{2*\SquareL}
            \pgfmathsetmacro{\upptx}{\Lcenterx + \radius*sin(67.5)}
            \pgfmathsetmacro{\uppty}{\Lcentery + \radius*cos(67.5)}
            \pgfmathsetmacro{\centralptx}{\Lcenterx + \radius*sin(112.5)}
            \pgfmathsetmacro{\centralpty}{\Lcentery + \radius*cos(112.5)}
            \pgfmathsetmacro{\Leftptx}{\Lcenterx + \radius*sin(157.5)}
            \pgfmathsetmacro{\Leftpty}{\Lcentery + \radius*cos(157.5)}
            \pgfmathsetmacro{\Rightptx}{\Rcenterx + \radius*sin(202.5)}
            \pgfmathsetmacro{\Rightpty}{\Rcentery + \radius*cos(202.5)}
            
            \draw[line width = \lineW pt] (\upptx,\uppty) -- (\centralptx,\centralpty);
            \draw[line width = \lineW pt] (\upptx+0.05,\uppty) -- (\centralptx+0.05,\centralpty);
            \draw[line width = \lineW pt] (\Leftptx,\Leftpty) -- (\centralptx,\centralpty);
            \draw[line width = \lineW pt] (\Rightptx,\Rightpty) -- (\centralptx,\centralpty);
            
            \node [] at (\Leftptx+0.2, \Lcentery+0.1) {$j$};
            \node [] at (\Rightptx-0.18, \Lcentery+0.1) {$\ell$};
            \node [] at (\Leftptx, \Lcentery-0.38) {$i$};
            \node [] at (\Rightptx, \Lcentery-0.38) {$k$};
        \end{scope}
    \end{tikzpicture}
    \Bigg \rangle
    = \delta_{ijk} \delta_{(j,\ell)}
    \Bigg \vert
    \begin{tikzpicture}[baseline={([yshift=-.5ex]current bounding box.center)}, every node/.style={scale=1}]
        \begin{scope}
            \pgfmathsetmacro{\lineW}{0.8}
            \pgfmathsetmacro{\radius}{0.7}
            \pgfmathsetmacro{\SquareL}{2*\radius*cos(22.5)}
            \pgfmathsetmacro{\Lcenterx}{0}
            \pgfmathsetmacro{\Lcentery}{2*\SquareL}
            \pgfmathsetmacro{\Rcenterx}{\SquareL}
            \pgfmathsetmacro{\Rcentery}{2*\SquareL}
            \pgfmathsetmacro{\upptx}{\Lcenterx + \radius*sin(67.5)}
            \pgfmathsetmacro{\uppty}{\Lcentery + \radius*cos(67.5)}
            \pgfmathsetmacro{\centralptx}{\Lcenterx + \radius*sin(112.5)}
            \pgfmathsetmacro{\centralpty}{\Lcentery + \radius*cos(112.5)}
            \pgfmathsetmacro{\Leftptx}{\Lcenterx + \radius*sin(157.5)}
            \pgfmathsetmacro{\Leftpty}{\Lcentery + \radius*cos(157.5)}
            \pgfmathsetmacro{\Rightptx}{\Rcenterx + \radius*sin(202.5)}
            \pgfmathsetmacro{\Rightpty}{\Rcentery + \radius*cos(202.5)}
            
            \draw[line width = \lineW pt] (\upptx,\uppty) -- (\centralptx,\centralpty);
            \draw[line width = \lineW pt] (\upptx+0.05,\uppty) -- (\centralptx+0.05,\centralpty);
            \draw[line width = \lineW pt] (\Leftptx,\Leftpty) -- (\centralptx,\centralpty);
            \draw[line width = \lineW pt] (\Rightptx,\Rightpty) -- (\centralptx,\centralpty);
            
            \node [] at (\Leftptx+0.2, \Lcentery+0.1) {$j$};
            \node [] at (\Rightptx-0.18, \Lcentery+0.1) {$\ell$};
            \node [] at (\Leftptx, \Lcentery-0.38) {$i$};
            \node [] at (\Rightptx, \Lcentery-0.38) {$k$};
        \end{scope}
    \end{tikzpicture}
    \Bigg \rangle,
\end{equation}
where we have used the doubled line to represent a principal edge in the state $\ket{j \ell}$, and
\begin{equation}
    \delta_{(j,\ell)} =
    \begin{cases}
        1 \quad &\text{for } (j,\ell) \in \mathcal{I}\\
        0 \quad & \text{otherwise}
    \end{cases}
    \label{eq:cage-net_DOF_deltafunction}
\end{equation}
with the index set $\mathcal{I} = \{(0,0),(1,1),(0,2),(2,0),(2,2)\}$.

\begin{table}[h]
\begin{center}
\begin{tabular}{c | c | c c c c }
\toprule 
\multirow{1}{*}{Mobility}& \multirow{1}{*}{Type} & \multicolumn{4}{c}{Excitations}\\
\colrule
\multirow{2}{*}{Planons} & Abelian & $(\psi\bar{\psi})_{\mu\nu}$ &  $\psi_{\mu\nu}$ & $\bar{\psi}_{\mu\nu}$\\
\cline{2-6}
 & Non-abelian  & \multicolumn{4}{c}{$(\sigma\bar{\sigma})_{\mu\nu}$} \\
\hline 
\multirow{3}{*}{Lineons} & Abelian & \multicolumn{4}{c}{---}\\
\cline{2-6}
& \multirow{2}{*}{Non-abelian} &$\sigma_{\mu\nu}\sigma_{\mu\rho}$ & $\bar{\sigma}_{\mu\nu}\sigma_{\mu\rho}$ & $\sigma_{\mu\nu}\bar{\sigma}_{\mu\rho}$ & \\
& & $\bar{\sigma}_{\mu\nu}\bar{\sigma}_{\mu\rho}$\\
\botrule
\end{tabular}
\captionsetup{justification=Justified}
\caption{Excitations in the Ising cage-net for each $\mu\nu$-plane, written in terms of the doubled-Ising excitations. Amongst, the only composite excitation, $(\psi\bar{\psi})_{\mu\nu}$, is a fracton dipole, a planon in the $\mu\nu$-plane. A lineon can only move along the line specified by the repeated position index. For example, $\sigma_{\mu\nu}\sigma_{\mu\rho}$ is mobile along a line in the $\mu$ direction. Moreover, pairs of lineons from different planes can form a lineon dipole, which is a planon.}
\label{table:Ising_CageNet_Excitations}
\end{center}
\end{table}

\begin{figure}[h]
    \centering
    \includegraphics[scale = 0.09]{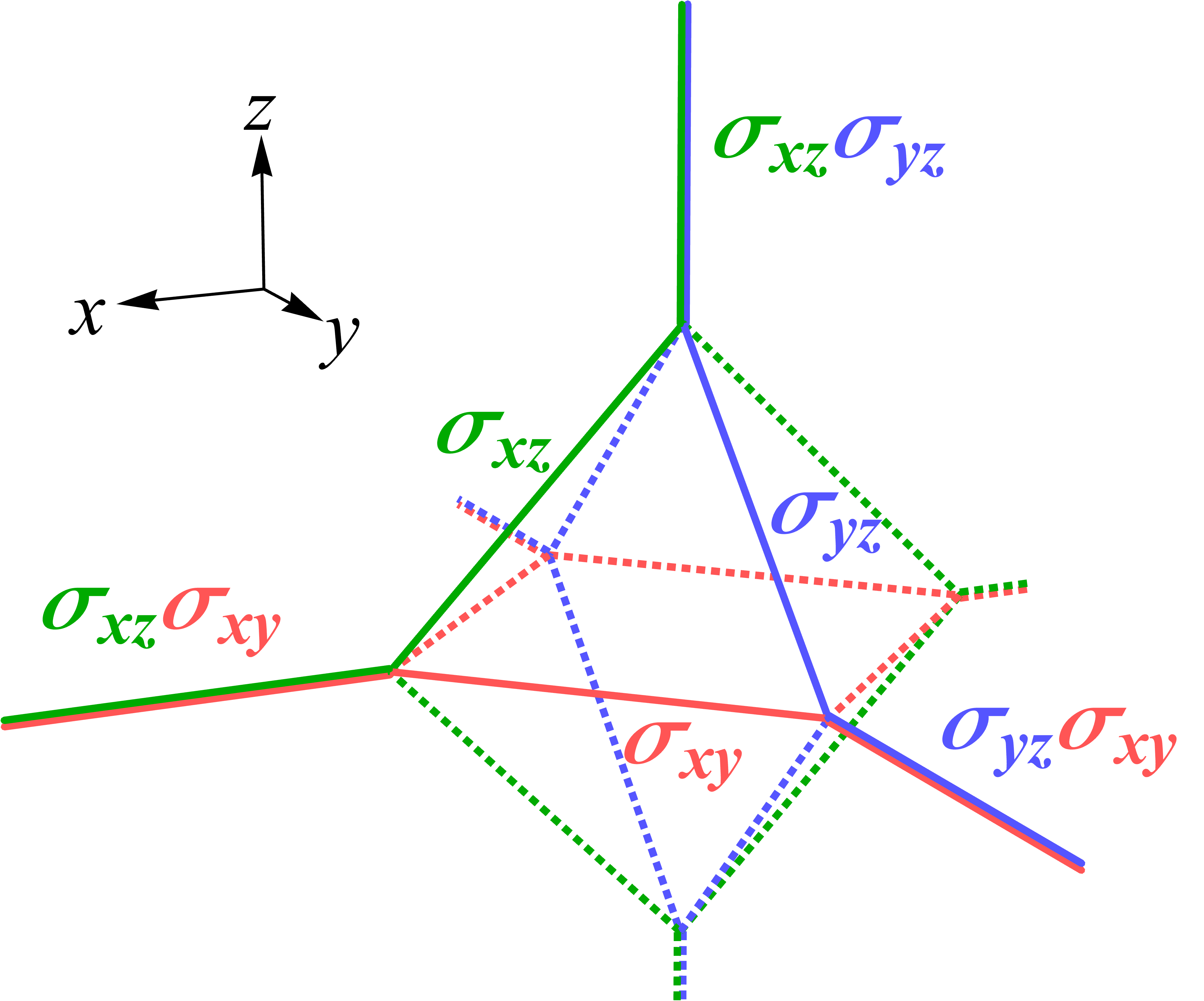}
    \captionsetup{justification=Justified}
    \caption{The vacuum fusion channel of three different lineons at an octahedron (Ref.~\onlinecite{Prem_2019}). The colored solid lines represent the labeled string-operators in the same color. The dashed lines represent edges without string-operators.}
    \label{fig:Lineons_turn_corner}
\end{figure}

The quasi-particle excitations of the Ising cage-net follow directly from the constituent doubled-Ising layers. Excitations that survive the condensation must have string-operators that commute with $V_{l_\mu}$. Thus, some of the doubled-Ising planons must now exist together with some other doubled-Ising planons from a perpendicular plane, hence the emergence of lineons. A lineon can turn at a corner and become another lineon at the cost of emitting a third one (Fig.~\ref{fig:Lineons_turn_corner}). The $\psi\bar{\psi}$, on the other hand, splits into two fractons, where each fracton is immobile as there is no operator that can annihilate an individual fracton and create it at a different location. We summarize the excitations in Table~\ref{table:Ising_CageNet_Excitations}. 

\begin{figure}[h]
    \centering
    \includegraphics[scale = 0.08]{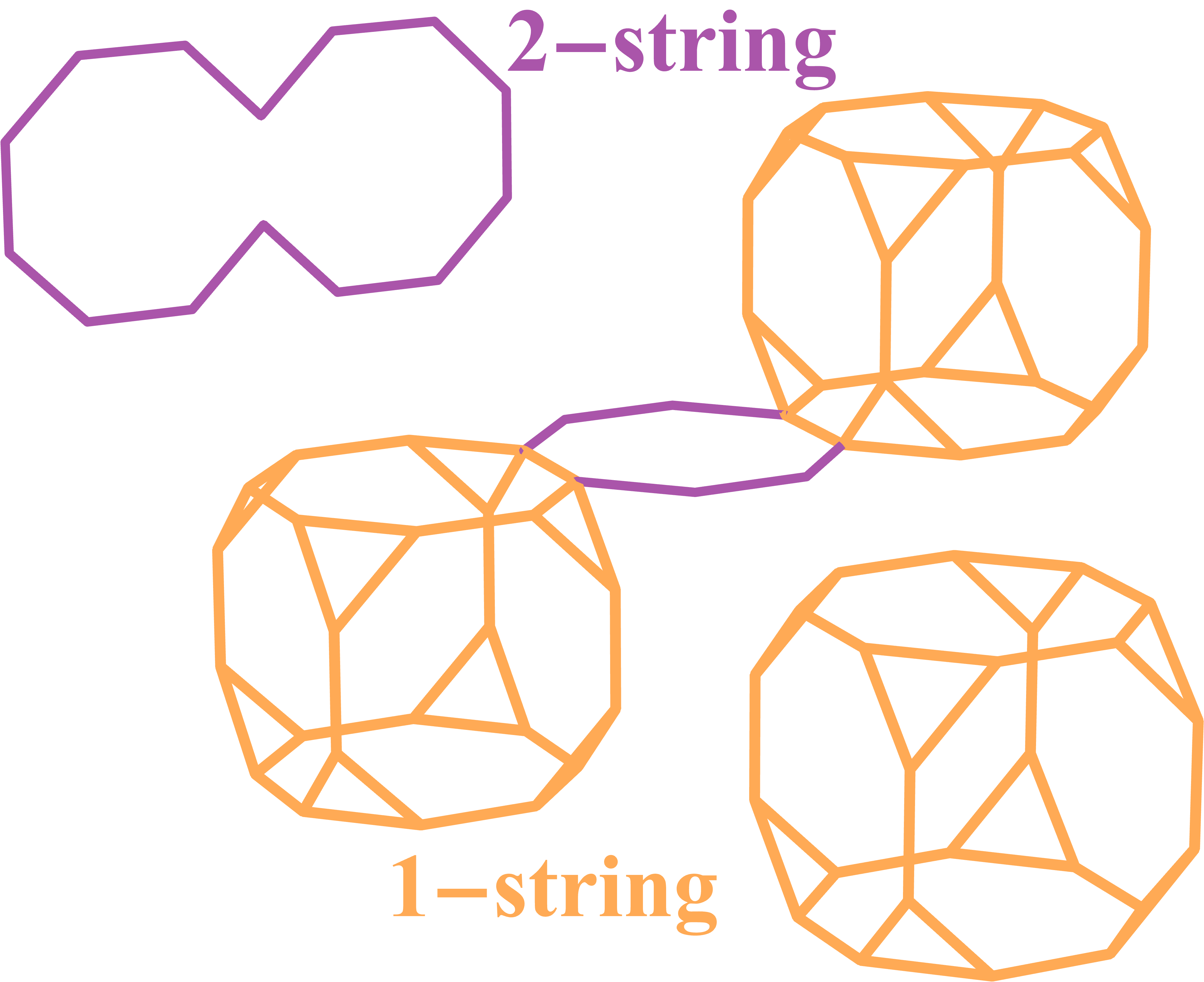}
    \captionsetup{justification=Justified}
    \caption{A cage configuration, as dictated by $A_v$. The orange colored cage is formed by a loop of $1$ on each octagonal face. The purple lines represent strings of $2$.}
    \label{fig:Cage_structure}
\end{figure}

A ground state of the Ising cage-net is a superposition of different configurations of cages, as illustrated in Fig.~\ref{fig:Cage_structure}. $B_{p_\text{s}}$, $B^0_{p_\text{o}}$, $B^2_{p_\text{o}}$, and $B_c$ all have the eigenvalue of $1$ on the ground state. In a separate paper \cite{GSD_IsCN}, we find the GSD of a $L_x\times L_y\times L_z$ Ising cage-net to be
\begin{equation} \label{eq:ising_cn_gsd}
   \text{GSD}(L_x,L_y,L_z) = \frac{1}{8} \Big(A + B + 5C + 45\Big),
\end{equation}
where $A = 9^{L_x+L_y+L_z}$, $B = 9^{L_x+L_y}+9^{L_y+L_z}+9^{L_z+L_x}$, and $C = 9^{L_x}+9^{L_y}+9^{L_z}$. We immediately see that $\text{GSD}(L_x,L_y,L_z+1)/\text{GSD}(L_x,L_y,L_z)$ is not an integer. Thus, the Ising cage-net is `not' foliated according to the foliation\cite{Shirley_2019_excitation,Shirley_2019} introduced previously for the X-cube and other models. Nevertheless, as we will see, the Ising cage-net is foliated in a generalized sense.

\section{Generalizing the notion of foliation} \label{sec:gfoliation}

The calculation of the GSD for Ising cage-net model shows that it is not foliated in the usual sense. However, from its construction in terms of stacks of 2D topological orders, it is reasonable to expect that it may be foliated in some generalized sense. Indeed, once we examine the original defnition of foliation in more detail, we can uncover two parallel ways in which it is unnaturally restrictive.

First, let us formulate the original foliated RG process purely in terms of quantum circuits. Recall that foliated RG in the X-cube model involves adding a topologically ordered layer and then coupling it to the X-cube bulk with a finite-depth quantum circuit. The topological layer cannot itself be created with a finite-depth circuit from a product state. However, it is now well-understood that it can be created with a linear-depth circuit \cite{Satzinger2021,FrankPollmann_SN_sim}. Therefore, if we view foliated RG as a generalization of usual entanglement RG \cite{VidalRG,Xie10}, in which one is allowed to add ancillary degrees of freedom in a product state and then apply finite-depth circuits, moving to foliated RG corresponds to additionally allowing linear-depth circuits within a 2D subsystem of the 3D model. However, from this perspective, the current definition of foliated RG is restricted, in that we only allow the linear-depth circuit to act on the ancillae qubits and not on the 3D bulk. A more natural definition would be to allow the linear-depth circuit to act arbitrarily within a 2D layer on both the ancillae and the bulk. We remark that the kinds of linear-depth circuits involved here have a special structure that preserves the area law of entanglement, as discussed in more detail in Sec.~\ref{sec:condvcirc}.

Second, we can also view foliated RG in terms of condensation. Namely, suppose we want to implement the inverse process of removing a single layer from the X-cube model, reducing its size in one direction. This can be achieved by condensing a planon within a single layer, corresponding to disentangling the toric code layer and then trivializing that layer by condensing a boson. In this case, the planon which we condense is very special: it can be viewed as being part of a 2D theory that is decoupled from the rest of the excitation spectrum of the 3D bulk. To be more general, if we allow condensation of planons in RG, we should allow condensation of arbitrary planons, not only those that are part of decoupled 2D theories.

In light of the above, there are two natural ways to extend the notion of foliated RG: linear-depth circuits and planon condensation. In what follows, we will show that both approaches lead to a generalized foliated RG that is applicable to the Ising cage-net model. Then, in Sec.~\ref{sec:condvcirc}, we argue that these two approaches, while seemingly distinct, are in fact very closely related to each other.

\section{RG via Condensation}
\label{sec:RG_cond}

How can the system size of the Ising cage-net model be increased / decreased? In this section, we show that it can be changed through condensation and un-condensation of bosonic planons. This is closely tied to the topic of anyon condensation in 2D systems, which we briefly review in Appendix~\ref{sec:Anyon_condensation_review}. For a comprehensive review, we refer the reader to Ref.~\onlinecite{Burnell2018} and references therein.

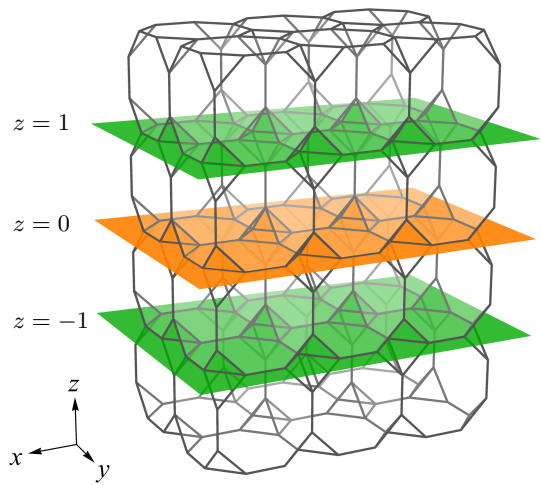
\begin{figure}
    \centering
    \begin{tikzpicture}
        \node[inner sep=0pt] (pt) at (0,0) {\includegraphics[scale = 0.2]{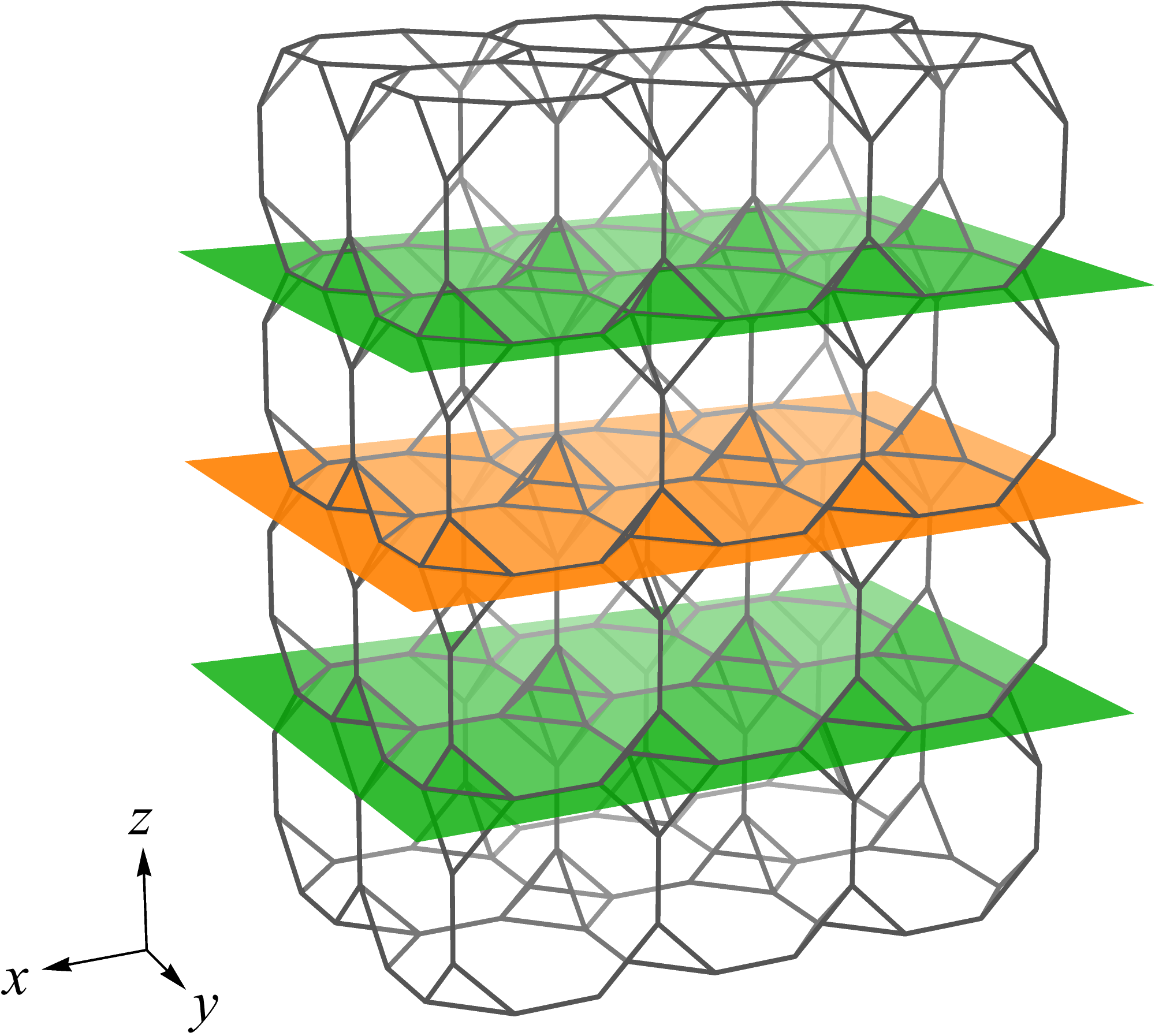}};
        \node[right] at (-3.6,-1) {$z=-1$};
        \node[right] at (-3.6,0.3) {$z=0$};
        \node[right] at (-3.6,1.65) {$z=1$};
    \end{tikzpicture}
    \captionsetup{justification=Justified}
    \caption{An illustration of the relevant $xy$-planes of a $L_x \times L_y \times L_z$ Ising cage-net. Via the condensation process described in the text, we remove the $z=0$ plane and obtain a $L_x \times L_y \times (L_z-1)$ Ising cage-net.}
    \label{fig:Condensation_layers}
\end{figure}

Let us begin by considering the process of condensing planons in an $xy$-plane to decrease the system size in the $z$ direction by one (Fig.~\ref{fig:Condensation_layers}). Recall from the last section that for each $xy$-plane there is a bosonic planon $\psi\bar{\psi}$ which can be condensed. When $\psi\bar{\psi}$ in plane $z=0$ is condensed, the quasi-particle content of the model changes as follows:
\begin{enumerate}
    \item Since $\psi\bar{\psi}$ is the fracton dipole, fractons between planes $z=0$ and $z=1$ are identified with the corresponding fracton between planes $z=-1$ and $z=0$.
    \item The planons $\psi$ and $\bar{\psi}$ on the $z=0$ plane are identified.
    \item The $\sigma\bar{\sigma}$ planon on the $z=0$ plane splits into two abelian bosonic planons $e$ and $m$ with a mutual $-1$ braiding statistics.
    \item The lineons in the $z=0$ plane composed of $\sigma_{xy}\sigma_{xz}$, $\bar{\sigma}_{xy}\sigma_{xz}$, $\sigma_{xy}\bar{\sigma}_{xz}$, and $\bar{\sigma}_{xy}\bar{\sigma}_{xz}$ are all confined.
    \item Planons and lineons on other planes are unchanged. 
\end{enumerate}   

After this step, we can further condense either $e$ or $m$. This gets rid of the remaining planons on the $z=0$ plane without affecting other quasi-particle excitations. Now, we see that the quasi-particle content of the model is the same as that of an Ising cage-net model with the $z=0$ plane removed. The planons and lineons on planes other than $z=0$ are left intact. Moreover, the fracton between $z=0$ and $z=1$, which is now identified with the fracton between $z=-1$ and $z=0$, becomes the new fracton between $z=-1$ and $z=1$. Therefore, the size of the Ising cage-net model can be decreased by one in the $z$
direction by first condensing the $\psi\bar{\psi}$ planon in a plane, and then by condensing one of the split channels of the $\sigma\bar{\sigma}$ planon on the same plane. 

We see that if we allow condensation of bosonic planons as a RG operation, we obtain a generalized foliated RG under which the Ising cage-net model is a fixed point. As noted in Sec.~\ref{sec:gfoliation}, the original foliated RG for the X-cube model can also be viewed in terms of such condensation. 

The condensation of planons is, of course, a singular process where the bulk gap needs to close and then reopen, corresponding to a phase transition between different standard phases (see Appendix~\ref{app:foliated-phases} for the definition of standard phases). This means that, similar to the original foliated RG, the generalized foliated RG operations can move across certain phase boundaries. However, only certain phase boundaries can be crossed; the singularity involved in planon condensation is localized to a selected plane and is hence a ``subsystem" singularity, not one in the full 3D bulk.

A useful way to think about the condensation process is to use the fact that the Ising cage-net model can be obtained by gauging the planar $Z_2$ symmetries of a subsystem symmetry protected topological (SSPT) model protected by the planar symmetries\footnote{to be discussed in future work}. Note that, subsystem symmetries usually contain generators associated with rigid subsystems like $x$, $y$, $z$ planes in the 3D bulk. They are different from higher-form symmetries\cite{gaiotto_generalized_2015} with generators associated with deformable subsystems. The planons being condensed correspond to the symmetry charges of the planar symmetries in the SSPT model. Hence the condensation of the planons in a given plane corresponds to breaking / removing that planar symmetry and reducing the size of the model. On the other hand, if we want to increase the size of the system by adding a plane at $z=0$, we need to add the planar symmetry and the corresponding planar state back to the SSPT model and `re-gauge' the planar symmetry.

\section{RG via planar Linear Depth Circuit}
\label{sec:RG_circ}

The planar linear depth circuit we construct for the Ising cage-net model is a direct generalization of a RG scheme that maps product states to ground states of a string-net model, introduced by Liu Y. \textit{et al.}\cite{FrankPollmann_SN_sim}. In Sec.~\ref{sec:SN_RG}, we review this RG procedure for the string-net models. We describe carefully an initialization step that is nontrivial for non-abelian string-net models, which was not discussed in detail in Ref.~\onlinecite{FrankPollmann_SN_sim}. In Sec.~\ref{sec:CN_RG}, we describe the RG scheme as a linear depth circuit for the Ising cage-net model. We will see that the initialization step is also important and nontrivial.

\subsection{String-net RG}
\label{sec:SN_RG}

In this section, we will first describe an important step in the RG procedure -- the `controlled gate' which adds a plaquette to the string-net wave-function. After that, we will describe the full RG procedure starting from the string-net wave-function on the minimal lattice on a torus and then adding plaquettes row by row. A brief review of the string-net models is given in Appendix~\ref{sec:SN_models}.

\subsubsection{Adding plaquettes via the controlled gate}
\label{sec:control_gate}

The controlled gate can be used to add a plaquette to the string-net wave-function. We present the definition and properties of the gate in this sub-section. Computational details of the results discussed here can be found in Appendix~\ref{sec:ControlledGate_details}. 

Suppose that on a trivalent lattice, a plaquette is added by adding an edge (the red edge in the diagrams below), and we want to extend the string-net wave-function from the original lattice to that including this new plaquette. When the edge is added, it is not entangled with the rest of the lattice and is in the state $|0\rangle$. To merge the added edge into the lattice, first, map it to $\sum_s \frac{d_s}{\sqrt{D}}|s\rangle$ where $D$ is the total quantum dimension of the string-net. 
\begin{equation}
  |0\rangle \mapsto  \sum_s \frac{d_s}{\sqrt{D}}|s\rangle 
\end{equation}
Then, we use this edge as the control to draw loops around the added plaquette. More specifically, we can represent the controlled gate $G_p = \sum_s G_p^s$ graphically as in Eq.~\eqref{eq:Gsp_def}. The action of $G^s_p$ is similar to the action of $B^s_p$ which adds a loop $s$ to a plaquette, but for the graphical evaluation of $G^s_p$, we treat the control edge as if it is in the state $\ket{0}$, i.e.
\begin{widetext}
    \begin{equation}
    \vcenter{\hbox{\includegraphics[scale = 1]{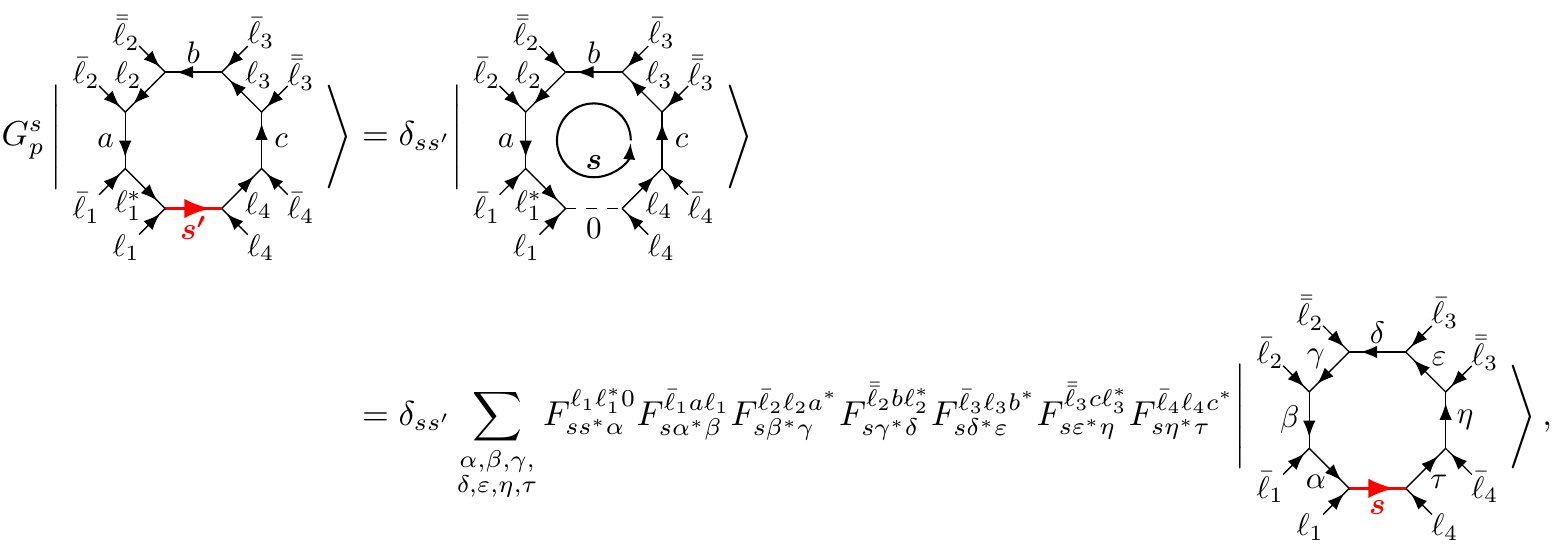}}}
    \label{eq:Gsp_def}
    \end{equation}
\end{widetext}
where the red line with an arrow marks the control edge. We carry out the explicit graphical evaluation in Appendix~\ref{sec:ControlledGate_graphicalDef}. Note that $G^s_p$ can be defined on any polygonal plaquette.

$G^s_p$ is not a unitary on the full Hilbert space, but only between subspaces. More specifically, it is an isometry from $\mathcal{V}^\text{SN}_{p,s}$ to $\mathcal{H}^\text{SN}_{p,s}$, both of which involve the DOF around a plaquette $p$. In $\mathcal{V}^\text{SN}_{p,s}$, the control edge is set to $|s\rangle$ while the other edges come from the string-net wave-function on the lattice with the control edge missing (pretending that it is set to $|0\rangle$). The vertices containing the control edge, then, involve configurations like 
\begin{equation}
    \vcenter{\hbox{\includegraphics[scale = 1]{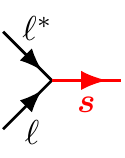}}}
\end{equation}
In $\mathcal{H}^\text{SN}_{p,s}$, all edges, including the control edge, come from the string-net wave-function with the control edge set to $|s\rangle$. 

In Appendix~\ref{sec:appen_isometry}, we prove that $G^s_p$ is an isometry from $\mathcal{V}^\text{SN}_{p,s}$ to $\mathcal{H}^\text{SN}_{p,s}$ by demonstrating
\begin{equation}
    \vcenter{\hbox{\includegraphics[scale = 1]{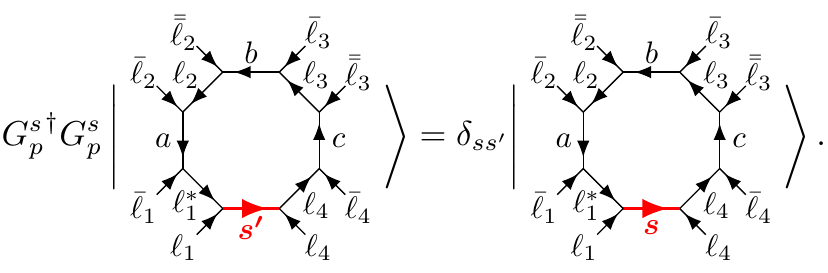}}}
\end{equation}

The controlled gates commute with each other
\begin{equation}
    \begin{aligned}
        \left[G^s_p,G^{s'}_{p'}\right] = &0 =\left[{G^s_p}^\dagger,G^{s'}_{p'}\right],
    \end{aligned}
    \label{eq:Gsp_commute1}
\end{equation}
as long as they do not act on each other's controlled edge. Moreover, we can show
\begin{equation}
    \left[G^s_p,B^{s'}_{p'}\right] = 0 =\left[{G^s_p}^\dagger,B^{s'}_{p'}\right],
    \label{eq:Gsp_commute2}
\end{equation}
provided that $B^{s'}_{p'}$ does not act on the control edge of $G^s_p$. We prove these commutation relations in Appendix~\ref{sec:appen_commutation}.

In Appendix~\ref{sec:Proof_centralequation}, we prove a useful equation, which we call \textit{the central equation}
\begin{equation}
    G^s_p \left(\ket{s}\bra{s'}\right)_\text{ct} {G^{s'}_p}^{\dagger} = P^s_\text{ct} \left(\sum_k \frac{d_k}{d_s d_{s'}}  B^k_p \right) P^{s'}_\text{ct},
    \label{eq:TheCentralEquation}
\end{equation}
where $\left(\ket{s}\bra{s'}\right)_\text{ct}$ acts on the control edge and $P^s_\text{ct} = \ket{s}\bra{s}$ is a projector on the control edge. With the central equation, we can show that the controlled gate does what we claimed -- it adds a plaquette to the string-net wave-function. In particular, we show below that under conjugation by $G_p = \sum_s G^s_p$, the projector on the control edge $P_\text{ct}=\sum_{s,s'} \frac{d_sd_{s'}}{D}|s\rangle\langle s'|$ is mapped to the plaquette projector $B_p = \sum_s \frac{d_s}{D}B^s_p$.
\begin{equation}
    \begin{aligned}
       G_p P_\text{ct} G^{\dagger}_p & =  \sum_{s,s'} \frac{d_sd_{s'}}{D}G^s_p \left(|s\rangle\langle s'|\right)_\text{ct} {G^{s'}_p}^{\dagger} \\
    & = \sum_{s,s',k} \frac{d_k}{D} P^s_\text{ct} B^k_p P^{s'}_\text{ct} \\
    & = \sum_k\frac{d_k}{D}B^k_p = B_p
    \end{aligned}
\end{equation}

\subsubsection{The RG circuit}

Using the controlled gate as a building block, we can construct the full linear depth circuit that maps a product state to the string-net wave-function. We present the linear depth circuit in two steps: 1. from a product state to a string-net wave-function on the minimal lattice on torus; 2. from the string-net wave-function on the minimal lattice to the full  lattice by adding plaquettes. We are going to focus on the trivalent square-octagon lattice, although the general procedure applies to other trivalent graphs as well. 

\begin{figure}[ht]
    \centering
    \includegraphics[scale=1]{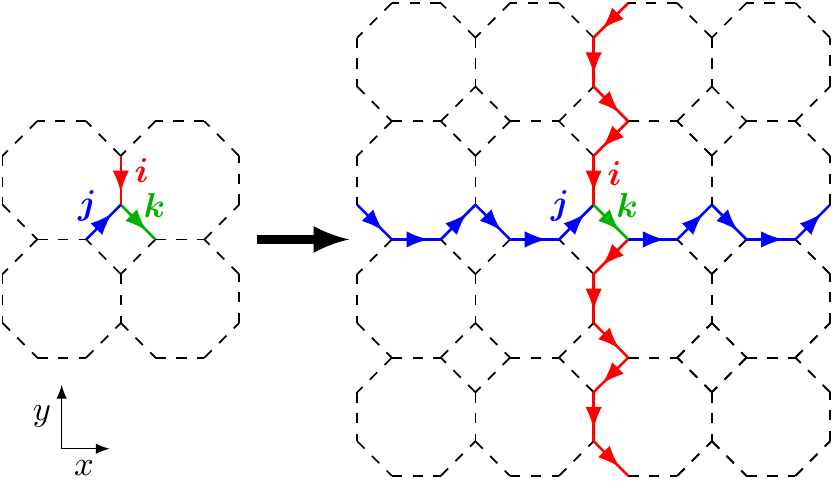}
    \captionsetup{justification=Justified}
    \caption{The initialization step in the RG circuit for generating the string-net wave-function. Left: pick three edges around a vertex and map them into one of the ground states of the string-net on the minimal lattice. Right: grow the minimal structure by copying the string states $|i\rangle$ and $|j\rangle$ along non-contractible loops so that they reach the full extent of the lattice.
    }
    \label{fig:SN_initialization}
\end{figure}

The minimal lattice on the torus consists of three edges, two vertices, and one plaquette, as shown in Fig.~\ref{fig:MinimalLattice_torus}. On the square-octagon lattice, we start from the product state $\otimes_l |0\rangle_l$. Pick three edges around a vertex as shown in Fig.~\ref{fig:SN_initialization}. Apply a local unitary transformation on the three edges so that they become one of the ground states on the minimal lattice. Note that for abelian string-net states, the ground states can be chosen to be a product state of the three edges. In fact, the $\otimes_l|0\rangle_l$ state is a legitimate state already, because it satisfies the vertex term while the plaquette term is trivial for abelian strings on the minimal lattice (for proof see Appendix~\ref{sec:SN_minimalLattice_torus}). However, for non-abelian string-nets, the $B^s_p$ term for a non-abelian $s$-string acts non-trivially in the stable vertex subspace, and the ground states generally become entangled. In the case of the doubled-Ising on the minimal lattice, ten configurations satisfy the vertex constraints. Of this ten-dimensional space, only nine dimensions belong to the ground space, where $B^0_p=1$, $B^1_p=\sqrt{2}$, and $B^2_p=1$. The remaining one dimension carries a $\psi\bar{\psi}$ fluxon excitation such that $B^0_p=1$, $B^1_p=-\sqrt{2}$, and $B^2_p=1$.  One possible choice of the nine doubled-Ising ground states on the minimal lattice is given in Appendix~\ref{sec:MinimalLattice_DI_SN}.

Now, we need to grow this minimal structure so that it reaches the full extent of the lattice. To do this, we `copy' the states on the $i$ and $j$ edges along the non-contractible loops in the $y$ and $x$ directions. To achieve this, we use controlled gates of the form $\sum_{i} |i\rangle|i\rangle \langle i|\langle 0|$, and apply them sequentially along the non-contractible loops, as shown in Fig.~\ref{fig:SN_initialization}. As this step has to be done sequentially along the loop, its depth increases linearly with the size of the lattice. This completes step 1 of the linear depth circuit, which we call initialization.

\begin{figure}[ht]
    \centering
    \includegraphics[scale = 1]{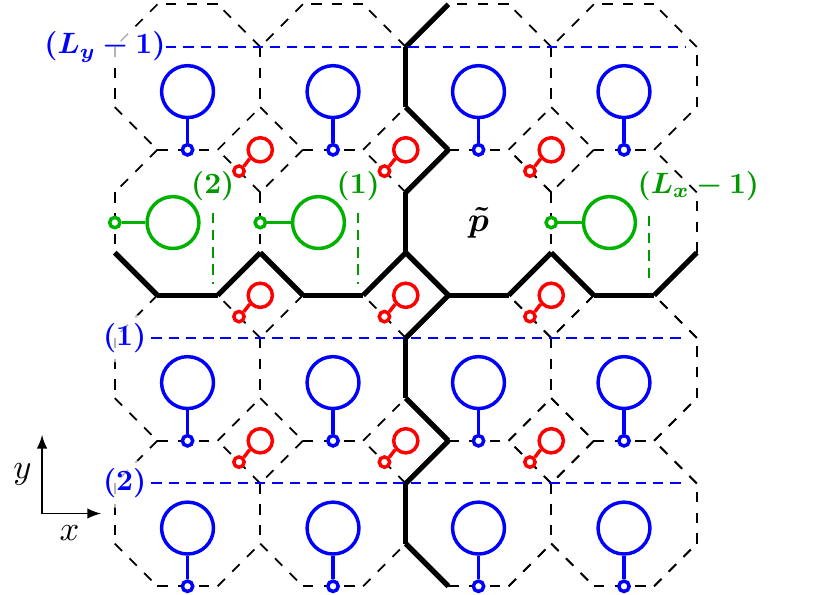}
    \captionsetup{justification=Justified}
    \caption{Adding loops to plaquettes in step 2 of the RG circuit for generating the string-net wavefunction. The state has been initialized into one of the ground states on the minimal lattice (black lines). First, loops are added to the square plaquettes (shown in red) in a single step. Then, loops are added to octagon plaquettes in row (1), (2), ... ($l_y-1$) sequentially. For the last row, loops are added to octagon plaquette in column (1), (2), ...., ($L_x-1$) sequentially. No action is needed in the last plaquette $\tilde{p}$.
    }
    \label{fig:SN_RG_Sgrow}
\end{figure}

Step 2 is also of linear depth. The minimal lattice has only one plaquette. In step 2, we add more plaquettes to the lattice using the controlled gate introduced in Sec.~\ref{sec:control_gate}. The plaquettes cannot be added all at once, because the controlled gates commute only when they do not act on each other's control edge. A linear depth circuit is hence needed to add all the plaquettes to the square-octagon lattice. A particular sequence for adding these plaquettes is shown in Fig.~\ref{fig:SN_RG_Sgrow}. Firstly, all the square plaquettes (red circles) can be added at the same time because they do not overlap with each other. The small circle indicates the control edge while the big circle indicates the action of $G^s_p$. Secondly, we add the square-octagon lattice in row one (labeled $(1)$ in Fig.~\ref{fig:SN_RG_Sgrow}). All controlled gates in row one commute with each other so they can be added in one step. Then we add row two, row three, etc., until the next to last row (labeled $(L_y-1)$ in Fig.~\ref{fig:SN_RG_Sgrow}). For the last row, we need to choose the control edges side ways because we need un-entangled edges to be used as control edge. Due to this change, the plaquettes in the last row need to be added sequentially as the controlled gates do not commute any more. As shown in the figure, we can add them in the order of (green labels) $(1)$, $(2)$, ..., $(L_x-1)$. We do not need to act in the last plaquette (labeled $\tilde{p}$) as the constraint due to the last plaquette is already implied by that of the largest plaquette that we started from combined with all the small plaquettes added so far. Therefore, at this point, we have finished the linear depth RG procedure that starts from a product state and maps it to the the string-net wave-function on the square-octagon lattice.

\subsection{Ising cage-net}
\label{sec:CN_RG}

In this section, we use the controlled gate of Eq.~\eqref{eq:Gsp_def} to build up the RG circuit to enlarge an Ising cage-net ground state on the three-torus by one layer. We will start, in Sec.~\ref{sec:Ising_CN_Growing_cages}, by introducing finite depth circuits that grow cages on the cage-net ground state. They serve as the building blocks of the full planar linear depth RG circuit, which we discuss in Sec.~\ref{sec:IsingCN_RG_circuit}.

\subsubsection{Adding cages via the controlled gate}
\label{sec:Ising_CN_Growing_cages}

In $2$D, we have seen that a plaquette can be added to the string-net wave function, via the controlled gates, after an edge is added to the lattice. We can extend this procedure to $3$D cage-net states. 

\begin{figure}[ht]
    \centering
    \includegraphics[scale = 0.07]{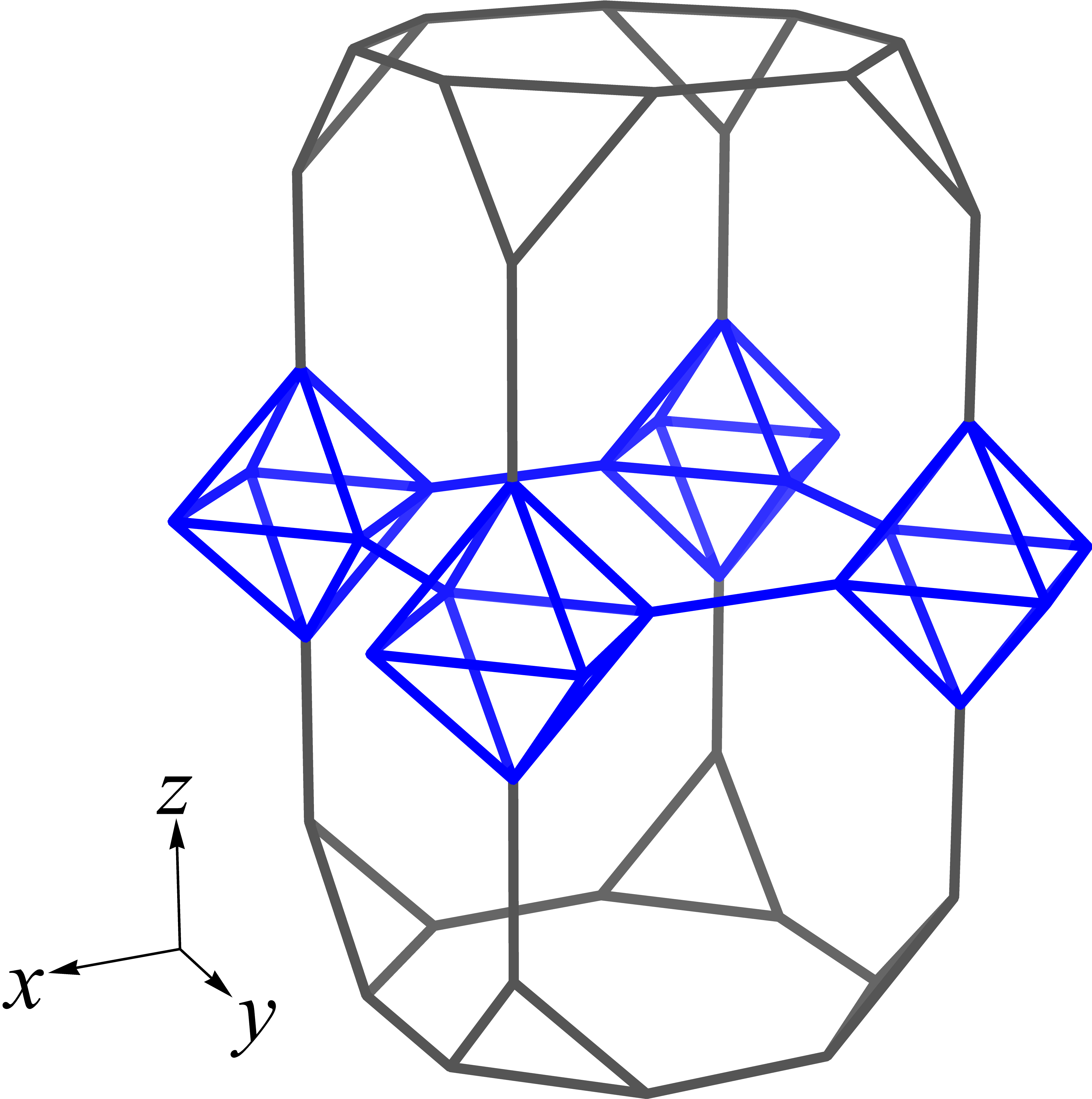}
    \captionsetup{justification=Justified}
    \caption{Insertion of an $xy$-plane bisects a cube in the original cage-net lattice into two cubes. Each intersection point between the $xy$-plane and the $z$-principal edges is expanded into an octahedron to preserve the trivalent structure in the $xy$, $yz$ and $zx$ planes.
    }
    \label{fig:Ising_CN_adding_plaquette}
\end{figure}

Suppose that we start with the Ising cage-net ground state on the truncated cubic lattice (Fig.~\ref{fig:Truncated_cubic_lattice2}) and add a plane in the $xy$ direction. At each point where the added plane bisects the $z$ direction edges, an octahedron is added, as shown in Fig.~\ref{fig:Ising_CN_adding_plaquette}, to ensure the trivalent structure in each of the coupled planes. In the added plane, octagonal plaquettes fill in the space between the octahedrons. Every edge of the added octahedrons carries a three dimensional Hilbert space spanned by $\{|0\rangle, |1\rangle, |2\rangle \}$. We start with these edges all set to the state $|0\rangle$. The principal edges on the octagons each carry a five dimensional Hilbert space spanned by $\{|00\rangle, |02\rangle, |20\rangle, |22\rangle, |11\rangle \}$, which is a subspace of the tensor product Hilbert space of two three dimensional DOFs $\{|0\rangle, |1\rangle, |2\rangle\} \otimes \{|0\rangle, |1\rangle, |2\rangle\}$ that come from the two intersecting planes. We start with these principal edges in the state $|00\rangle$. 

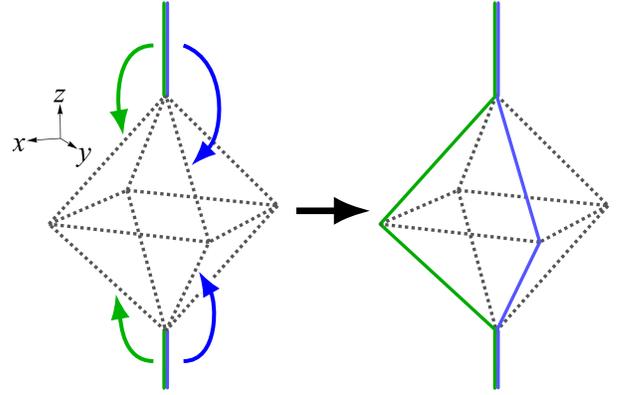
\begin{figure}[ht]
    \centering
    \begin{tikzpicture}
        \node[inner sep=0pt] (pt1) at (-2.2,0) {\includegraphics[scale = 0.05]{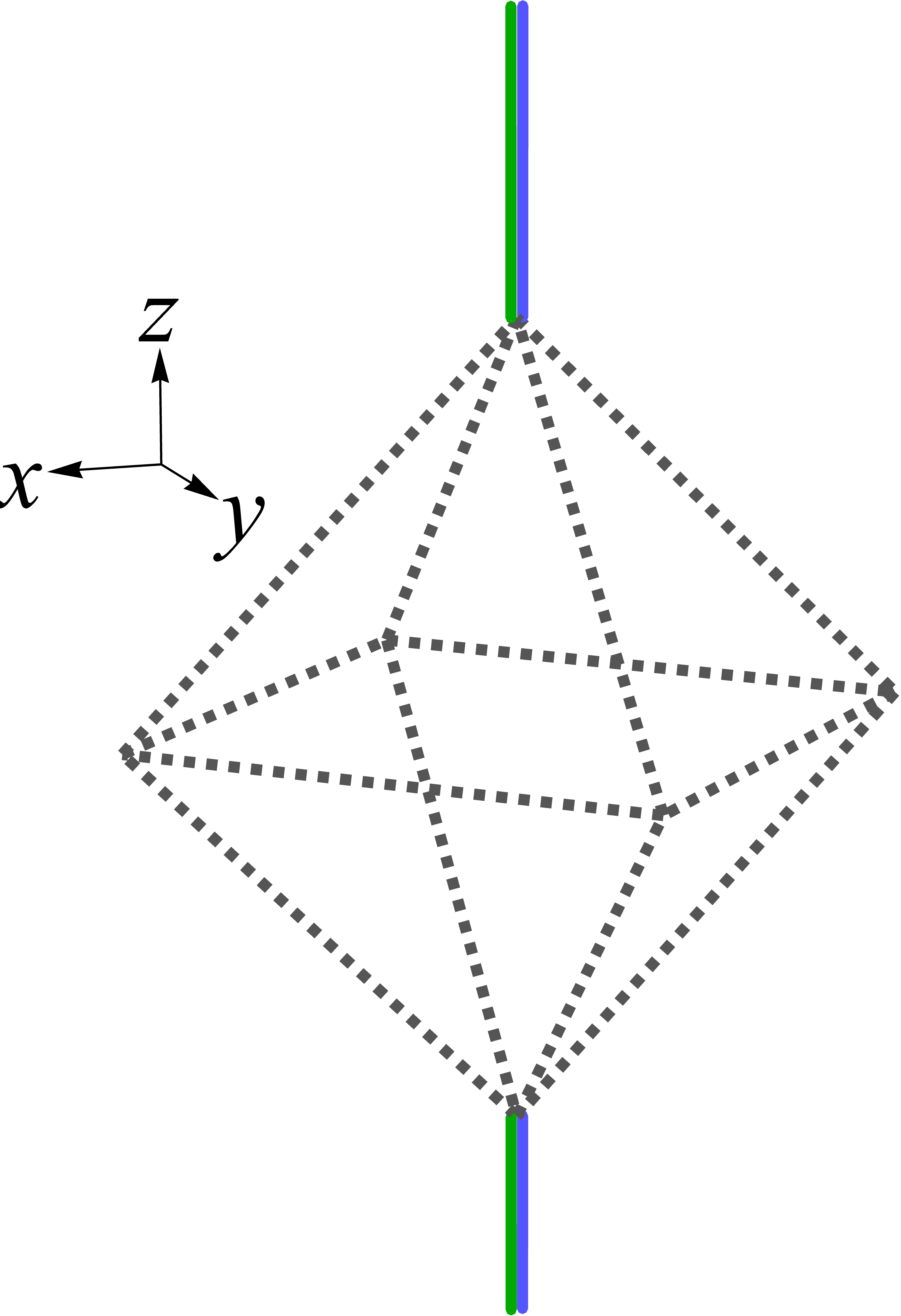}};
        \node[inner sep=0pt] (pt2) at (2.2,0) {\includegraphics[scale = 0.05]{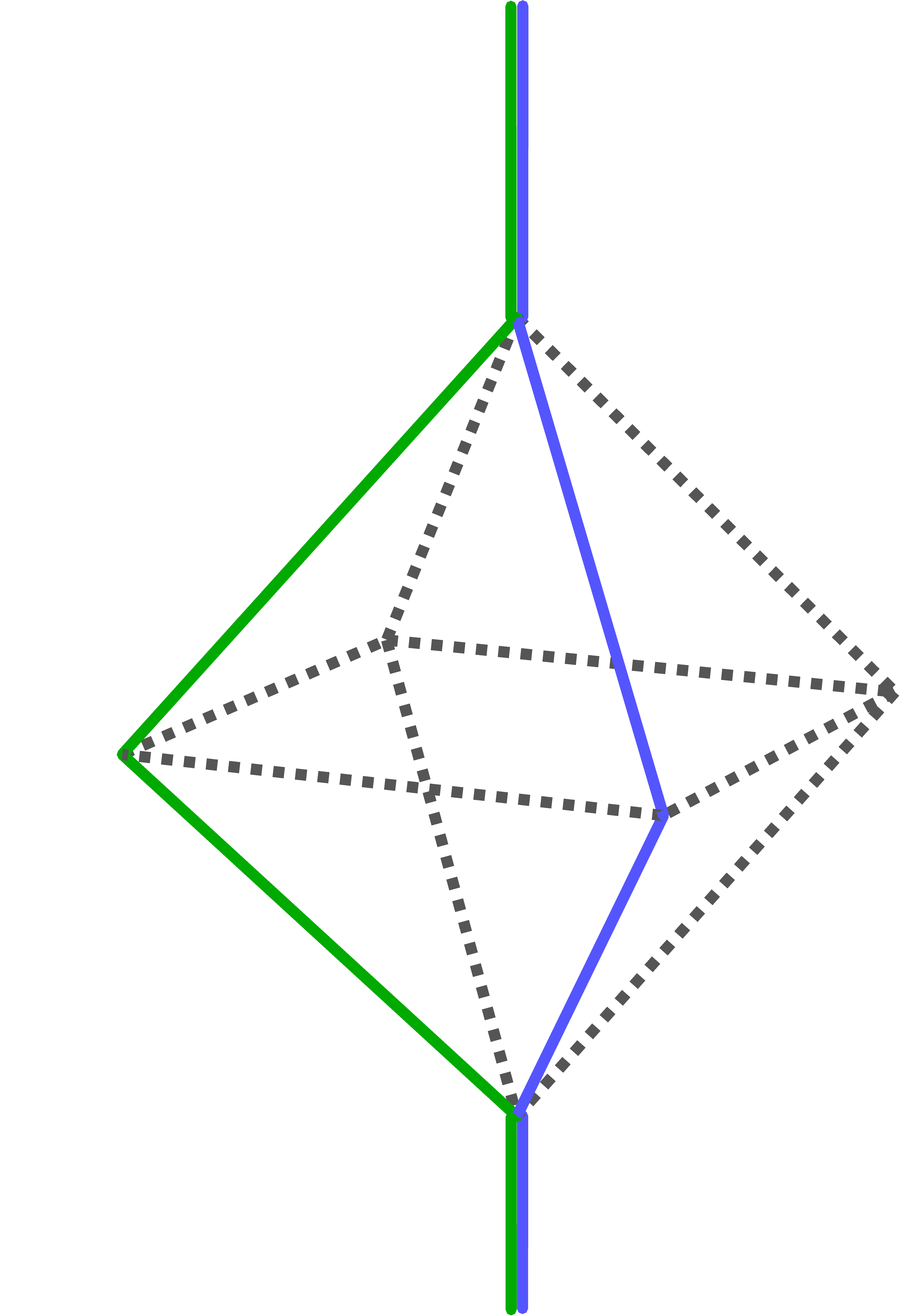}};
        \draw[draw = black, line width = 2.5 pt, -Latex] (-0.2,-0.2) -- ++(1,0);
        \draw[draw = green!70!black, line width = 1.5 pt, -Latex] (-2.1,2) to[out = 180, in = 100] (-2.5,0.8);
        \draw[draw = green!70!black, line width = 1.5 pt, -Latex] (-2.1,-2.2) to[out = 180, in = 280] (-2.6,-1.3);
        \draw[draw = blue, line width = 1.5 pt, -Latex] (-1.7,2) to[out = 340, in = 45] (-1.6,0.4);
        \draw[draw = white, line width = 4 pt] (-1.45,-1.3) -- (-1.35,-1.1);
        \draw[draw = blue, line width = 1.5 pt, -Latex] (-1.7,-2.2) to[out = 0, in = 300] (-1.5,-1);
    \end{tikzpicture}
    \captionsetup{justification=Justified}
    \caption{`Copying' the states on the bisected $z$-principal edges onto edges of the added octahedron to satisfy vertex rules in the $xz$ and $yz$ planes. The copying process can be performed by controlled gates of the form $\sum_{xy}|xy\rangle \langle xy| \otimes |x\rangle \langle 0|$ and $\sum_{xy} |xy\rangle \langle xy| \otimes |y\rangle \langle 0|$, indicated by the arrows pointing from the control to the target.}
    \label{fig:Ising_CN_add_plaquette_diamond_vertices}
\end{figure}

We describe first the process to add one cube into the new layer, which consists of two steps: 1. add the octahedrons to the cage-net wave-function; 2. grow a cage structure in the upper truncated cube of Fig.~\ref{fig:Ising_CN_adding_plaquette}. 
In step one, we first need to copy the state of the bisected $z$-principal edge onto some of the octahedron edges so that the vertex rules are satisfied at the octahedrons' vertices. Suppose the bisected edge is in the state $|xy\rangle$. The copying process can be achieved with the controlled gates $\sum_{xy}|xy\rangle \langle xy| \otimes |x\rangle \langle 0|$ and $\sum_{xy} |xy\rangle \langle xy| \otimes |y\rangle \langle 0|$ as indicated by the blue and green arrows in Fig.~\ref{fig:Ising_CN_add_plaquette_diamond_vertices}. Then, we add the square plaquettes to the cage-net wave-function. This can be done as described in the previous section on how to add a square plaquette to the doubled-Ising string-net wave function, as the square plaquettes remain unaffected when the doubled-Ising layers are coupled into Ising cage-net. More specifically, for each square plaquette, we pick an edge in the state $|0\rangle$ as the control edge, map it to $\sum_s\frac{d_s}{\sqrt{D}}|s\rangle$, and use it as the control in the controlled gate $G_p$ that adds loops into the plaquette.   

\begin{figure}[ht]
    \centering
    \begin{subfigure}[b]{0.23\textwidth}
        \centering
        \includegraphics[scale = 0.05]{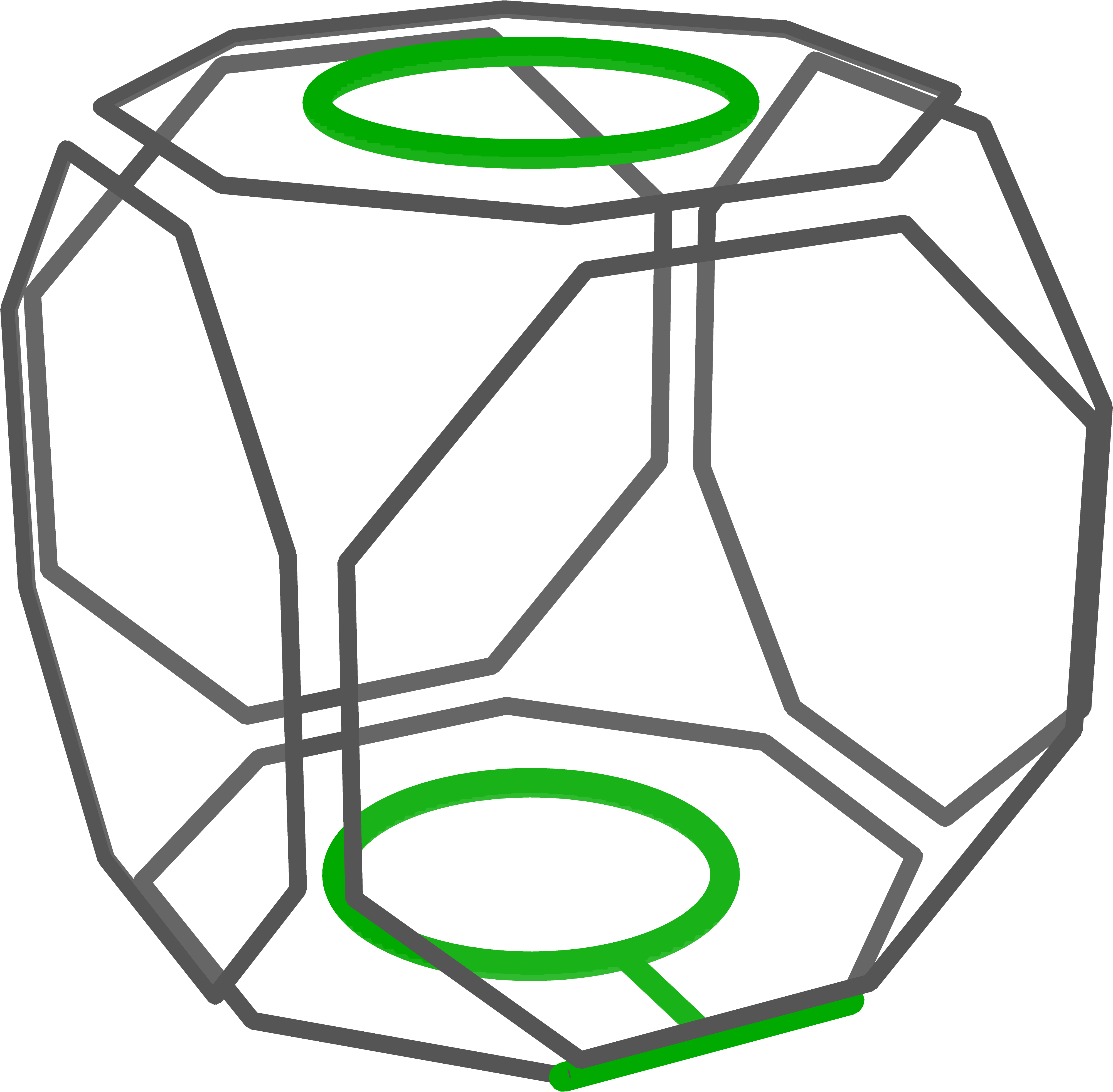}
        \caption{}
        \label{fig:Ising_CN_cube_circuit_1}
    \end{subfigure}
    \begin{subfigure}[b]{0.23\textwidth}
        \centering
        \includegraphics[scale = 0.05]{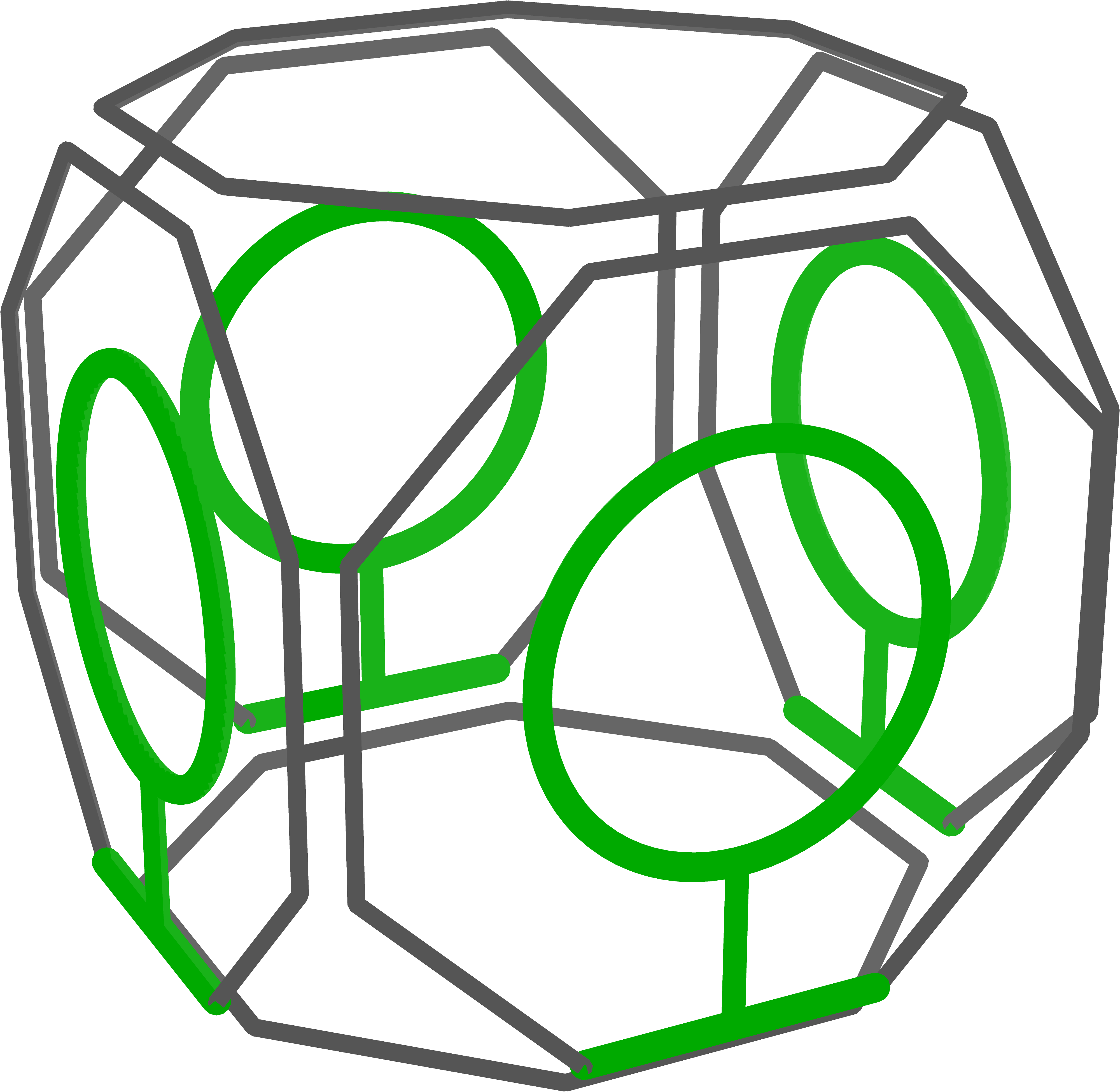}
        \caption{}
        \label{fig:Ising_CN_cube_circuit_2}
    \end{subfigure}
    \captionsetup{justification=Justified}
    \caption{Growing a cage structure in an added cube. (a) First, using an edge from the bottom face (colored green) as control, add loops to the bottom and top faces, (b) then use the edges on the side faces (colored green) as control to add loops to the side face.
    }
    \label{fig:Ising_CN_cube_circuit}
\end{figure}

Step 2, which adds a cage structure to the cube, is more complicated. As shown in Fig.~\ref{fig:Ising_CN_cube_circuit}, first we add loops to the bottom and top faces and then to the side faces. More specifically, first we pick a principal edge on the bottom face in the state $|00\rangle$ as the control. We will use the convention where the first $|0\rangle$ comes from the $xy$ plane while the second $|0\rangle$ comes from the vertical $xz$ and $yz$ planes. Map the control edge as
\begin{equation}
    \ket{00} \mapsto \sum_s \frac{d_s}{\sqrt{D}}  \ket{s 0},
\end{equation}
Note that this takes the controlled edge out of the five dimensional subspace of $\{|00\rangle, |02\rangle, |20\rangle, |22\rangle, |11\rangle\}$ but keeps it in the nine dimensional space of $\{|0\rangle, |1\rangle, |2\rangle\}^{\otimes 2}$. This will also happen to other principal edges as we implement the procedure, but at the end of the process of growing a cube, all principal edges will be back to the five dimensional subspace. 

Now, using the $|s\rangle$ state as the control, apply the controlled gate to the bottom face $p_b$ and top face $p_t$ as
\begin{equation}
    G^0_{p_b}+G^2_{p_b}+ \frac{1}{\sqrt{2}}G^1_{p_b} B^1_{p_t}
    \label{eq:tbfaces}
\end{equation}
as shown in Fig.~\ref{fig:Ising_CN_cube_circuit} (a). Note that $G^s_{p_b}$ and $B^s_{p_t}$ act on the first part of the principal edges (the part that comes from horizontal planes). After these controlled gates, the projector on the control edge $|0\rangle\langle 0|$ (the first part) gets mapped to
\begin{equation}
  \begin{aligned}
 \left(|0\rangle\langle 0|\right)_\text{ct} &\mapsto \sum_{ss'} \frac{d_sd_{s'}}{D}\left(|s\rangle\langle s'|\right)_\text{ct} \\
 &\mapsto B^0_{p_b}+B^2_{p_b}+B^1_{p_b} B^1_{p_t},
  \end{aligned}
  \label{eq:cube_term0}
\end{equation}
where in deriving the last line, we used the fact that the top face is part of the original cage-net wave-function and $B^0_{p_t}=B^2_{p_t}=1$. Note that it might seem that the operator in Eq.~\eqref{eq:tbfaces} is not unitary as $B^1_p$ is not. But since $B^1_{p_t}{B^1}^{\dagger}_{p_t} = B^0_{p_t}+B^2_{p_t} = 2$, the action of the operator restricted to the ground space of the original cage-net model is indeed unitary.

Next, we need to add loops to the side faces. To do this, we take the principal edges on the bottom face, which are now in the states $|s0\rangle$ and send them to $|s\alpha_s\rangle$, where $\alpha_s$ comes from the $xz$ or $yz$ planes and $\alpha_s = 0$ if $s$ is even, $\alpha_s=1$ if $s$ is odd. This brings the principal edges on the bottom face back to the five dimensional Hilbert space. Then map the $|\alpha_s\rangle$ states to
\begin{equation}
    |0\rangle \mapsto \frac{1}{\sqrt{2}}\left(|0\rangle + |2\rangle\right), \ |1\rangle \mapsto |1\rangle
\end{equation}
Use the $|\alpha_s\rangle$ states as the control to draw loop on the side faces by applying $\sum_{\alpha_s} G^{\alpha_s}_{p_s}$ as shown in Fig.~\ref{fig:Ising_CN_cube_circuit} (b) to each side face. Let us see how the Hamiltonian terms in Eq.~\eqref{eq:cube_term0} transforms. We show the step by step calculation for the third term $B^1_{p_b}B^1_{p_t}$. The $B^1_{p_t}$ part is not affected by the transformation and will be omitted from the following equation. Let us focus on the transformation induced by on principal edge. We label the two three-dimensional DOFs on the principal edge as $1$ and $2$ respectively, where $1$ comes from the bottom face whose state is labeled by $s$ and $2$ comes from the side face whose state is labeled by $\alpha_s$.
\begin{equation}
\begin{aligned}
  &  \left[(P^0_1+P^2_1)B^1_{p_b}P^1_1 + P^1_1B^1_{p_b}(P^0_1+P^2_1)\right] \otimes \left(|0\rangle\langle 0|\right)_2 \\
\mapsto & \frac{1}{\sqrt{2}}(P^0_1+P^2_1)B^1_{p_b}P^1_1\otimes \left(|0\rangle_2+|2\rangle_2\right){}_2\langle 1| \\
& + \frac{1}{\sqrt{2}}P^1_1B^1_{p_b}(P^0_1+P^2_1)\otimes|1\rangle_2\left({}_2\langle 0| + {}_2\langle 2|\right) \\
\mapsto & \frac{1}{\sqrt{2}}(P^0_1+P^2_1)B^1_{p_b}P^1_1\otimes \left(P^0_2+P^2_2\right)B^1_{p_s}P^1_2 \\
& + \frac{1}{\sqrt{2}}P^1_1B^1_{p_b}(P^0_1+P^2_1)\otimes P^1_2B^1_{p_s}\left(P^0_2+P^2_2\right)
\end{aligned}
\end{equation}
The result is the product of $B^1_{p_b}$ and $B^1_{p_s}$ projected onto the five dimensional subspace of the principal edge, as promised. This works for all side faces. Similar calculations can be carried out for the first two terms in Eq.~\eqref{eq:cube_term0}. If we put everything together and omit the projection onto the five dimensional subspace of the principal edges, we see the Hamiltonian terms in Eq.~\eqref{eq:cube_term0} becomes
\begin{equation}
    \left(B^0_{p_b}+B^2_{p_b}\right)\prod_{p_s}\left(B^0_{p_s}+B^2_{p_s}\right) + B^1_{p_b}B^1_{p_t}\prod_{p_s}B^1_{p_s}, 
\end{equation}
which is a sum over the desired plaquette terms on the bottom and side faces as well as the cube term on the cube.

\begin{figure}[ht]
    \centering
    \includegraphics[scale = 0.09]{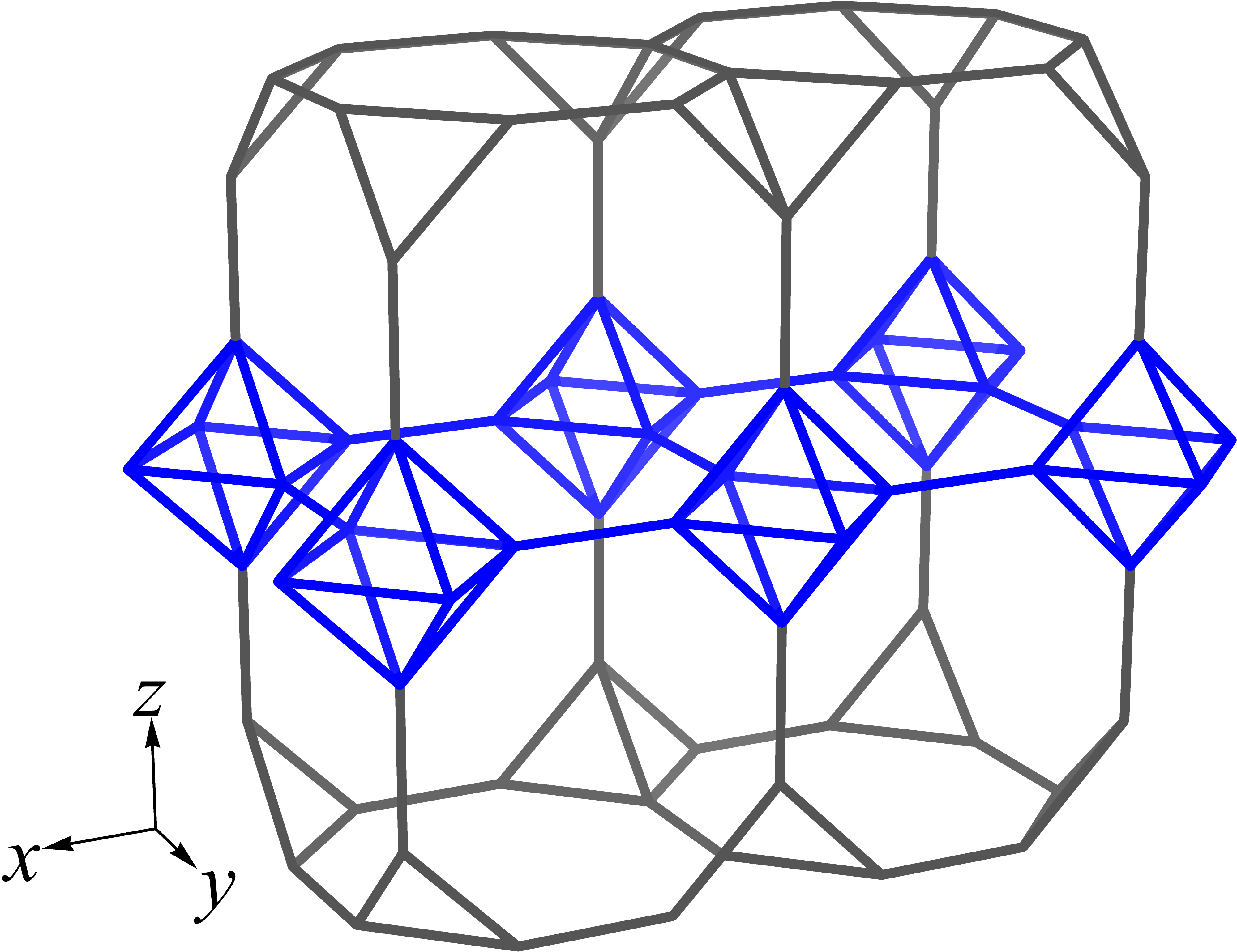}
    \captionsetup{justification=Justified}
    \caption{Adding a row of cubes to the cage-net state, step 1: the inserted $xy$-plane bisects the cubes into two; octahedrons are added at the intersection point.
    }
    \label{fig:Ising_CN_adding_two_plaquettes}
\end{figure}

\begin{figure}[ht]
    \centering
    \begin{subfigure}[b]{0.48\textwidth}
        \centering
        \includegraphics[scale = 0.08]{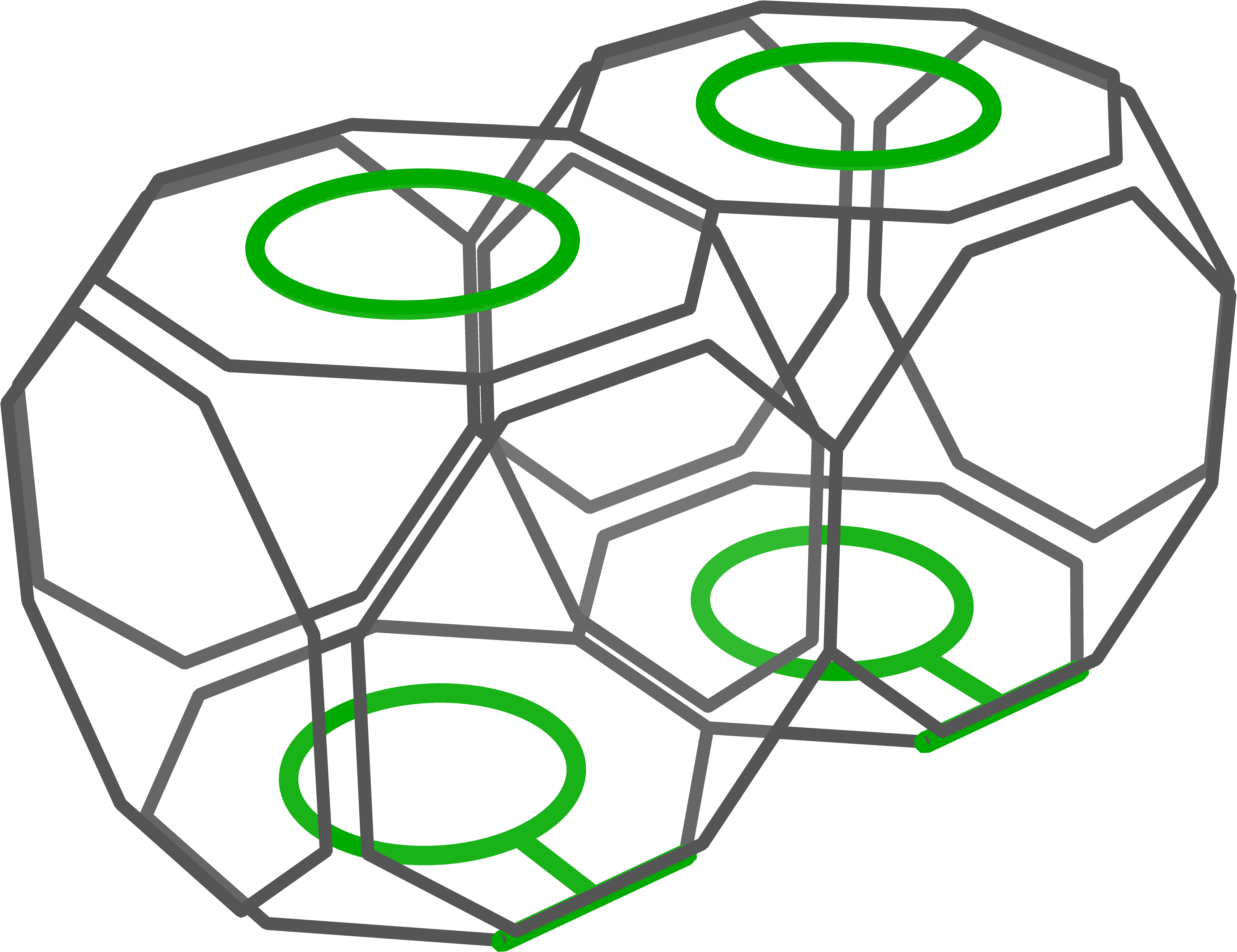}
        \caption{}
        \label{fig:Ising_CN_rectanguloid_circuit_1}
    \end{subfigure}
    \begin{subfigure}[b]{0.48\textwidth}
        \centering
        \includegraphics[scale = 0.08]{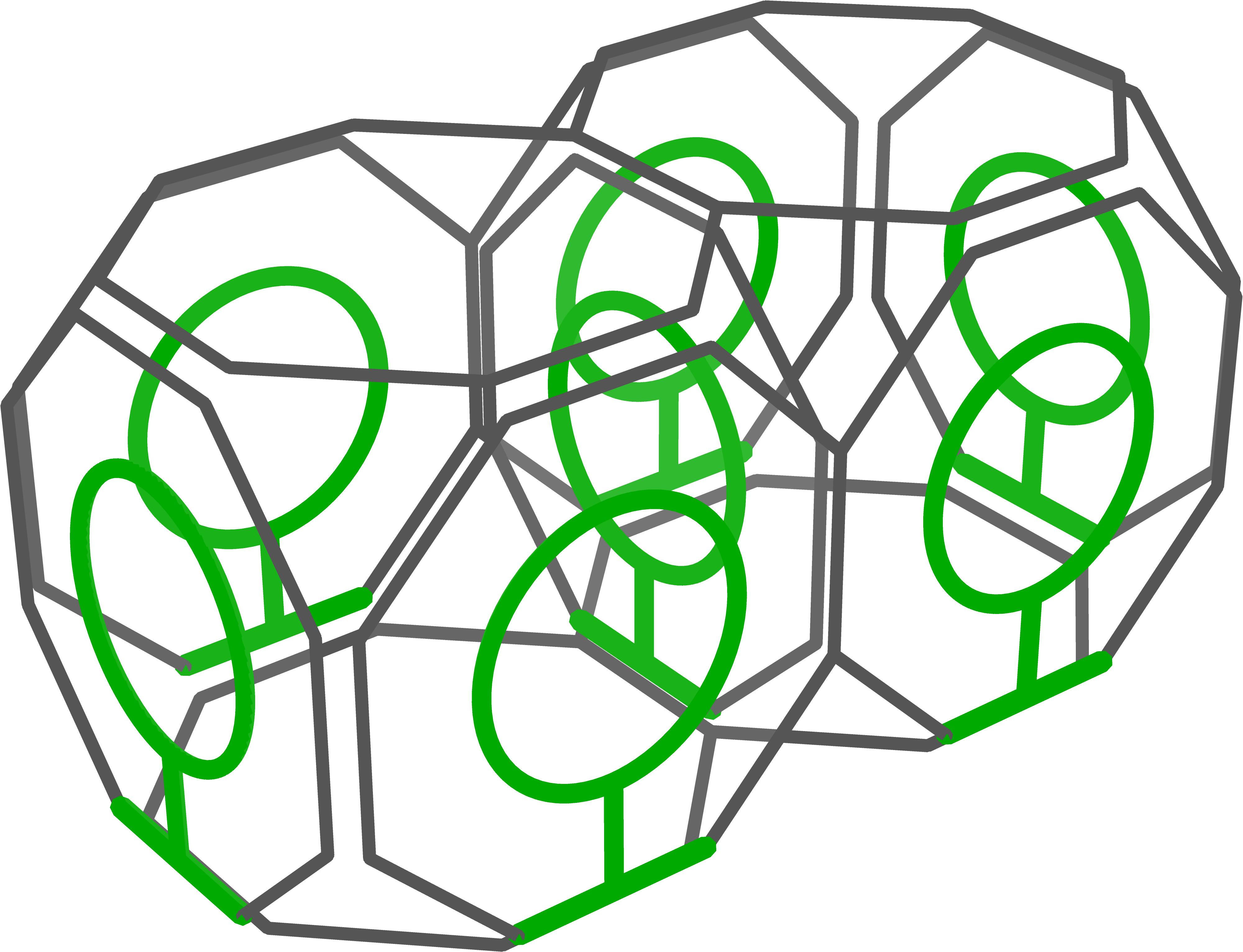}
        \caption{}
        \label{fig:Ising_CN_rectanguloid_circuit_2}
    \end{subfigure}
    \captionsetup{justification=Justified}
    \caption{Adding a row of cubes to the cage-net state, step 2: (a) first, we simultaneously add loops to the bottom and the top faces of all cubes in the row; (b), use the edges on the side face (colored green) as control to add loops to all the side faces at the same time.}
    \label{fig:Ising_CN_rectanguloid_circuit}
\end{figure}

In the RG circuit to be discussed in the next section, we need to grow cubes in the same row at the same time. This works in a similar way as growing a single cube and we describe the procedure here. First, as shown in Fig.~\ref{fig:Ising_CN_adding_two_plaquettes} which illustrates the situation with two cubes in the row, a new plane is added which bisects the row of cubes into two. Octahedrons are added to the intersection points to preserve the trivalent structure in the coupled $xy$, $yz$ and $zx$ planes. The `copying' process illustrated in Fig.~\ref{fig:Ising_CN_add_plaquette_diamond_vertices} is then used to restore vertex rules at the vertices of the octahedrons and then the square plaquettes in the octahedrons are added to the cage-net wave-function. The next step is illustrated in Fig.~\ref{fig:Ising_CN_rectanguloid_circuit}, which adds cage structures to a whole row of cubes at the same time. This is done by first picking the principal edge in, for example, the $x$ direction and use them as controls to add loops in the bottom and top faces as described above for each cube in the row (Fig.~\ref{fig:Ising_CN_rectanguloid_circuit} (a)). The operations in each cube commute with that in another cube, and hence they can be done all at the same time. Next, loops are added to the side faces using the principal edges on the bottom face as control, as shown in Fig.~\ref{fig:Ising_CN_rectanguloid_circuit} (b). Again, the operations on each side face commute with each other, so they can be done at the same time. As a result of this process, all the cubes in the row are now added to the cage-net wave-function. Note that the process illustrated in Fig.~\ref{fig:Ising_CN_rectanguloid_circuit} applies to the first row in the added plane. When we try to add subsequent rows, some of the side faces would have been added to the cage-net state already. Those side faces can be treated in the same way as the top face. That is, apply $B^1_{p_s}$ in step Fig.~\ref{fig:Ising_CN_rectanguloid_circuit} (a) when the $x$-principal edge is in the state $|10\rangle$, instead of applying $\sum_{\alpha_s}G^{\alpha_s}_{p_s}$ controlled by the bottom principal edge of the side face in the state $|s\alpha_s\rangle$. A similar procedure applies to the cubes in the last row of the added plane as well, which have to be added one by one.

\subsubsection{RG circuit -- Ising cage-net}
\label{sec:IsingCN_RG_circuit}

The processes for adding single cubes and a row of cubes are building blocks for the full RG circuit that adds a full plane to the cage-net state. Similar to the case of the doubled-Ising, we first need to initialize the added plane into proper eigenstates of the non-local logical operators before adding the local structures of cubic cages (plaquettes in the case of doubled-Ising). 

A commuting set of logical operators of the Ising cage-net ground space can be chosen to be generated by the string-operators of $\psi,\bar{\psi}$ planons in each $\mu\nu$ plane along the $\mu$ and $\nu$ directions respectively. We can choose the original cage-net state (before adding the plane) to be an eigenstate of all such logical operators. The added $xy$ plane can be initialized into an eigenstate of $\psi^x$, $\psi^y$, $\bar{\psi}^x$ and $\bar{\psi}^y$ on that plane. The circuit described in the last section on how to add cubic cages and plaquette terms to the wave-function does not affect these nonlocal logical operators. Therefore, the resulting cage-net state after the RG circuit remains an eigenstate of all the  $\psi,\bar{\psi}$ logical operators.

But the choice of the eigenvalue for the $\psi,\bar{\psi}$ logical operators is not arbitrary as the operators are related to each other and hence their eigenvalues are constrained. In Ref.~\onlinecite{GSD_IsCN}, we study carefully the relations among these operators, which allowed us to derive the ground state degeneracy of the Ising cage-net model. The relations are listed below. For derivation, see the discussion in section VII of Ref.~\onlinecite{GSD_IsCN}. For $\{\mu, \nu, \lambda\} = \{x,y,z\}$
\begin{align}
    \prod_i \left(\psi\bar{\psi}\right)^{\mu}_{\mu\lambda}(\nu=i) \prod_j \left(\psi\bar{\psi}\right)^{\nu}_{\nu\lambda}(\mu=i) = 1 \nonumber \\ 
    r_{\mu\nu}(\lambda=i)\bar{r}_{\mu\nu}(\lambda=i) = 1, \forall i, \forall \{\mu,\nu\} \nonumber \\ 
    r_{\mu\nu}(\lambda=i)r_{\mu\nu}(\lambda=i+1) = 1, \forall i, \forall \{\mu,\nu\}
\label{eq:relation_psi}
\end{align}
where $r_{\mu\nu} = \frac{1}{2}\left(1+\psi^{\mu}_{\mu\nu}+\psi^{\nu}_{\mu\nu}-\psi^{\mu}_{\mu\nu}\psi^{\nu}_{\mu\nu}\right)$, $\bar{r}_{\mu\nu} = \frac{1}{2}\left(1+\bar{\psi}^{\mu}_{\mu\nu}+\bar{\psi}^{\nu}_{\mu\nu}-\bar{\psi}^{\mu}_{\mu\nu}\bar{\psi}^{\nu}_{\mu\nu}\right)$. As we started from a ground state of the cage-net model, the original set of $\psi,\bar{\psi}$ operators satisfy the relations in Eq.~\eqref{eq:relation_psi}. When we add a new $xy$-plane, we need to make sure that after the new $\psi^{x}_{xy}$, $\psi^y_{xy}$, $\bar{\psi}^{x}_{xy}$, $\bar{\psi}^y_{xy}$ operators are added to the original set, the total set still satisfy the relations in Eq.~\eqref{eq:relation_psi}. This can be guaranteed when the added string-operators satisfy
\begin{align}
    \psi^{x}_{xy}\bar{\psi}^{x}_{xy}=1, \ \psi^y_{xy}\bar{\psi}^y_{xy}=1 \label{eq:relation_added1}\\
    r_{xy} = \bar{r}_{xy} = \pm 1 \label{eq:relation_added2}
\end{align}
The choice of $\pm 1$ in the last relation depends on whether $r_{xy}(z=i) = 1$ or $-1$ in the original set. Compared to the eigenstates listed in Appendix~\ref{sec:MinimalLattice_DI_SN}, $\ket{\Psi^\text{D.I.}_\text{min}}_1$, $\ket{\Psi^\text{D.I.}_\text{min}}_5$, $\ket{\Psi^\text{D.I.}_\text{min}}_9$ satisfy the relations in Eq.~\eqref{eq:relation_added1} and $r_{xy}=1$ while $\ket{{\psi\bar{\psi}}^\text{D.I.}_\text{min}}$ satisfies the relations in Eq.~\eqref{eq:relation_added1} and $r_{xy}=-1$. Therefore, we can initialize the added layer into one of these states.

\begin{figure}[ht]
    \centering
    \includegraphics[scale = 0.1]{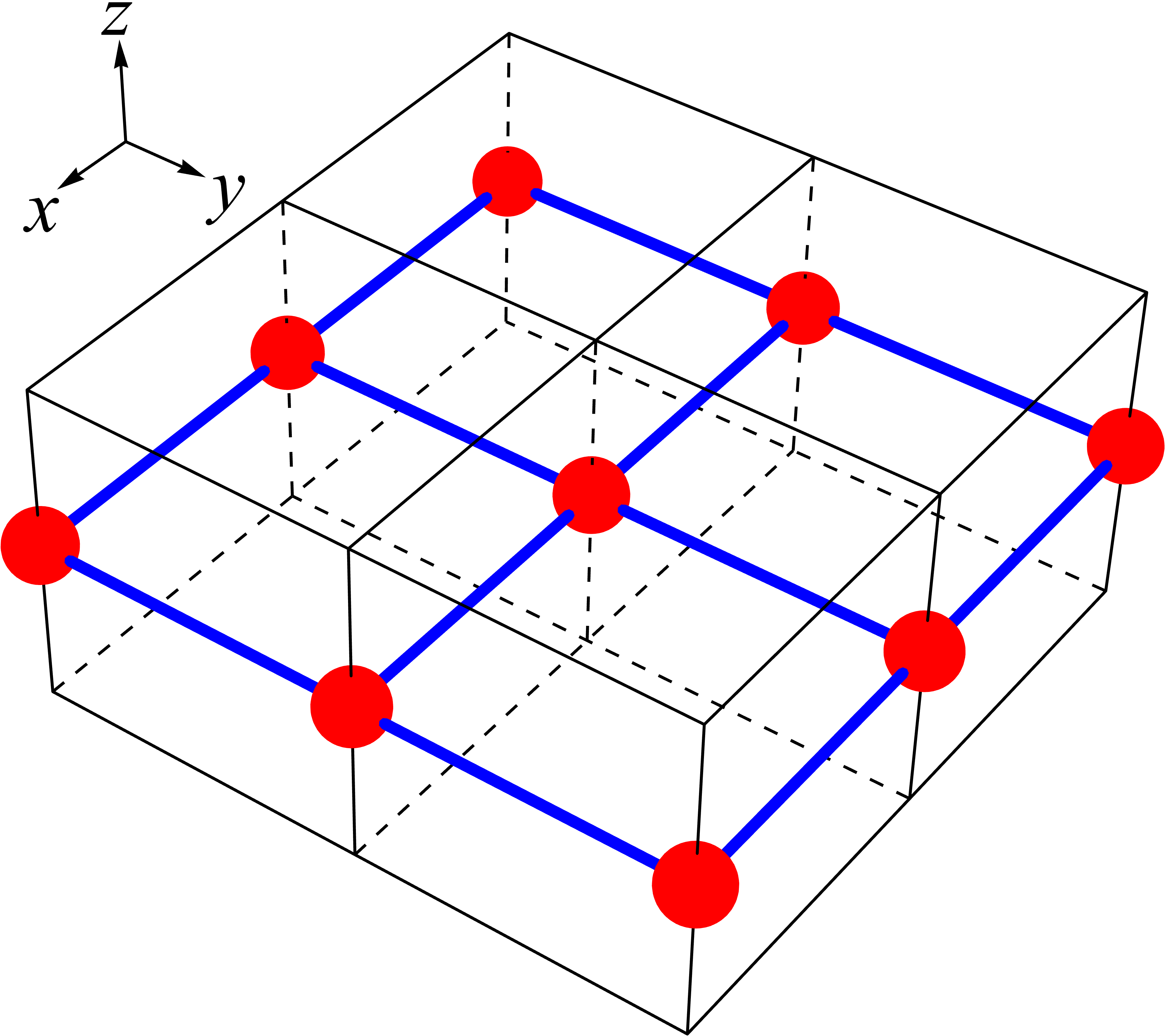}
    \captionsetup{justification=Justified}
    \caption{Inserting an $xy$-plane into the original cage-net lattice. Each red ball represents an octahedron. The new principal edges are shown in blue.
    }
    \label{fig:Ising_CN_add_layer}
\end{figure}

In particular, consider the added $xy$-plane in Fig.~\ref{fig:Ising_CN_add_layer}. Each red ball represents an octahedron. The added DOF are initially set to be either in state $|0\rangle$ (on edges of the octahedron) or $|00\rangle$ (on principal edges). Now initialize the trivalent lattice in the $xy$-plane into one of $\ket{\Psi^\text{D.I.}_\text{min}}_1$, $\ket{\Psi^\text{D.I.}_\text{min}}_5$, $\ket{\Psi^\text{D.I.}_\text{min}}_9$ and $\ket{{\psi\bar{\psi}}^\text{D.I.}_\text{min}}$ following the procedure described in Fig.~\ref{fig:SN_initialization}. This linear depth process set up the stage for the next step of the RG circuit: adding cage structures to the cubes.

\begin{figure}[ht]
    \centering
    \includegraphics[scale = 0.12]{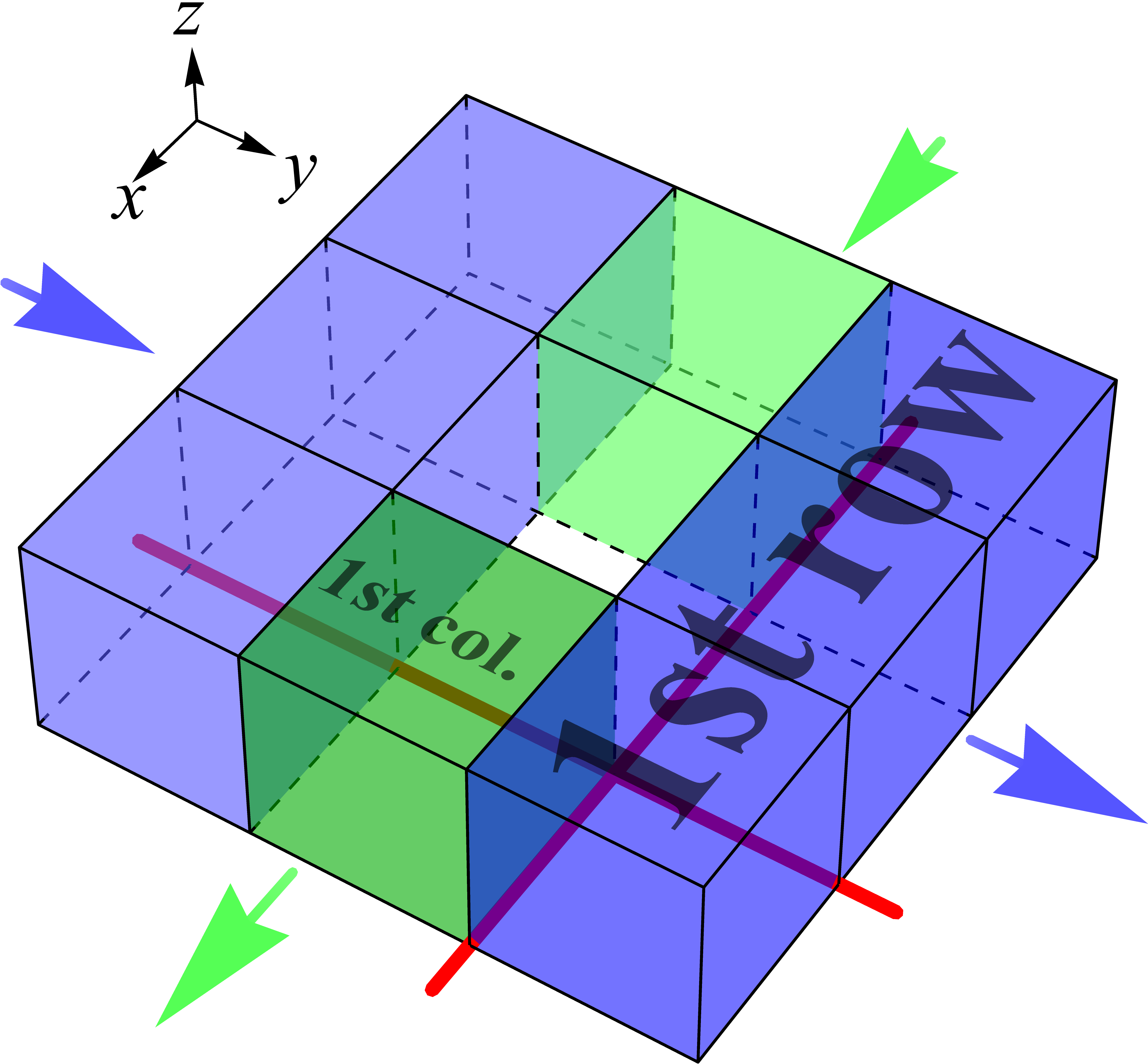}
    \captionsetup{justification=Justified}
    \caption{Adding cage structures to the cubes in step 2 of the RG circuit for the cage-net state. The red lines indicate the minimal lattice state determined by the initialization step. Cage structures are added to the cubes in the 1st row, the 2nd row, ... the $(L_y-1)$th row in each step. In the last row, cage structures are added to the cube in the 1st column, 2nd column, ..., $(L_x-1)$th column in each step. No action is required in the last cube.
   }
    \label{fig:Ising_CN_add_layer_RG_circuit}
\end{figure}

Now we can use the procedure described in the last section to add cage structures to the cubes. As shown in Fig.~\ref{fig:Ising_CN_add_layer_RG_circuit}, on top of the minimal structure set up in the initialization step (red lines), cage structures are added to the cubes in the 1st row, the 2nd row, ... the $(L_y-1)$th row in each step. In the last row, cage structures are added to the cube in the 1st column, 2nd column, ..., $(L_x-1)$th column in each step. No action is required in the last cube. This process has depth $\sim (L_x+L_y)$ and completes the addition of a new layer into the cage-net wave-function.

\section{Relating condensation and linear-depth circuits via gapped boundaries} \label{sec:condvcirc}

\subsection{General discussion}

In Sec.~\ref{sec:RG_cond}, we discussed the RG process in terms of condensation of planons. In Sec.~\ref{sec:RG_circ}, we discussed the RG process in terms of a linear depth circuit. In this section, we show that these two are closely related to each other by understanding each in terms of gapped boundaries.

We first consider a gapped boundary between a 2D topological order and vacuum. If an excitation moves from the bulk to the boundary, it may become trivial in the sense that it can be destroyed by a local operator on the boundary. This phenomenon is referred to as condensation at the boundary. On the other hand, some excitations remain non-trivial as they approach the boundary. These phenomena can be characterized precisely in a category-theoretic language\cite{Bais2009,kitaev2012models,KONG2014436,hung2015generalized}; in the abelian case, this amounts to specifying a maximal subset of bosons that can simultaneously condense at the boundary\cite{KAPUSTIN2011393,PhysRevB.88.241103,PhysRevB.88.235103,PhysRevX.3.021009}. It is believed the universality class of a gapped boundary is fully determined by its category-theoretic characterization.

The above discussion allows us to \emph{define} distinct types of anyon condensation (to vacuum) in a precise way, as distinct types of gapped boundaries (to vacuum). Such a definition is natural if we view the vacuum as a condensate of certain anyons in the 2D topological order. For instance, creating a puddle of anyon condensate within the bulk 2D topological order amounts to creating a puddle of trivial state (vacuum) separated from the bulk by a gapped boundary. This discussion, and the  definition of anyon condensation in terms of gapped boundaries, can be generalized to gapped boundaries between arbitrary 2D topological orders.

In the context of generalized foliated RG, we consider condensation of planons. Condensation of a single planon can similarly be associated with -- and defined in terms of -- certain gapped boundaries between two fracton orders, with the property that the boundary should be transparent to mobile excitations away from the selected plane where the condensation occurs. It will be an interesting problem for future work to fully characterize those boundaries between fracton phases that correspond to planon condensation. We note that there has been some related prior work discussing gapped boundaries of fracton models in terms of condensation \cite{Bulmash2019a,Luo2022}.

It turns out that the kind of linear-depth circuits considered here can also be associated with a type of gapped boundary. A linear depth circuit has the general form $\mathcal{U} = \prod_{\ell=1}^K U_\ell$ where each layer $U_\ell$ consists of a number of local unitary gates with non-overlapping support, and the number of layers $K$ is proportional to the linear system size $L$. In general, $U_\ell$ can contain gates acting across the entire system. However, for the circuits we employed for RG, each layer $U_\ell$ only contains gates acting in a lower dimensional subsystem of the entire system, such as the rows in Figs.~\ref{fig:SN_RG_Sgrow} and \ref{fig:Ising_CN_add_layer_RG_circuit}. Such circuits are much more restrictive than generic dense linear-depth circuits, particularly because they preserve the area law when acting on a state. We call this class of circuits \textit{sequential circuits}.

Again we first focus on the 2D case, where as we have discussed, sequential circuits can be used to generate topologically ordered ground states from an initial product state (the topological ``vacuum'').  In order to avoid complications associated with periodic boundary conditions, we make a simplification as compared to the circuits discussed in Sec.~\ref{sec:RG_circ}; namely, we work with an infinite system and consider circuits that generate a disc of 2D topological order from vacuum.  If desired, the size of the disc can later be taken to infinity.  This allows us to drop the initialization step, whose role is to take care of the non-trivial ground state degeneracy on a 2-torus.  We can also drop the final linear-depth sequence of gates needed to stitch two gapped boundaries together in a manner consistent with periodic boundary conditions.

With these simplifications, the circuits operate in the following way. We slice the 2D space into 1D concentric circles surrounding the center of the disc, and order these subspaces according to their radial coordinate. The $\ell$th layer of the circuit is assumed to be supported near (but not entirely within) the $\ell$th circle. After applying some number of layers of the circuit, one is left with a disc of topological order which has a gapped boundary to the vacuum region which has not yet been acted on by the circuit. Then, the next layer in the circuit acts only within the vicinity of the one-dimensional gapped boundary between the topological order and the vacuum. The action of the unitary in this layer is to ``grow'' the topological order by a small amount, pushing the gapped boundary further into the vacuum region. Continuing in this way allows one to grow the topologically ordered region arbitrarily. 

Based on the above, given a sequential circuit, we can associate the universality class of the gapped boundary to vacuum which emerges when the circuit is truncated at some radius. This association is well-defined in the following sense. We can define a truncation of the circuit $\bar{\mathcal{U}}=\sum_{\ell=1}^{K_0}U_\ell$ where $K_0<K$. This will create a disc of topological order with a particular gapped boundary to vacuum. Now, consider a different truncation $\bar{\mathcal{U}}'=\sum_{\ell=1}^{K_0}V_\ell$ where each $V_\ell$ again consists of non-overlapping gates such that $V_\ell=U_\ell$ for $\ell$ sufficiently less than $K_0$, but the layers near the boundary may differ. By definition, the two truncated circuits differ only by a finite-depth circuit near the boundary. But a 1D finite depth circuit cannot change the universality class of the gapped boundary, \emph{i.e.} it cannot change the set of anyons which can condense on the boundary. So the gapped boundary type is independent of how the sequential circuit is truncated.  We note this conclusion only holds for truncations that are compatible with the 1D layer structure of concentric circles; the key property is that the truncation only cuts through a finite number of 1D layers, which is bounded above as the size of the disc increases.

We emphasize that this discussion can be generalized to gapped boundaries between two different 2D topological orders. That is, given two topological orders referred to as A and B that admit a gapped boundary, an A-ground-state can be converted into a B-ground-state by applying a sequential circuit. Or, if we apply a truncated version of the same sequential circuit, we can create a puddle of B within the bulk topological order A, separated by a gapped boundary whose universality class does not depend on how the circuit is truncated.

In formulating the generalized foliated RG in terms of quantum circuits, we apply sequential circuits within 2D layers of a 3D fracton model. Truncating such a sequential circuit (along its 1D layer structure) results in a gapped boundary between two different fracton orders, where some of the mobile excitations may condense along the layer where the circuit is applied. This is how we described planon condensation above, and thus we propose that planon condensation and applying 2D sequential circuits are different ways to realize the same operation in generalized foliated RG.

\subsection{Condensation in the Ising cage-net circuit}
\label{sec:circu_&_condensation}

In accordance with the above discussion, we now identify the type of gapped boundary that is associated with the sequential circuits used to create Ising cage-net model. To accomplish this, we are going to apply the circuit only to a finite disc-shaped region within a plane; we will not take the limit that the size of the disc goes to infinity.  Inside the region, we get the fracton order as expected. Outside of the region, the added degrees of freedom remain unentangled. There is a gapped boundary between the two sides. We show that the gapped boundary and the region outside can be obtained by condensing bosonic planons starting from a complete fractonic state.

\begin{figure}[h]
    \centering
    \begin{tikzpicture}[baseline={([yshift=-.5ex]current bounding box.center)}, every node/.style={scale=1}]
        \begin{scope}
            \node[inner sep=0pt] (pt1) at (0,0) {\includegraphics[scale = 0.116]{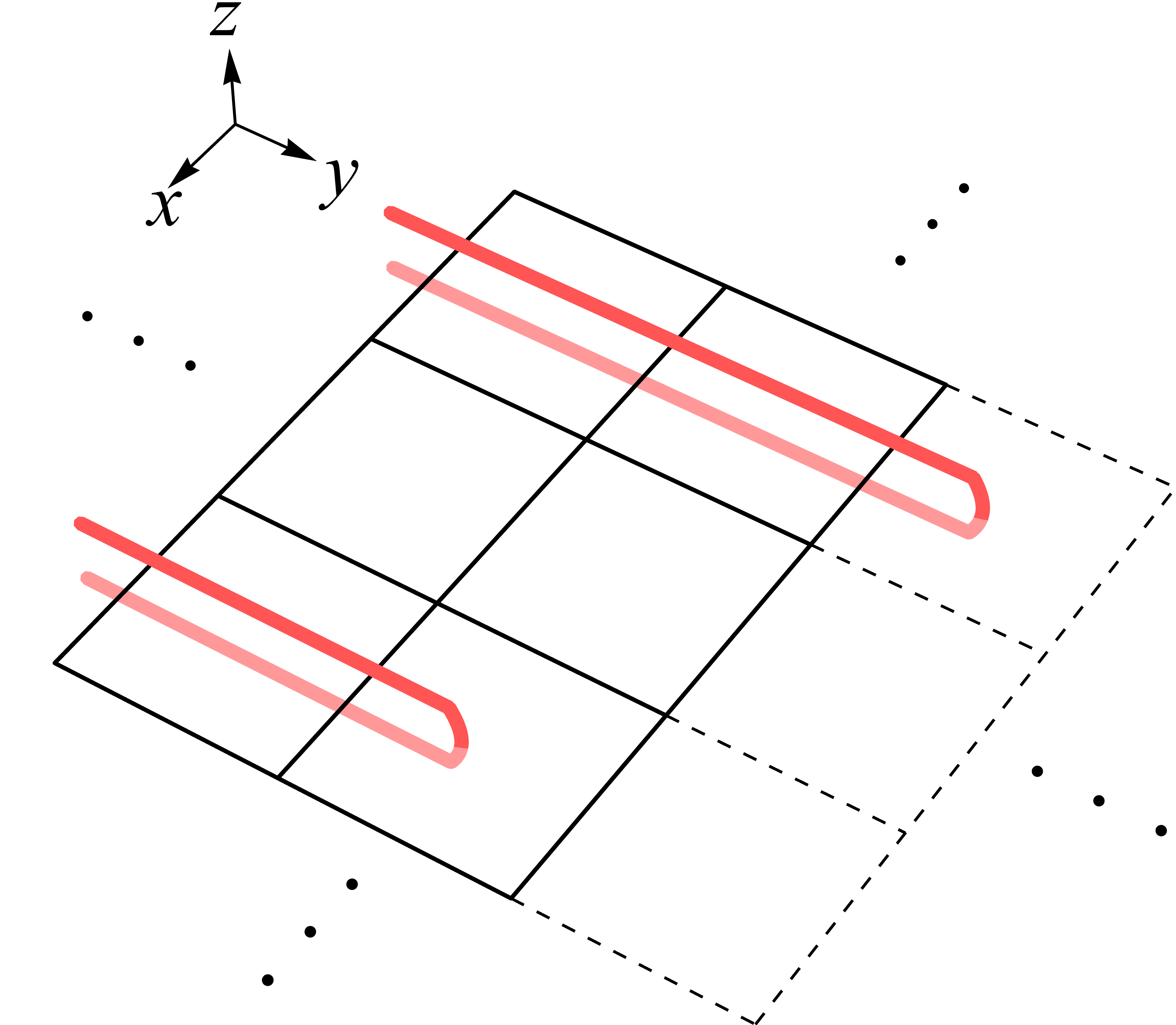}};
            \draw[draw = red, line width = 1.2 pt, -latex](-3.9,-1.7) to[out = 120, in = 220] (-3.5,-0.2);
            \node [] at (-3.5,-2) {\scalebox{1.15}{$\textcolor{red}{\bm{s\textbf{-loop}}}$}};
            \node [] at (1.7,1.6) {\scalebox{1.15}{$\bm{i}$}};
            \node [] at (0.5,2.2) {\scalebox{1.15}{$\bm{i-1}$}};
            \node [] at (3.5,0.9) {\scalebox{1.15}{$\bm{i+1}$}};
            \node [] at (-0.5,-1.8) {\scalebox{1.15}{$\bm{p}$}};
            \node [] at (3.2,-0.1) {\scalebox{1.15}{$\bm{p'}$}};
        \end{scope}
    \end{tikzpicture}
    \captionsetup{justification=Justified}
    \caption{Condensation of the $\psi\bar{\psi}$ and the $\sigma\bar{\sigma}$ fluxons on the smooth boundary of the doubled-Ising model. The vertex details are omitted. The dashed lines represent the unentangled edges. An open ended fluxon string-operator is constructed from a loop of $s$-string that passes through the lattice plane vertically at a plaquette. If the plaquette (for example, the one labeled $p$) lies within the doubled-Ising region, it creates a fluxon excitation. If the plaquette (for example, the one labeled $p'$) falls outside the string-net region, then no excitation is generated. Thus, all fluxons condense on the smooth boundary. For computational details on the condensation, see Appendix~\ref{sec:DI_fluxon_condensation}.}
    \label{fig:FluxonCondensation_SN}
\end{figure}
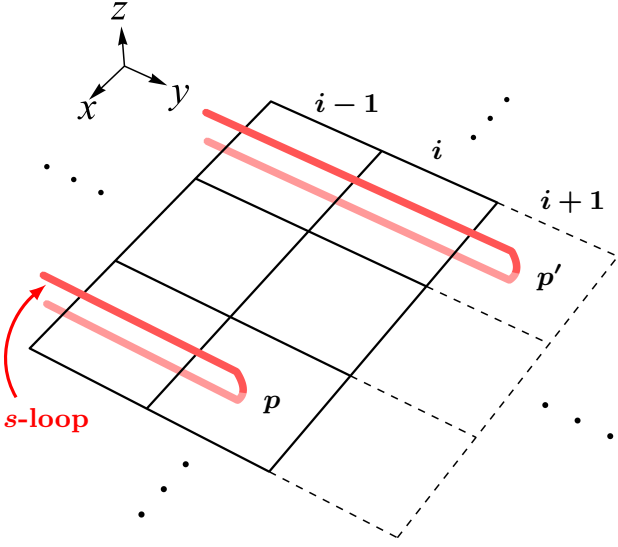

First, let's see how a similar relation works in the doubled-Ising string-net state. We imagine a very large disc of string-net state, and we ignore the curvature of the disc's boundary to simplify the following discussion.  Recall that in the RG circuit, the plaquettes are added row by row. Suppose that we stop the process at row $i$. The boundary between row $i$ and row $i+1$ is a smooth boundary on the lattice. As the Hamiltonian terms remain commuting throughout the process, the boundary is gapped.

The gapped boundary can be induced by the condensation of `fluxon excitations'\cite{ExtendedStringnet_Hu} $\psi\bar{\psi}$ and $\sigma\bar{\sigma}$ on the boundary and beyond. To see that, consider a string-operator of the form shown in Fig.~\ref{fig:FluxonCondensation_SN}, which consists of a string segment above the lattice, a parallel segment under the lattice and the two are connected by segments that vertically go through the lattice plane. Note that, while embedded in the 3D space, the string-operator is a closed loop, from the 2D perspective, it ends at the locations where the string goes through the lattice plane and can create excitations at those points. In particular, such string-operators in general violate the plaquette term at their ends, as the plaquette terms correspond to a loop operator that links with the string-operator and the linking generates nontrivial action. Therefore, in the bulk of the string-net state, the string-operator generates `fluxon excitations' at its ends. In the doubled-Ising model, there are two string-operators of this type, corresponding respectively to a loop of string type 1 and a loop of string type 2. The two string-operators generate the $\psi\bar{\psi}$ and $\sigma\bar{\sigma}$ excitations, respectively. If the string-operator ends (goes vertically through the lattice plane) outside of the smooth boundary (Fig.~\ref{fig:FluxonCondensation_SN}), there are no more plaquette terms to violate and the string-operator does not generate any excitations. Detailed calculations can be found in Appendix~\ref{sec:DI_fluxon_condensation}. Therefore, the $\psi\bar{\psi}$ and $\sigma\bar{\sigma}$ excitations condense on the boundary and beyond, thus demonstrating the connection between anyon condensation and the linear depth circuit for the doubled-Ising string-net state.

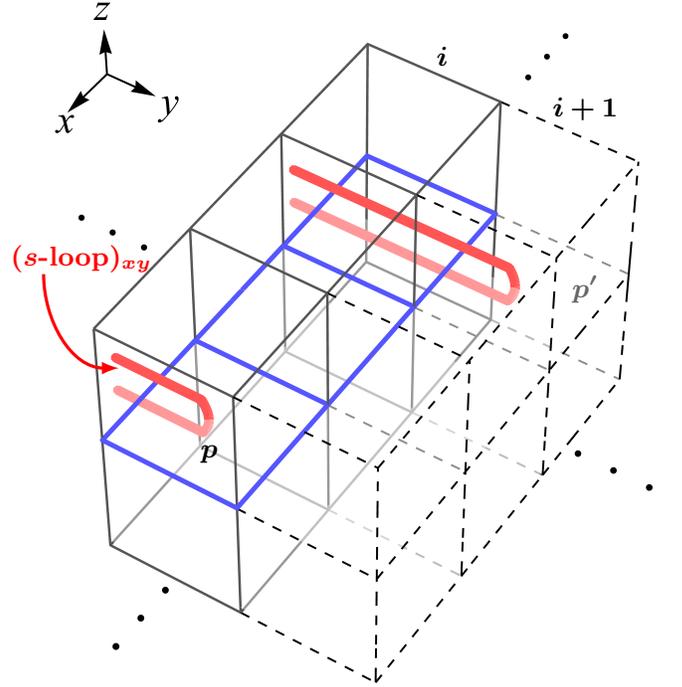
\begin{figure}[h]
    \centering
    \begin{tikzpicture}[baseline={([yshift=-.5ex]current bounding box.center)}, every node/.style={scale=1}]
        \begin{scope}
            \node[inner sep=0pt] (pt1) at (0,0) {\includegraphics[scale = 0.116]{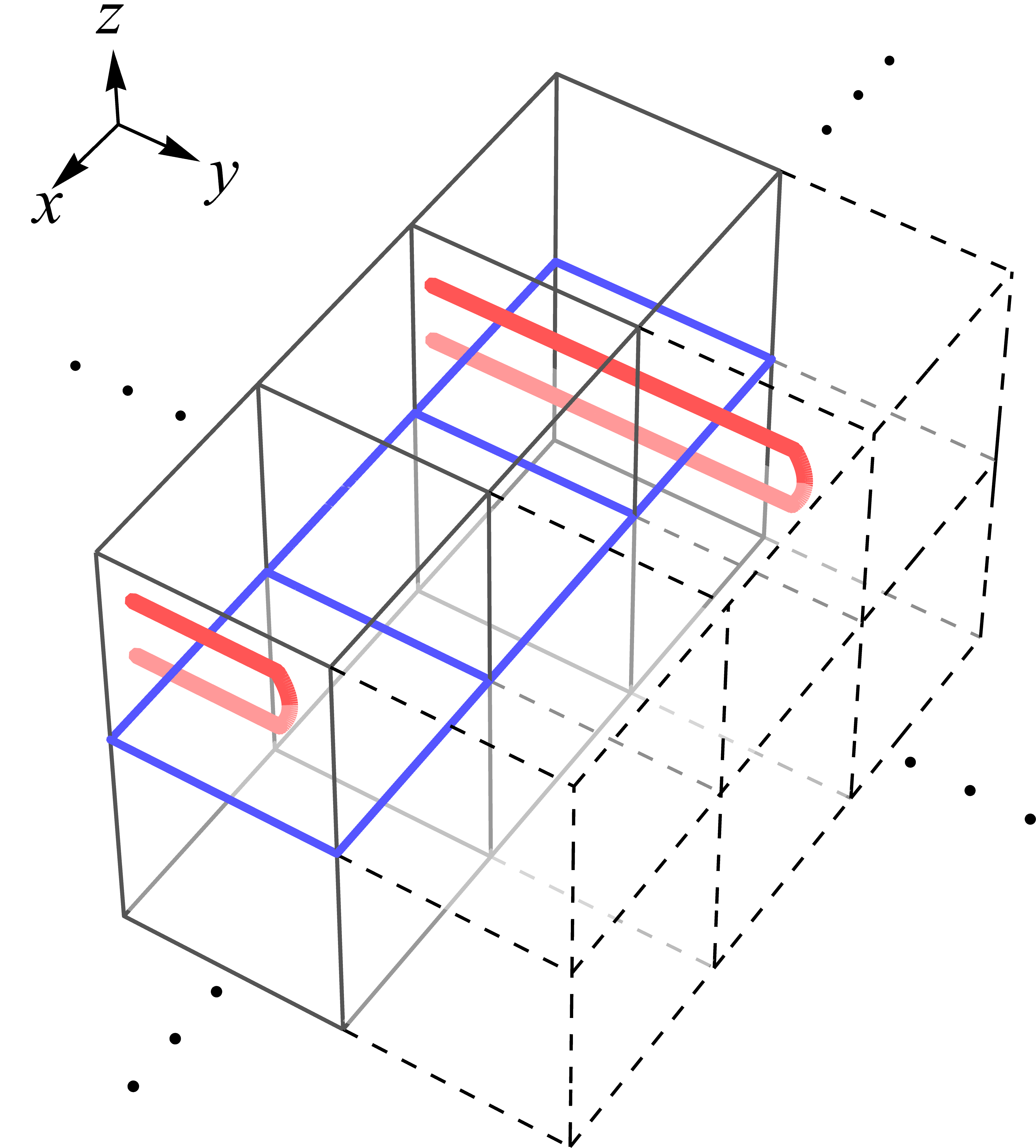}};
            \draw[draw = red, line width = 1.2 pt, -latex](-4,0.9) to[out = 270, in = 180] (-3,-0.35);
            \node [] at (-3.5,1.1) {\scalebox{1.15}{$\textcolor{red}{\bm{(s\textbf{-loop})_{xy}}}$}};
            \node [] at (1.3,3.8) {\scalebox{1.15}{$\bm{i}$}};
            \node [] at (3.2,3.1) {\scalebox{1.15}{$\bm{i+1}$}};
            \node [] at (-1.8,-1.5) {\scalebox{1.15}{$\bm{p}$}};
            \node [] at (3.2,0.7) {\scalebox{1.15}{$\textcolor{black!60!white}{\bm{p'}}$}};
        \end{scope}
    \end{tikzpicture}
    \captionsetup{justification=Justified}
    \caption{Condensation of the $\psi\bar{\psi}$ and the $\sigma\bar{\sigma}$ fluxon excitations in the half $xy$-plane (shown in blue) in the Ising cage-net. If the end of the the fluxon string operator falls within the Ising cage-net region (for example at the plaquette $p$), a fluxon excitation is created. If the end falls outside of the Ising cage-net region (for example at the plaquette $p'$), then no excitation is generated. Therefore, both $\psi\bar{\psi}$ and $\sigma\bar{\sigma}$ planons condense on the boundary.
    }
    \label{fig:FluxonCondensation_IsingCN}
\end{figure}

The situation is very similar in the Ising cage-net model. The RG circuit is again implemented row by row in a sequential manner. Suppose that we stop the process at row $i$, there will be a gapped boundary between row $i$ and row $i+1$. As shown in Fig.~\ref{fig:FluxonCondensation_IsingCN}, like for the string-nets, a vertical loop operator that goes through the lattice plane at two points generates planon excitations $\psi\bar{\psi}$ and $\sigma\bar{\sigma}$ in the bulk of the cage-net state (in rows $j\le i$). Beyond row $i$, however, it does not generate any excitations and hence the $\psi\bar{\psi}$ and $\sigma\bar{\sigma}$ are condensed. This agrees with the RG procedure driven by condensation described in Sec.~\ref{sec:RG_cond}. Therefore, the process of sequential application in the linear depth circuit can be interpreted as moving the boundary between the cage-net state and the condensed state, hence enlarging or shrinking the fracton order in the plane.

\section{Summary and discussion}
\label{sec:summary}

In this paper, we studied the renormalization group transformation for the Ising cage-net model and found that the system size of the Ising cage-net model can be decreased / increased by condensing / uncondensing planon excitations near a 2D plane, or correspondingly through a so-called sequential circuit which preserves the area law and whose depth scales with the linear size of the plane. We argued that these two ways of carrying out the RG are closely related through gapped boundaries. 

We call this procedure the generalized foliated RG, because the previously defined foliated RG, under which the X-cube and related models are fixed points,\cite{Shirley_2018} fits into this new definition as a special case. On the one hand, the system size of the X-cube can be decreased / increased by condensing / uncondensing a lineon dipole or fracton dipole on a given plane (both these excitations are planons). Or, the RG procedure can be carried out with a linear depth circuit in the same plane. One way to construct the linear depth circuit is to use the finite depth circuit discussed for the original foliation scheme\cite{Shirley_2018} to decouple a layer of toric code out of the X-cube model, and then disentangled the toric code into product state with a linear depth circuit. Altogether this is a linear depth circuit. Alternatively, we can use a circuit similar to that discussed in Sec.~\ref{sec:RG_circ} to remove cage structures in a plane row by row and hence removing a plane from the X-cube model. 

On the other hand, the generalized foliated RG allows a broader class of RG operations. Indeed, the Ising cage-net model is not a fixed point of the original foliated RG as can be seen from its ground state degeneracy calculation\cite{GSD_IsCN}.
We recall that the original foliated RG led to an associated notion of foliated fracton phases (see Appendix~\ref{app:foliated-phases} for a definition), with the key property that two systems related by a foliated RG operation lie within the same foliated fracton phase. Similarly, we expect that there exists a notion of \emph{generalized foliated fracton phase} (GFF phase), again with the key property that two systems related by a generalized foliated RG operation lie in the same GFF phase. GFF phases should be a coarser equivalence relation on quantum systems than foliated fracton phases, because a broader class of RG operations are allowed. We do not currently know how to give a definition of GFF phases along the lines of those in Appendix~\ref{app:foliated-phases}; however, one possibility is to give a definition based on circuit equivalence of ground states, where one allows certain linear depth circuits supported on planes.  

In Sec.~\ref{sec:gfoliation}, we pointed out that the original foliated RG contains certain unnatural restrictions, while the generalized foliated RG seems to be more natural. Therefore, we expect that GFF phases are correspondingly a more natural concept than foliated fracton phases as originally defined, so it will be important to revisit what we have learned about foliated fracton phases. In particular, several invariants have been devised for foliated fracton phases as originally defined, including those based on fractional excitations and entanglement entropy \cite{Shirley_2019_excitation,Shirley_2019}. Now, with a new notion of GFF phases, we need to reconsider the question of what quantities remain invariant under the new equivalence relation, and which models belong to the same GFF phase and which do not. For example, we can ask whether the twisted foliated fracton model proposed in Ref.~\onlinecite{shirley2019twisted} is still in a different phase than the X-cube model or not under the new definition.

Finally, we want to comment that the generalized foliation defined in this paper makes the discussion of type I fracton models more in-line with that of Subsystem Symmetry Protected Topological (SSPT) phases with planar symmetry in e.g. Ref.~\onlinecite{You2018, Devakul2018, Devakul2020}. In the definition of `strong SSPT' in these papers, when a decoupled layer with planar symmetry is added to the bulk of the system, the planar symmetry can be combined with an existing planar symmetry in the system, which corresponds to the condensation of the composite of the symmetry charges from the decoupled plane and a planar symmetry 
charge in the bulk of the system. The `strong SSPT' orders discussed in these papers hence may become nontrivial (twisted) foliated fracton orders when the planar symmetries are gauged.

\begin{acknowledgments}
We are indebted to inspiring discussions with Dave Aasen, Kevin Slagle, Nathan Seiberg, and Dominic Williamson, and helpful correspondence with Fiona Burnell and Michael Levin. Z.W., X.M. and X.C. are supported by the National Science Foundation under award number DMR-1654340, the Simons Investigator Award (award ID 828078) and the Institute for Quantum Information and Matter at Caltech. X.C. is also supported by the Walter Burke Institute for Theoretical Physics at Caltech. The research of MH is supported by the U.S. Department of Energy (DOE), Office of Science, Basic Energy Sciences (BES) under Award number DE-SC0014415. This work is also partly supported by the Simons Collaboration on Ultra-Quantum Matter, which is a grant from the Simons Foundation (651438, XC and ZW; 651440, MH and DTS). The work of MH on general aspects of the generalized foliated RG (Sections~\ref{sec:gfoliation}, \ref{sec:RG_cond} and~\ref{sec:condvcirc}) was supported by the DOE BES project, while his work on the RG in the Ising cage-net model (Sec.~\ref{sec:RG_circ}) was supported by the Simons Foundation. X.C. wants to thank the Institute for Advanced Study at Tsinghua University for hospitality when the paper was written.
\end{acknowledgments}

\bibliography{references}

\appendix

\section{Definition of foliated fracton phases}
\label{app:foliated-phases}

Here we give a definition of foliated fracton phases, and in the process provide a framework for thinking about the relationship among fracton phases, universality and RG fixed points. Here we focus on foliated fracton phases as introduced in earlier works; in Sec.~\ref{sec:summary}, we briefly comment on a possible definition of generalized foliated fracton phases associated with the generalized foliated RG. An important point is that notions of standard and foliated phases both play important, but different, roles in fracton physics. For ease of presentation, we do not consider symmetry in this discussion. 

First we recall the definition of standard gapped phases and make some comments on the physical basis of this definition. Phases are equivalence classes of systems; by a system we mean a specification of the degrees of freedom on some $d$-dimensional spatial lattice, together with a local Hamiltonian $H$. In a slight abuse of notation we use $H$ to denote the system and not just its Hamiltonian. Two systems $H$ and $H'$ are in the same phase (considered equivalent) if there exist resource systems $R$ and $R'$ so that there is a continuous path between the Hamiltonians for the systems $H \otimes R$ and $H' \otimes R'$. Here ``$\otimes$'' denotes the operation of stacking two systems. Each ``trivial'' resource system is a collection of gapped, decoupled zero-dimensional systems arranged in $d$-dimensional space, \emph{i.e.} $R$ and $R'$ have product ground states. The energy gap is required to remain open along the continuous path, which must also avoid first-order phase transitions.\footnote{While it may seem the latter condition is redundant, it is needed if one defines the energy gap as the gap to local excitations.} A special case of the above definition is that $H$ and $H'$ are equivalent if there is a continuous path between their Hamiltonians (without stacking with resource systems).

Typically we are interested in the universal properties of phases, which we define simply as those properties that are the same everywhere within a phase. In many cases, a standard phase contains within it a representative system that is a RG fixed point under some (conventional) scheme for carrying out the RG. When this occurs, the universal properties of a phase are encapsulated in the properties of the RG fixed point. This holds because two (infinite) systems related by a RG step are in the same phase; for instance, this property is clear in ``entanglement RG'' schemes.

Why do we consider stacking with trivial resource systems? This certainly leads to nice mathematical properties, and has the advantage of allowing for comparison between systems with different local degrees of freedom. However, there is a more fundamental reason, namely to obtain a definition of phases that can be distinguished in experiments, at least in principle. The key point is that lattice models are always idealizations of continuum systems, where some degrees of freedom are deemed unimportant and left out of the model (\emph{e.g.} atomic core levels). Any physically measurable notion of phases cannot depend on which degrees of freedom we choose to include or ignore in a theoretical model, and this issue is addressed by including stacking with trivial resource systems in the equivalence relation. 

More generally, we emphasize that the equivalence relation used to define standard phases is not arbitrary. Given the idea that gapped phases should be connected components of parameter space where the gap remains open, standard phases are the finest equivalence classes that can be distinguished (in principle) by experiments. Therefore, it is always physically relevant to consider standard phases, even when the universal properties of a standard phase are not captured in a RG fixed point, as occurs in fracton models. Put another way, we cannot achieve a complete understanding of fracton physics if we ignore standard phases. However, this does not preclude the relevance of other notions of phases to fracton physics.

Before defining foliated fracton phases, we first introduce the closely related notion of $F$-phases. In order to talk about $F$-phases (and foliated fracton phases), we need to introduce a foliation of $3$D space, which is a certain geometrical structure. In particular, a foliation consists of one or more decompositions of space into parallel $2$D layers. If we have $k$ separate decompositions, we sometimes speak more specifically about a $k$-foliation. An important example of a 3-foliation is given by the sets of all $xy$, $yz$ and $xz$ planes.

Similar to standard phases, two systems $H$ and $H'$ are considered to be in the same $F$-phase (or to be $F$-equivalent) if there is a continuous path between the Hamiltonians for $H \otimes R$ and $H' \otimes R'$. The difference from standard phases is that the resource systems $R$ and $R'$ are allowed to consist of decoupled gapped $2$D systems on any layers of the foliation structure. For instance, $R$ can consist of decoupled $2$D toric codes lying on a set of $xy$ planes (as long as the foliation structure includes $xy$ planes). Here, as in the foliated RG, we view $2$D layers as a free resource, analogous to product states in the definition of standard phases.

The relation of $F$-equivalence is obviously coarser than standard phase equivalence, because the set of allowed resource systems contains those allowed for standard phases. Moreover, it is strictly coarser; for instance, a stack of $2$D topologically ordered layers is non-trivial as a standard phase, but is in the trivial $F$-phase (with appropriate foliation structure). Therefore each $F$-phase can contain multiple standard phases.

The reason we define $F$-phases is to be able to define foliated fracton phases, which are those $F$-phases that contain a representative system that is a fixed point of the foliated RG. Because two (infinite) systems related by a foliated RG step are in the same $F$-phase, it is expected that the fixed point of a foliated fracton phase captures certain universal properties that are the same throughout the foliated fracton phase, and are referred to as its foliated fracton order. It is important to emphasize that the foliated fracton order consists of properties that are the same even within different standard phases, so long as these standard phases are $F$-equivalent and belong to a foliated fracton phase.

Foliated fracton phases are a useful concept in the study of fracton physics because, in some fracton systems, they restore a connection between universal properties of a phase and a RG fixed point. This connection fails when we study fracton models using standard phases.

It should be noted that not every $F$-phase is a foliated fracton phase. For instance, the $F$-phase containing Haah's cubic code model is not a foliated fracton phase for any choice of foliation structure. It is not clear that such $F$-phases are interesting objects of study.

\section{Review of string-net models}
\label{sec:Review_SN}

In Appendix \ref{sec:SN_models}, we review the basics of the string-net models that are relevant for our purposes. We follow the original construction as introduced in Ref.~\onlinecite{Stringnet}. For more comprehensive introductions, we refer the readers to Ref.~\onlinecite{Stringnet,ExtendedStringnet_Hu,TianLan_stable,SN_thorough}. In Appendix \ref{sec:SN_minimalLattice_torus}, we discuss the string-net models on the minimal lattice on the torus.

\subsection{String-net models}
\label{sec:SN_models}

The input data of a string-net model is a unitary fusion category\cite{ZH_Wang_topologicalQcomputation}, which includes an index set $\{0,1,...,N\}$ and the associated data set $(\delta_{ijk},d_s,F^{ijm}_{k\ell n})$. A string-net model is defined on a trivalent lattice, where the local DOF live on the edges. Each edge has a Hilbert space of $\text{span}_\mathbb{C}\{\ket{0},\ket{1}...,\ket{N}\}$. Usually, an edge of the string-net is represented by a directed line. For a directed edge, $i^*$ represents the edge in the state $i$ pointing in the opposite direction. That is
\begin{equation}
    \begin{tikzpicture}[baseline={([yshift=-1.5ex]current bounding box.center)}, every node/.style={scale=1}]
        \begin{scope}[decoration={markings, mark=at position 0.8 with {\arrow{Latex}}}]
            \pgfmathsetmacro{\lineW}{0.8}
            \pgfmathsetmacro{\radius}{0.7}
            \pgfmathsetmacro{\halfradius}{\radius/2}
            
            \draw[line width = \lineW pt, postaction={decorate}] (0,0) -- (\radius,0);
            \draw[line width = \lineW pt, postaction={decorate}] (3.5*\radius,0) -- (2.5*\radius,0);
            
            \node [] at (\halfradius, 0.3) {$i^*$};
            \node [] at (\halfradius+2.5*\radius, 0.3) {$i^\text{ }$};
            \node [] at (1.75*\radius, 0) {$=$};
        \end{scope}
    \end{tikzpicture} \quad .
\end{equation}
In particular, $0^*= 0$. 

The $\delta$-symbol specifies the vertex rules. $\delta_{ijk}$ takes values in $\{0,1\}$ and it is symmetric under permutation of the indices. $\delta_{ijk}$ determines the allowed states on edges at a trivalent vertex. A vertex is stable\cite{TianLan_stable} if
\begin{equation}
    \begin{tikzpicture}[baseline={([yshift=-.5ex]current bounding box.center)}, every node/.style={scale=1}]
        \begin{scope}[decoration={markings, mark=at position 0.8 with {\arrow{Latex}}}]
            \pgfmathsetmacro{\lineW}{0.8}
            \pgfmathsetmacro{\radius}{0.7}
            \pgfmathsetmacro{\halfradius}{\radius/2}
            \pgfmathsetmacro{\centerx}{0}
            \pgfmathsetmacro{\centery}{0}
            \pgfmathsetmacro{\Rightptx}{\centerx + \radius}
            \pgfmathsetmacro{\Rightpty}{0}
            \pgfmathsetmacro{\Upptx}{\centerx + \radius*sin(315)}
            \pgfmathsetmacro{\Uppty}{\centery + \radius*cos(315)}
            \pgfmathsetmacro{\Lowptx}{\centerx + \radius*sin(225)}
            \pgfmathsetmacro{\Lowpty}{\centery + \radius*cos(225)}
            
            \draw[line width = \lineW pt, postaction={decorate}] (\Upptx,\Uppty) -- (\centerx,\centery);
            \draw[line width = \lineW pt, postaction={decorate}] (\Lowptx,\Lowpty) -- (\centerx,\centery);
            \draw[line width = \lineW pt, postaction={decorate}]  (\Rightptx,\Rightpty) -- (\centerx,\centery);
            
            \node [] at (\Upptx+\halfradius, 1.5*\halfradius) {$i$};
            \node [] at (\Upptx+\halfradius, -1.5*\halfradius) {$j$};
            \node [] at (\halfradius, \centery-0.25) {$k$};
        \end{scope}
    \end{tikzpicture}
\end{equation}
satisfies $\delta_{ijk} = 1$. A vertex is not stable if $\delta_{ijk} = 0$.

The $d$- and $F$-symbols define the graphical rules. The $d$-symbols evaluate loops to real numbers as
\begin{equation}
    \vcenter{\hbox{\includegraphics[scale=1]{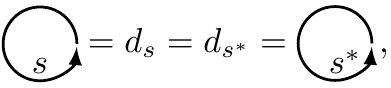}}}
    \label{eq:graphical_rule_loops}
\end{equation}
where $d_0 = 1$. They satisfy the equation
\begin{equation}
    d_i d_j = \sum_{k}\delta_{ijk^*} d_k.
    \label{eq:The_d_symbol_relation}
\end{equation}

The $F$-symbols define the transformations
\begin{equation}
    \vcenter{\hbox{\includegraphics[scale=1]{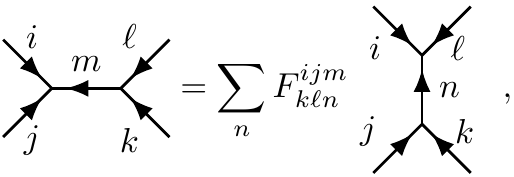}}}
    \label{eq:graphical_rule_F}
\end{equation}
where the $F$-symbols are nonzero if all the vertices satisfy the vertex rules. They are normalized as
\begin{equation}
    F^{ijk}_{j^* i^* 0} = \sqrt{\frac{d_k}{d_i d_j}}\delta_{ijk} \, .
    \label{eq:F_normalization}
\end{equation}
For the cases of interest in this paper, the $F$-symbols satisfy the tetrahedral symmetry
\begin{equation}
    F^{ijm}_{k\ell n} = F^{\ell k m^*}_{j i n} = F^{j i m}_{\ell k n^*} = F^{i m j}_{k^* n \ell} \sqrt{\frac{d_m d_n}{d_j d_\ell}} ,
    \label{eq:tetrahedral_symmetry}
\end{equation}
as well as \textit{the pentagon equation}
\begin{equation}
    \sum_n F^{i j m}_{k \ell n} F^{p s \ell^*}_{i n q} F^{p q n}_{j k r^*} = F^{p s \ell^*}_{m^* k r^*} F^{r s m^*}_{i j q} \, .
    \label{eq:pentagon_equation}
\end{equation}
From the pentagon equation, we can derive the orthogonality relation of the $F$-symbols that
\begin{equation}
    \sum_n F^{i j m'}_{k\ell n} \left(F^{i j m}_{k\ell n}\right)^* = \delta_{m m'},
    \label{eq:F_orthogonality}
\end{equation}
where the complex conjugation on the $F$-symbol is given by
\begin{equation}
    \left(F^{i j m}_{k\ell n}\right)^* = F^{i^* j^* {m}^*}_{k^* \ell^* n^*}.
\end{equation}

The ground state wave-function of the string-net model $\ket{\Psi}$ is given by
\begin{equation}
    \ket{\Psi} = \sum_{\ket{X} \in \mathcal{H}^\text{SN}_{Q_v}} \Psi(X)\ket{X},
\end{equation}
where $\Psi(X)= \braket{X|\Psi}$, and $\ket{X}$ denotes a string-net configuration in the stable vertex subspace $\mathcal{H}^\text{SN}_{Q_v}$. A vector $\ket{X}$ is a product state. Note that the set of all different $\ket{X}$'s form an orthonormal basis for this subspace. The graphical rules define a set of relations between the amplitudes
\begin{equation}
    \vcenter{\hbox{\includegraphics[scale=1]{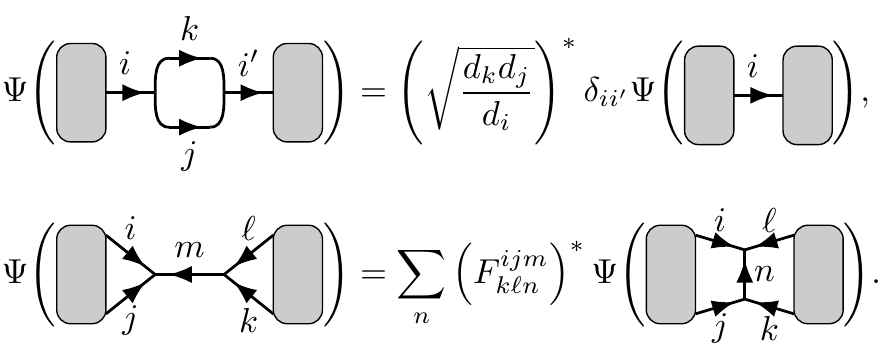}}}
    \label{eq:RG_wavefunction}
\end{equation}
Moreover, the graphical rules can be used to define transformations for a generic string-net configuration ket-vector. Eq.~\eqref{eq:Double_Ising_Bsp_def} and Eq.~\eqref{eq:Gsp_def} are examples. 

The commuting projector Hamiltonian, which has the above wave-function as the ground state, is given by
\begin{equation}
    H_\text{SN} = - \sum_v Q_v - \sum_p B_p,
\end{equation}
where $Q_v$ is the vertex projector enforcing the vertex rules $\delta_{ijk}$, and $B_p = \sum_s (d_s /D) B^s_p$ with $D = \sum_s (d_s)^2$ being the total quantum dimension is the plaquette projector. Each $B^s_p$ adds a counter-clockwise directed loop of $s$ in the interior of a plaquette. Its action can be evaluated by the $F$-symbols as illustrated in Eq.~\eqref{eq:Double_Ising_Bsp_def}. Eq.~\eqref{eq:The_d_symbol_relation} implies that the $B^s_p$'s satisfy
\begin{equation}
    B^i_p B^j_p = \sum_k \delta_{ijk^*} B^k_p.
\end{equation}
The ground state satisfies
\begin{equation}
    Q_v \ket{\Psi} = \ket{\Psi}, \quad B_p \ket{\Psi} = \ket{\Psi}
\end{equation}
for all $v$ and $p$.

\subsection{The minimal lattice}
\label{sec:SN_minimalLattice_torus}

\begin{figure}[ht]
    \centering
    \includegraphics[scale = 1]{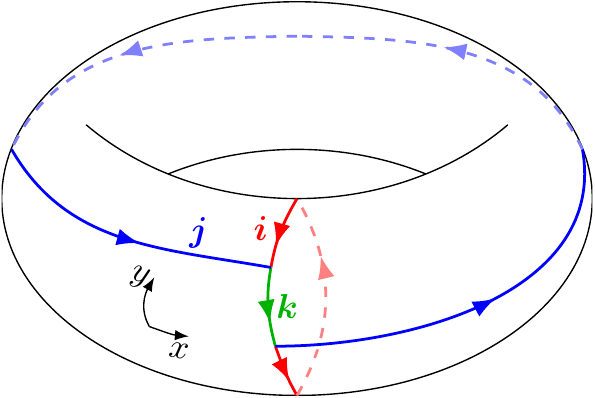}
    \captionsetup{justification=Justified}
    \caption{The minimal trivalent lattice on the torus. The lattice has three edges colored by red, blue, and green; two vertices; and one plaquette. $i$, $j$, and $k$ are state labels on the colored edges.}
    \label{fig:MinimalLattice_torus}
\end{figure}

\begin{figure}[ht]
    \centering
    \includegraphics[scale=1]{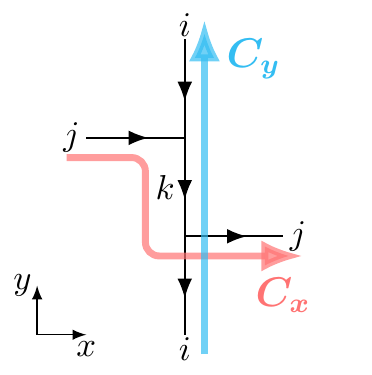}
    \captionsetup{justification=Justified}
    \caption{An illustration of the two non-contractible loops $C_x$ and $C_y$ on the minimal lattice, taken by the logical operators $W^{C_i}_\alpha$ where $\alpha$ is the excitation label.}
    \label{fig:Logical_operators_MinimalLattice}
\end{figure}

A string-net model can be defined on the minimal trivalent lattice on the torus. The minimal lattice consists of three edges, two vertices, and one plaquette as shown in Fig.~\ref{fig:MinimalLattice_torus}. The ground states first have to satisfy the vertex constraints $Q_v$. Recall that a vertex is called stable if the vertex constraint is satisfied. We denote a basis vector of the stable vertex subspace on the minimal lattice by
\begin{equation}
    \ket{ijk}
    = \vcenter{\hbox{\includegraphics[scale=1]{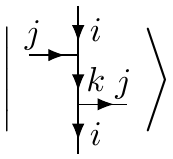}}}
    \,  \in \mathcal{H}^\text{SN}_{Q_v} .
        \label{eq:standardBasis_minimalLattice}
\end{equation}
The logical operators act within $\mathcal{H}^\text{SN}_{Q_v}$. Illustrated in Fig.~\ref{fig:Logical_operators_MinimalLattice} are the two different paths taken by the logical operators $\{W^{C_i}_{\alpha}\}$ where $i\in \{x,y\}$ and $\alpha\in \{\text{excitations}\}$. The action of the logical operators on a basis vector $\ket{ijk}$ can be computed by the method introduced in Ref.~\onlinecite{Stringnet}, which we will not discuss in this paper. We will review a string-operator construction for the double-Ising in Appendix~\ref{sec:DI_fluxon_condensation}.

Next, the ground states need to satisfy the one plaquette term on the minimal lattice. Consider the action of $B^s_p$ on a basis vector $\ket{abc}$. Instead of directly fusing the $s$-loop into the edges, we can first fuse different parts of the $s$-loop together and map it into a trivalent diagram 
\begingroup
\allowdisplaybreaks
\begin{align}
        &B^{s}_p
        \Bigg \vert 
 
        \Bigg \rangle ,
        \label{eq:Bsp_alternative_expression}
\end{align}
\endgroup
where in the second equality, we have brought the $s$-loop over the lattice. We can also bring the loop below the lattice. The choice does not matter. The action of the $B^s_p$ term can then be determined by fusing the trivalent diagram (orange) into the underlying lattice (black).

This will help us to show that for abelian string-net models, the plaquette term on the minimal lattice is trivial. For abelian string-net models, the fusion of $s$ and $s^*$ is $0$. So, the above equation reduces to
\begin{equation}
    \vcenter{\hbox{\includegraphics[scale=1]{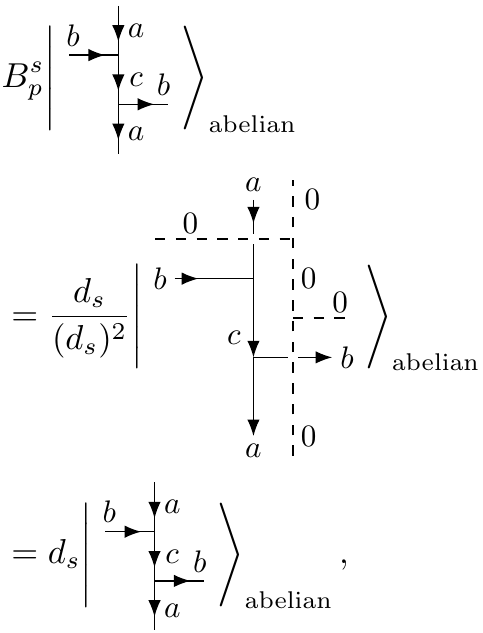}}}
\end{equation}
where to get to the last line, we have used $(d_s)^2=1$ for abelian models. Hence, we see that $B_p$ acts as identity on $\mathcal{H}^\text{SN}_{Q_v}$ in abelian models. In other words, on the minimal lattice, $\mathcal{H}^\text{SN}_{Q_v}$ is the ground space of the abelian models.

\section{Other details of the doubled-Ising}
\label{sec:OtherDetails_DI_SN}

In Appendix~\ref{sec:MinimalLattice_DI_SN}, we discuss the ground states of the doubled-Ising string-net on the minimal lattice (introduced in Appendix~\ref{sec:SN_minimalLattice_torus}). In Appendix~\ref{sec:DI_fluxon_condensation}, we show that when we put the doubled-Ising state on a lattice with smooth boundary, $\psi\bar{\psi}$ and $\sigma\bar{\sigma}$ are condensed on the boundary.

\subsection{Doubled-Ising on minimal lattice}
\label{sec:MinimalLattice_DI_SN}

The doubled-Ising string-net on the minimal lattice has a 10-dimensional stable vertex subspace $\mathcal{H}^\text{D.I.}_{Q_v}$, spanned by $\ket{000}$, $\ket{220}$, $\ket{202}$, $\ket{022}$, $\ket{110}$, $\ket{112}$, $\ket{101}$, $\ket{011}$, $\ket{211}$, and $\ket{121}$.

One of the dimensions is not part of the ground space, because the projector $B_p$ has the eigenvalue of $0$ on this state. To find the ground space, we calculate the action of $B^1_p$. Following from Eq.~\eqref{eq:Bsp_alternative_expression} (the actions of $B^0_p$ and $B^2_p$ are trivial), we find
\begin{widetext}
\begin{equation}
    \vcenter{\hbox{\includegraphics[scale=1]{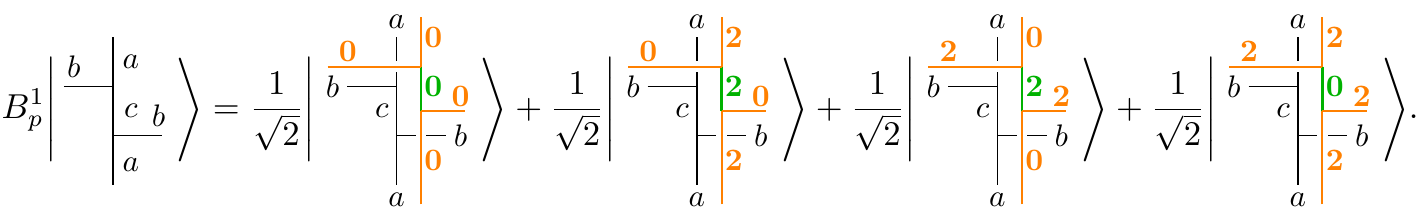}}}
    \label{eq:B1p_alternative_action_min}
\end{equation}
\end{widetext}
That is,
\begin{equation}
    \begin{aligned}
        B^1_p \ket{abc} &= \frac{1}{\sqrt{2}}\left(\mathbbm{I}+ W^{C_y}_{\psi}+W^{C_x}_{\psi}-W^{C_x}_{\psi}W^{C_y}_{\psi}\right)\ket{abc} \\
        &= \frac{1}{\sqrt{2}}\left(\mathbbm{I}+ W^{C_y}_{\bar{\psi}}+W^{C_x}_{\bar{\psi}}-W^{C_x}_{\bar{\psi}}W^{C_y}_{\bar{\psi}}\right) \ket{abc},
    \end{aligned}
\end{equation}
where we have identified the logical operators by the construction discussed in Appendix~\ref{sec:DI_fluxon_condensation}, and the second equality follows from that we can bring the $s$-loop below the lattice in Eq.~\eqref{eq:Bsp_alternative_expression}. We can further compute the action of $B^1_p$ by fusing the orange trivalent diagram into the underlying lattice. At this point, it is clear that a product state $\ket{abc}$ is generally not an eigenstate of $B^1_p$.

An explicit calculation shows that $\ket{{\psi\bar{\psi}}^\text{D.I.}_\text{min}}= -\frac{1}{2}\ket{000} + \frac{1}{2} \ket{202} + \frac{1}{2} \ket{022} + \frac{1}{2} \ket{220}$ has the eigenvalue $-\sqrt{2}$ under $B^1_p$. Hence, it is an excited state, which carries a $\psi\bar{\psi}$ fluxon. The other 9 dimensions have the eigenvalue $\sqrt{2}$ under $B^1_p$ and are, thus, ground states on the minimal lattice. 

An orthonormal basis for the nine-dimensional ground space can be chosen to be the common eigenstates of logical operators $W^{C_x}_{\psi}$, $W^{C_y}_{\psi}$, $W^{C_x}_{\bar{\psi}}$, and $W^{C_y}_{\bar{\psi}}$, which all commute with each other. The nine common eigenstates are
\begingroup
\allowdisplaybreaks
\begin{align*}
        \ket{\Psi^\text{D.I.}_\text{min}}_1 &= \frac{1}{2} \ket{000} + \frac{1}{2} \ket{202} + \frac{1}{2} \ket{022} - \frac{1}{2} \ket{220} , \\%
        \ket{\Psi^\text{D.I.}_\text{min}}_2 &= \frac{1}{\sqrt{2}} \ket{011} + \frac{i}{\sqrt{2}} \ket{211} , \\%
        \ket{\Psi^\text{D.I.}_\text{min}}_3 &= \frac{1}{\sqrt{2}} \ket{101} -\frac{i}{\sqrt{2}} \ket{121} , \\%
        \ket{\Psi^\text{D.I.}_\text{min}}_4 &= \frac{1}{\sqrt{2}} \ket{011} -\frac{i}{\sqrt{2}} \ket{211} , \\%
        \ket{\Psi^\text{D.I.}_\text{min}}_5 &= \frac{1}{2} \ket{000} - \frac{1}{2}\ket{202} + \frac{1}{2} \ket{022} + \frac{1}{2} \ket{220} , \\%
        \ket{\Psi^\text{D.I.}_\text{min}}_6 &= \frac{e^{-\frac{i \pi}{8}}}{\sqrt{2}} \ket{110} + \frac{i e^{-\frac{i \pi}{8}}}{\sqrt{2}} \ket{112} , \\%
        \ket{\Psi^\text{D.I.}_\text{min}}_7 &= \frac{1}{\sqrt{2}} \ket{101} + \frac{i}{\sqrt{2}} \ket{121} , \\%
        \ket{\Psi^\text{D.I.}_\text{min}}_8 &= \frac{e^{\frac{i \pi }{8}}}{\sqrt{2}} \ket{110} -\frac{i e^{\frac{i \pi }{8}}}{\sqrt{2}} \ket{112} , \\%
        \ket{\Psi^\text{D.I.}_\text{min}}_9 &= \frac{1}{2}\ket{000} + \frac{1}{2} \ket{202} - \frac{1}{2} \ket{022} + \frac{1}{2} \ket{220}.
\end{align*}
\endgroup
The 10th dimension $\ket{{\psi\bar{\psi}}^\text{D.I.}_\text{min}}$ is also a common eigenstate of $W^{C_x}_{\psi}$, $W^{C_y}_{\psi}$, $W^{C_x}_{\bar{\psi}}$, and $W^{C_y}_{\bar{\psi}}$. In this basis, the logical operators takes the diagonal form
\begingroup
\allowdisplaybreaks
\begin{align}
    W^{C_x}_{\psi} &= 
    \left[ \left(
    \begin{array}{ccc}
     1 & 0 & 0 \\
     0 & 1 & 0 \\
     0 & 0 & -1 \\
    \end{array}
    \right) \otimes \mathbb{I}_{3\times 3} \right] \oplus (-1) , \\ 
    W^{C_y}_\psi &=
    \left[ \left(
    \begin{array}{ccc}
     1 & 0 & 0 \\
     0 & -1 & 0 \\
     0 & 0 & 1 \\
    \end{array}
    \right) \otimes \mathbb{I}_{3\times 3} \right] \oplus (-1) , \\
    W^{C_x}_{\bar{\psi}} &=
    \left[ \mathbb{I}_{3\times 3} \otimes 
    \left(
    \begin{array}{ccc}
     1 & 0 & 0 \\
     0 & 1 & 0 \\
     0 & 0 & -1 \\
    \end{array}
    \right) \right] \oplus (-1) , \\
    W^{C_y}_{\bar{\psi}} &=
    \left[ \mathbb{I}_{3\times 3} \otimes 
    \left(
    \begin{array}{ccc}
     1 & 0 & 0 \\
     0 & -1 & 0 \\
     0 & 0 & 1 \\
    \end{array}
    \right) \right] \oplus (-1) .
\end{align}
\endgroup

\subsection{Condensation on smooth boundary}
\label{sec:DI_fluxon_condensation}

\begin{figure}[h]
    \centering
    \includegraphics[scale = 0.11]{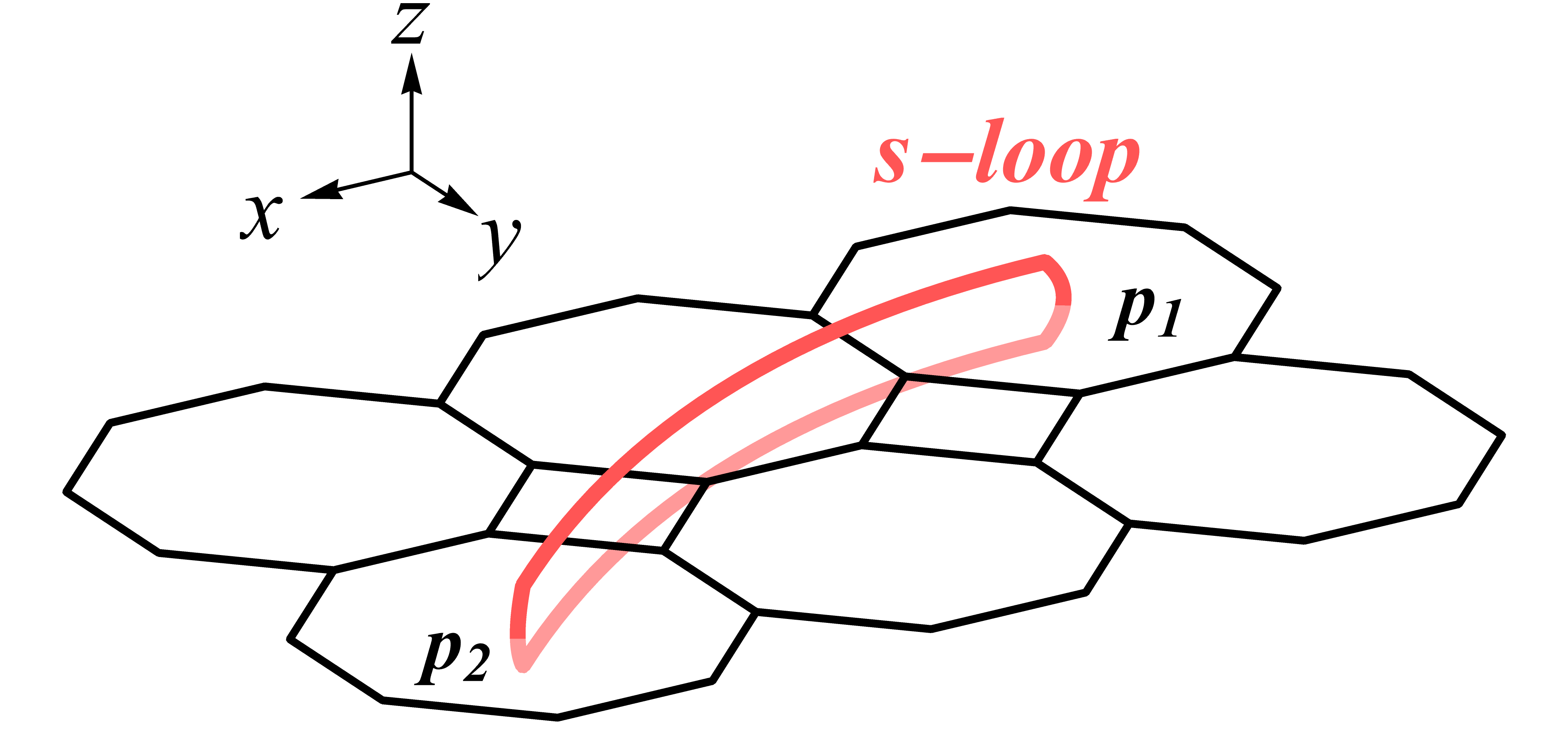}
    \captionsetup{justification=Justified}
    \caption{An illustration of an open ended fluxon string-operator $W^\text{path}_\text{fluxon}$ acting on a wave-function of the doubled-Ising string-net. $W^\text{path}_\text{fluxon}$ creates the fluxon and its antiparticle on the plaquettes $p_1$ and $p_2$. It does not create any excitations along the path. $W^\text{path}_\text{fluxon}$ is constructed by a loop of $s$-string that vertically penetrates the square-octagon lattice at the two plaquettes. The string-operators of $\psi\bar{\psi}$ and $\sigma\bar{\sigma}$ are given by a loop of $2$-string and a loop of $1$-string respectively. }
    \label{fig:DoubleIsing_fluxon_stringOperators}
\end{figure}

\begin{figure}[h]
    \centering
    \includegraphics[scale = 0.11]{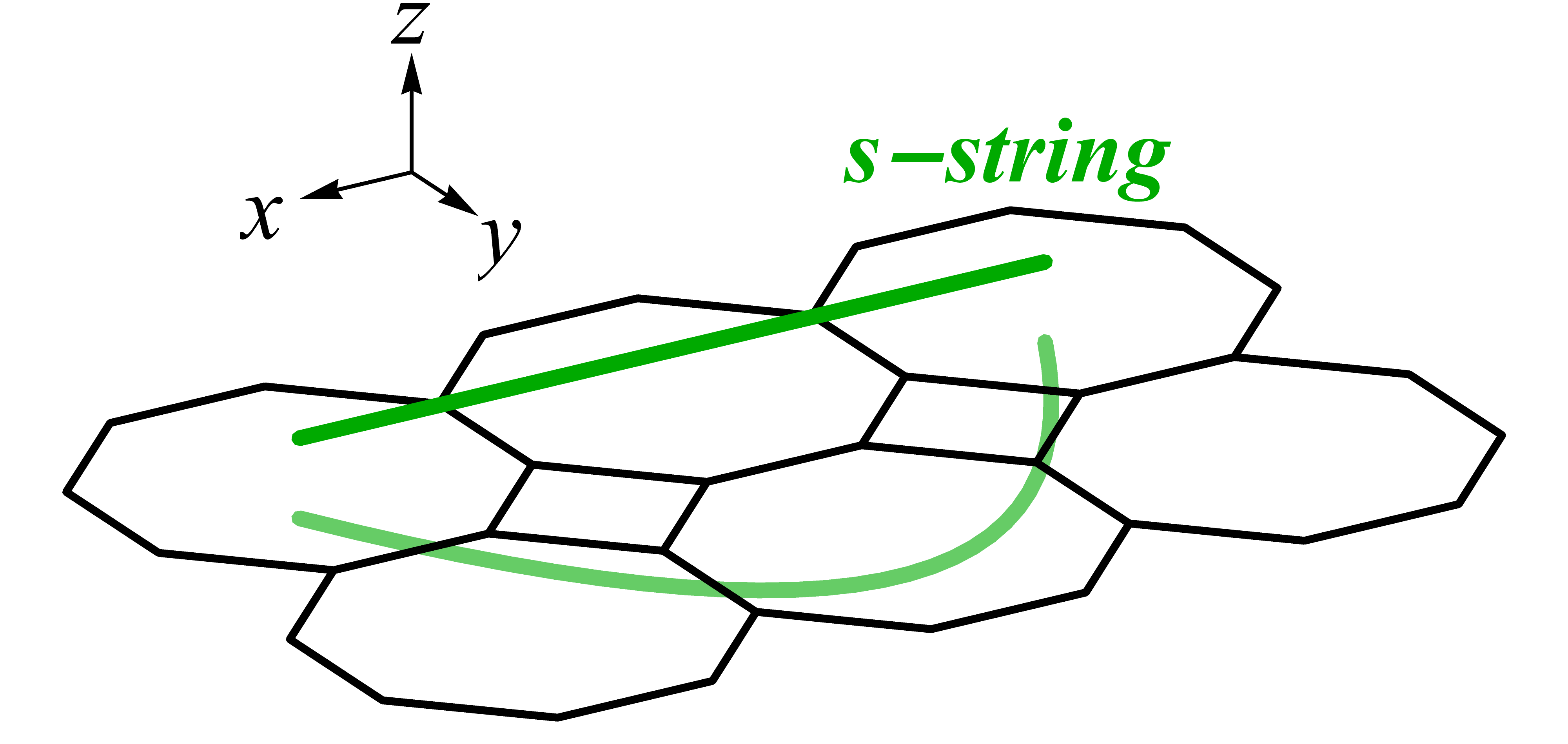}
    \captionsetup{justification=Justified}
    \caption{The open ended chargeon string-operators of the doubled-Ising string-net. They are constructed by line segements of the strings, which can be either above or below the lattice depending on the type of the chargeon excitations. On a ground state of the doubled-Ising, An open chargeon string-operator creates a vertex violation at each end. The $\psi$ and $\sigma$ string-operators correspond to a $2$-string and a $1$-string above the lattice respectively. The string-operators for $\bar{\psi}$ and $\bar{\sigma}$ correspond to those below the lattice.}
    \label{fig:DoubleIsing_chargeon_stringOperators}
\end{figure}

Using the string-operator construction discussed in Ref.~\onlinecite{Burnell_doubleIsing,ExtendedStringnet_Hu}, we can readily show that on the smooth boundary, the bosonic fluxons $\psi\bar{\psi}$ and $\sigma\bar{\sigma}$ condense.

Let us start with a review of the fluxon string-operators. Consider a large square-octagon lattice placed on the $xy$-plane as shown in Fig.~\ref{fig:DoubleIsing_fluxon_stringOperators}. An open ended fluxon string-operator is given by a loop of $s$-string which vertically penetrates the lattice through the center of the plaquettes $p_1$ and $p_2$. The fluxon excitations are created at $p_1$ and $p_2$ respectively. The $\psi\bar{\psi}$ string-operator is given by a loop of $2$-string, and that of $\sigma\bar{\sigma}$ is given by a loop of $1$-string.

An open ended string-operator of a chargeon is constructed by a line segement, which can be either above or below the lattice, as shown in Fig.~\ref{fig:DoubleIsing_chargeon_stringOperators}.

To compute the action of the string-operators, we need the $R$-symbols and the $S$-matrix\cite{KitaevAnyons,rowell_classification_2009}. The $R$-symbols define the braiding transformations
\begin{equation}
    \vcenter{\hbox{\includegraphics[scale=1]{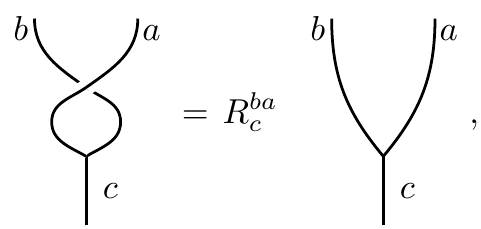}}}
    \label{eq:graphical_rule_R1}
\end{equation}
where $R^{ba}_c \in \mathbb{C}$. The inverse transformations are defined by
\begin{equation}
    \vcenter{\hbox{\includegraphics[scale=1]{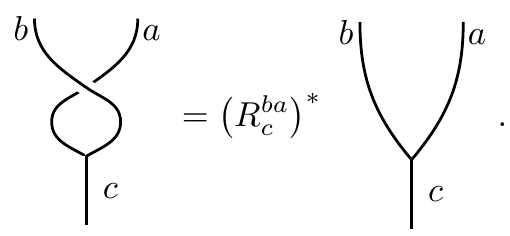}}}
    \label{eq:graphical_rule_R2}
\end{equation}
Same as the $F$-symbols, $R^{ba}_c \neq 0$ if $\delta_{bac} \neq 0$. Elements of the $S$-matrix are given by
\begin{equation}
    S_{ab} = \frac{1}{\sqrt{D}} \sum_c d_c R^{ba}_c R^{ab}_c,
\end{equation}
where $\sqrt{D} = \sqrt{\sum_s (d_s)^2} = 2$ is the total quantum dimension of the Ising unitary modular tensor category. The full $S$-matrix is
\begin{equation}
    S = \frac{1}{2} 
    \left(
\begin{array}{ccc}
     1 & \sqrt{2} & 1 \\
     \sqrt{2} & 0 & -\sqrt{2} \\
     1 & -\sqrt{2} & 1 \\
    \end{array}
    \right).
\end{equation}

Using the $R$- and $F$-symbols, we can always fuse the chargeon string-operators into the lattice at the cost of violating the vertex constraints at the ends. On the other hand, the fluxon string-operators can be fused into the lattice without introducing any vertex violations. For example, consider the action of a non-trivial fluxon string-operator on the string-net wave-function 
\begin{equation}
    \begin{aligned}
        &\Bigg \vert
        %
        \Bigg \rangle.
    \end{aligned}
\end{equation}
Note that the expression above differs from that of Ref.~\onlinecite{ExtendedStringnet_Hu} by a normalization factor, which is not important for our purposes. For an example of a chargeon string-operator, let us consider a line segment below the lattice. We compute
\begin{equation}
    \begin{aligned}
        &\Bigg \vert
        %
        \Bigg \rangle ,
    \end{aligned}
\end{equation}
where in the last step we have removed the $s$-strings. So, we see that an open chargeon string-operator always creates violations to the vertex constraints.

Moreover, different fluxons correspond to different sets of eigenvalues of $B^a_p$. To determine these eigenvalues, we need the graphical rule
\begin{equation}
    %
.
    \label{eq:graphical_rule_S_matrix}
\end{equation}
As an example, let us compute 
\begin{equation}
    \vcenter{\hbox{\includegraphics[scale=1]{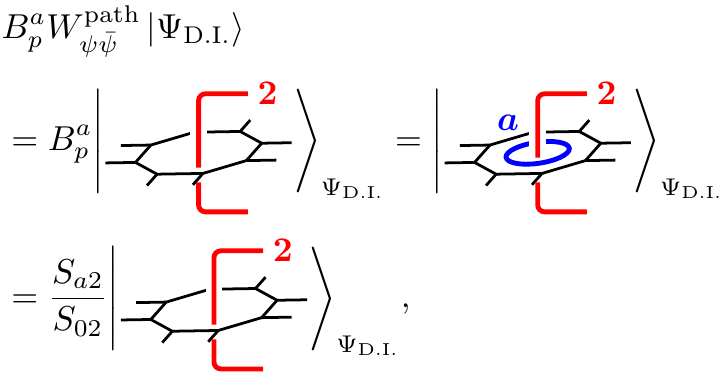}}}
\end{equation}
from which we see that the $\psi\bar{\psi}$ fluxon has the eigenvalues of $1$, $-\sqrt{2}$, and $1$ for $B^0_p$, $B^1_p$, and $B^2_p$ respectively. This result is exactly what we found for the $\psi\bar{\psi}$ fluxon state in Appendix~\ref{sec:MinimalLattice_DI_SN}. It is easy to see that the $\sigma\bar{\sigma}$ fluxon has $B^0_p = 1$, $B^1_p = 0$, and $B^2_p = -1$.

We now show that, on the smooth boundary, the fluxons $\psi\bar{\psi}$ and $\sigma\bar{\sigma}$ are condensed. Without loss of generality, let us consider the doubled-Ising string-net with a single plaquette and everywhere else is set to $\ket{0}$. Consider the action of an open ended fluxon string-operator passing through the lattice just outside the plaquette
\begin{equation*}
    \vcenter{\hbox{\includegraphics[scale=1]{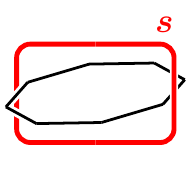}}}
\end{equation*}

Since the $s$-loop does not pass through the region enclosed by the plaquette, the string-operator does not create any fluxon excitation on the plaquette. Via the $F$- and $R$-symbols, we can fuse the $s$-loop into the edges of the plaquette without changing any of the edges outside the plaquette or introducing any vertex violations. Therefore, we see that the fluxons all condense on the smooth boundary. On the other hand, because the chargeon string-operators necessarily introduce vertex violations, the chargeons remain as excitations on the boundary. Thus, we reach the conclusion that, on the smooth boundary, the condensed excitations are the fluxons $\psi\bar{\psi}$ and $\sigma\bar{\sigma}$.

\section{Controlled gate details}
\label{sec:ControlledGate_details}

In this appendix, we present the details of the graphical definition of $G^s_p$, its inverse, the commutation relations, and the proof for the central equation.

\subsection{Graphical definition}
\label{sec:ControlledGate_graphicalDef}

We perform the graphical calculation that leads to the last line in  Eq.~\eqref{eq:Gsp_def}. We compute
\begin{widetext}
\begin{equation}
    \vcenter{\hbox{\includegraphics[scale=1]{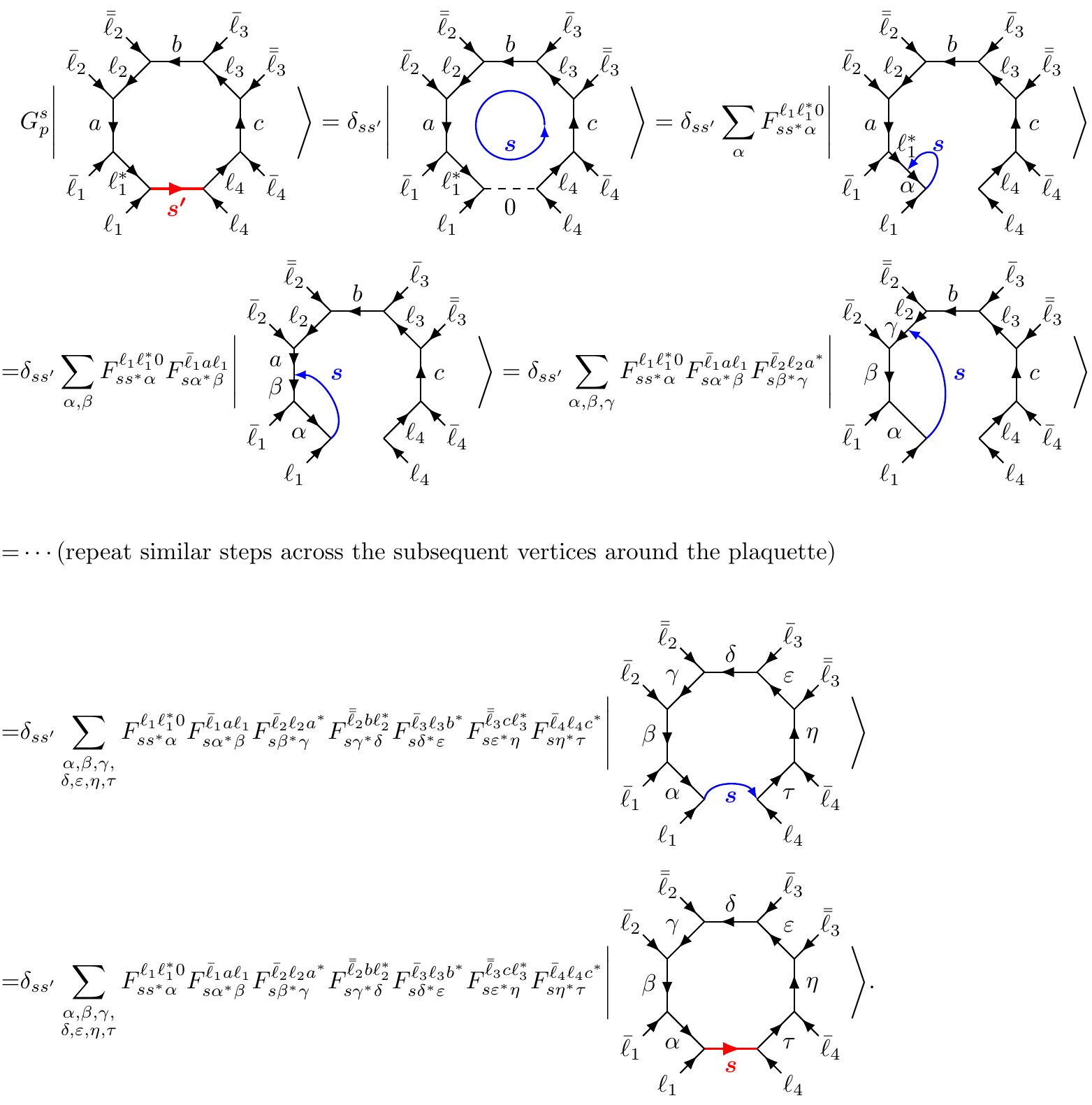}}}
\end{equation}
\end{widetext}

\subsection{Isometric property}
\label{sec:appen_isometry}

We now show that $G^s_p$ is an isometry. That is, ${G^s_p}^\dagger G^s_p$ equals identity on the input space, which we called $\mathcal{V}^{\text{SN}}_{p,s}$ in section~\ref{sec:control_gate}.

Graphically, ${G^s_p}^\dagger$ removes a string of $s$ from the edges of the plaquette when the controlled edge is in the state $\ket{s}$. Let us denote the matrix elements of $G^s_p$ by
\begin{equation}
    \begin{aligned}
        &{[G^s_p]}^{(s,\alpha,\beta,\gamma,\delta,\varepsilon,\eta,\tau)}_{(s',\ell^*_1,a,\ell_2,b,\ell_3,c,\ell_4)}(\ell_1, \bar{\ell}_1, \bar{\ell}_2,\bar{\bar{\ell}}_2, \bar{\ell}_3, \bar{\bar{\ell}}_3, \bar{\ell}_4, \ell_4)  \\
        &= \delta_{ss'} 
        F^{\ell_1 \ell_1^* 0}_{s s^* \alpha} F^{\bar{\ell}_1 a \ell_1}_{s \alpha^* \beta} F^{\bar{\ell}_2 \ell_2 a^*}_{s \beta^* \gamma} F^{\bar{\bar{\ell}}_2 b \ell_2^*}_{s \gamma^* \delta} F^{\bar{\ell}_3 \ell_3 b^*}_{s \delta^* \varepsilon} F^{\bar{\bar{\ell}}_3 c \ell_3^*}_{s \varepsilon^* \eta} F^{\bar{\ell}_4 \ell_4 c^*}_{s \eta^* \tau}.
    \end{aligned}
\end{equation}
Then, ${G^s_p}^\dagger$ has an algebraic expression of
\begin{equation}
        \begin{aligned}
        &{G_p^s}^\dagger  
        \Bigg \vert
        \begin{tikzpicture}[baseline={([yshift=-.5ex]current bounding box.center)}, every node/.style={scale=0.9}]
            %
            \pgfmathsetmacro{\radius}{0.7}
            \pgfmathsetmacro{\Axesscaling}{1.68}
            \pgfmathsetmacro{\OffAxesscaling}{1.25}
            \pgfmathsetmacro{\sinval}{sin(45)}
            %
            \octagonControlled{\radius}{0.8}{0.4};
            %
            \node [] at (-\OffAxesscaling*\radius*\sinval,-\OffAxesscaling*\radius*\sinval) {$\alpha'$};
            \node [] at (-\OffAxesscaling*\radius*\sinval,\OffAxesscaling*\radius*\sinval) {$\gamma'$};
            \node [] at (\OffAxesscaling*\radius*\sinval,\OffAxesscaling*\radius*\sinval) {$\varepsilon'$};
            \node [] at (\OffAxesscaling*\radius*\sinval,-\OffAxesscaling*\radius*\sinval) {$\tau'$};
            \node [] at (-\Axesscaling*\radius*\sinval,0) {$\beta'$};
            \node [] at (0,\Axesscaling*\radius*\sinval) {$\delta'$};
            \node [] at (\Axesscaling*\radius*\sinval,0) {$\eta'$};
            \node [] at (0,-\Axesscaling*\radius*\sinval) {\textcolor{red}{$\bm{s'}$}};
        \end{tikzpicture} 
        \Bigg \rangle  \\
        &\begin{aligned}
        =\sum_{\substack{t_1, t_2, t_3,\\ t_4, t_5}}
        \bigg( &{[G^s_p]}^{(s',\alpha',\beta',\gamma',\delta',\varepsilon',\eta',\tau')}_{(s,\ell^*_1,t_5,t_4,t_3,t_2,t_1,\ell_4)} (\ell_1, \bar{\ell}_1, ..., \ell_4)\bigg)^* \times \\
        & 
        \Bigg \vert
        \begin{tikzpicture}[baseline={([yshift=-.5ex]current bounding box.center)}, every node/.style={scale=0.9}]
            %
            \pgfmathsetmacro{\radius}{0.75}
            \pgfmathsetmacro{\Axesscaling}{1.68}
            \pgfmathsetmacro{\OffAxesscaling}{1.25}
            \pgfmathsetmacro{\sinval}{sin(45)}
            %
            \octagonControlled{\radius}{0.8}{0.4};
            %
            \node [] at (-\OffAxesscaling*\radius*\sinval,-\OffAxesscaling*\radius*\sinval) {$\ell_1^*$};
            \node [] at (-\OffAxesscaling*\radius*\sinval,\OffAxesscaling*\radius*\sinval) {$t_4$};
            \node [] at (\OffAxesscaling*\radius*\sinval,\OffAxesscaling*\radius*\sinval) {$t_2$};
            \node [] at (\OffAxesscaling*\radius*\sinval,-\OffAxesscaling*\radius*\sinval) {$\ell_4$};
            \node [] at (-\Axesscaling*\radius*\sinval,0) {$t_5$};
            \node [] at (0,\Axesscaling*\radius*\sinval) {$t_3$};
            \node [] at (\Axesscaling*\radius*\sinval,0) {$t_1$};
            \node [] at (0,-\Axesscaling*\radius*\sinval) {\textcolor{red}{$\bm{s}$}};
        \end{tikzpicture} 
        \Bigg \rangle.
        \end{aligned}
        \end{aligned}
    \label{eq:DefGspInverse}
\end{equation}

Using the orthogonality relation Eq.~\eqref{eq:F_orthogonality}, we find
\begin{equation}
    \vcenter{\hbox{\includegraphics[scale=1]{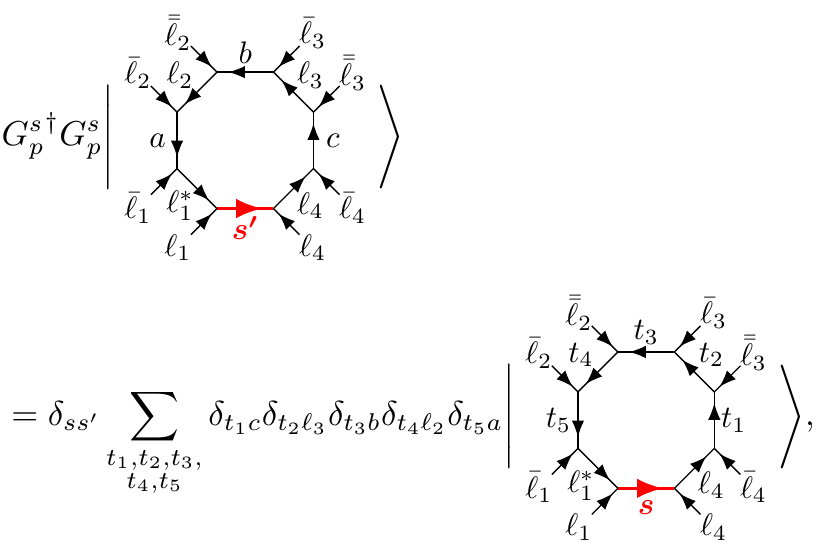}}}
\end{equation}
thereby establishing $G^s_p$ as an isometry. Hence, $G_p = \sum_s G^s_p$ is also an isometry.

\subsection{Commutation relations}
\label{sec:appen_commutation}

The commutation relations of the $G^s_p$ operators immediately follow from the graphical definition. Any two $G_p^s$ and $G_{p'}^{s'}$ commute, provided that they do not act on each other's controlled edge. When $p$ and $p'$ are the same plaquette, $G_p^s$ and $G_{p}^{s'}$ commute trivially, because they act on orthogonal spaces with the control edge in $|s\rangle$ and $|s'\rangle$ respectively. If the plaquettes $p$ and $p'$ are not next to each other, $G_p^s$ and $G_{p'}^{s'}$ obviously commute. When $p$ and $p'$ are adjacent, the proof of commutation amounts to showing the order, in which the string $s$ and $s'$ are fused into the bordering edges, does not matter.

Consider $G_p^s$ and $G_{p'}^{s'}$ acting on two adjacent plaquettes $p$ and $p'$. We focus on the bordering edges of these two plaquettes. We will show that the $F$-symbols associated with the two diagrams (the thickened red arrows indicate the direction of motion of the $s$- and $s'$-strings),
\begin{equation*}
    \vcenter{\hbox{\includegraphics[scale=1]{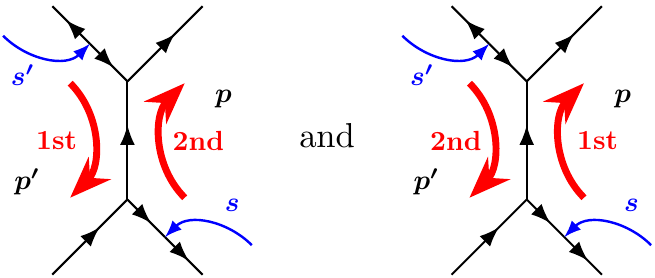}}}
\end{equation*}
are equal. The left diagram corresponds to computing $G^{s}_p G^{s'}_{p'}$ on a reference ket-vector, and the right diagram corresponds to $G^{s'}_{p'} G^{s}_p$.

In the case where $s'$ moves first, we find
\begin{equation}
    \vcenter{\hbox{\includegraphics[scale = 1]{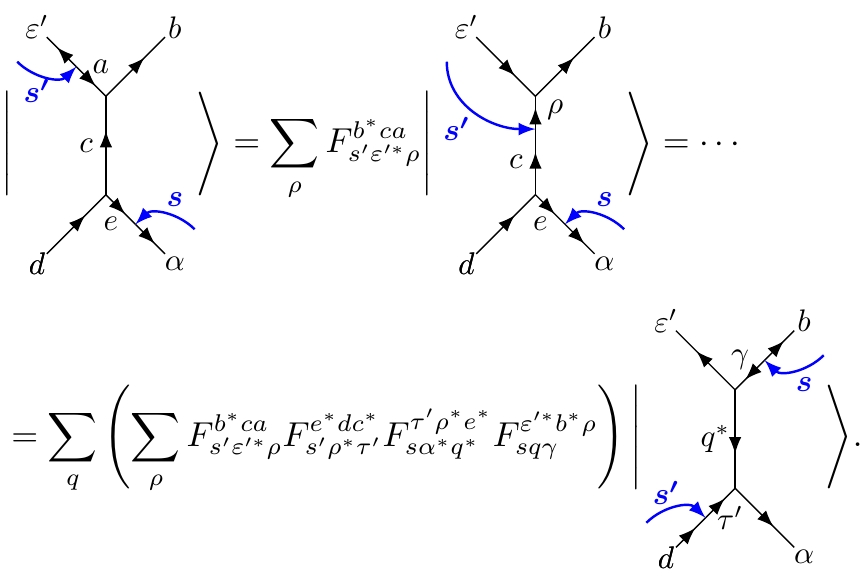}}}
    \label{eq:CommutationSPrimedFirst}
\end{equation}
In the case where $s$ moves first, we have
\begin{equation}
    \vcenter{\hbox{\includegraphics[scale=1]{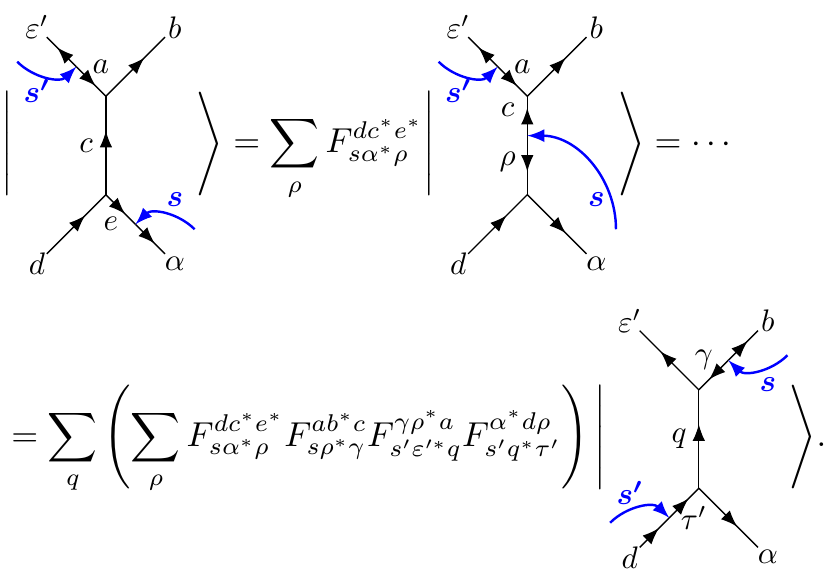}}}
    \label{eq:CommutationSFirst}
\end{equation}
For each fixed $q$, we want to show that the coefficients, i.e. free sums over $\rho$, in Eq.~\eqref{eq:CommutationSPrimedFirst} and Eq.~\eqref{eq:CommutationSFirst} are equal. To do this, we consider an alternative way of moving the strings $s$ and $s'$. We will show that this alternative expression can be simplified, via the pentagon equation, to produce either Eq.~\eqref{eq:CommutationSPrimedFirst} or Eq.~\eqref{eq:CommutationSFirst}. The alternative expression is obtained by moving both $s$ and $s'$ to the central edge, 
\begingroup
\allowdisplaybreaks
\begin{align*}
        &\Bigg \vert
        \begin{tikzpicture}[baseline={([yshift=-.5ex]current bounding box.center)}, every node/.style={scale=0.9}]
            \pgfmathsetmacro{\length}{0.5}
            \pgfmathsetmacro{\Axesscaling}{1.68}
            \pgfmathsetmacro{\OffAxesscaling}{1.18}
            \pgfmathsetmacro{\sinval}{sin(45)}
            \pgfmathsetmacro{\lineW}{0.4}
            \pgfmathsetmacro{\markingPos}{0.65}
            \pgfmathsetmacro{\branchScale}{1.7}
            %
            \begin{scope}[decoration={markings, mark=at position \markingPos with {\arrow{Latex}}}]
                \draw[line width = \lineW pt, postaction={decorate}] (0,-\length) -- (0,\length);
                %
                \draw[line width = \lineW pt, postaction={decorate}]
                (-\branchScale*\length*\sinval/2+0,\branchScale*\length*\sinval/2+\length) -- (-\branchScale*\length*\sinval+0,\branchScale*\length*\sinval+\length);
                \draw[line width = \lineW pt, postaction={decorate}] (-\branchScale*\length*\sinval/2+0,\branchScale*\length*\sinval/2+\length) -- (0,\length); 
                \draw[line width = \lineW pt, postaction={decorate}] (0,\length) -- (\branchScale*\length*\sinval+0,\branchScale*\length*\sinval+\length); 
                %
                \draw[line width = \lineW pt, postaction={decorate}] (-\branchScale*\length*\sinval+0,-\branchScale*\length*\sinval-\length) -- (0,-\length); 
                \draw[line width = \lineW pt, postaction={decorate}] (0,-\length) -- (\branchScale*\length*\sinval/2+0,-\branchScale*\length*\sinval/2-\length); 
                \draw[line width = \lineW pt, postaction={decorate}] (\branchScale*\length*\sinval/2+0,-\branchScale*\length*\sinval/2-\length) -- (\branchScale*\length*\sinval+0,-\branchScale*\length*\sinval-\length); 
            \end{scope}
            %
            \draw[line width = 0.7 pt, draw = blue,  -latex] (-\branchScale*\length*\sinval-0.3,\branchScale*\length*\sinval+0.5*\length) to[out = 315, in = 225] (-\branchScale*\length*\sinval/2+0,\branchScale*\length*\sinval/2+\length);
            %
            \node [] at (-\branchScale*\length*\sinval-0.1,\branchScale*\length*\sinval+0.5*\length-0.3) {\textcolor{blue}{$\bm{s'}$}};
            %
            \draw[line width = 0.7 pt, draw = blue,  -latex] (\branchScale*\length*\sinval+0.3,-\branchScale*\length*\sinval-0.5*\length) to[out = 135, in = 45] (\branchScale*\length*\sinval/2+0,-\branchScale*\length*\sinval/2-\length);
            %
            \node [] at (\branchScale*\length*\sinval+0.1,-\branchScale*\length*\sinval-0.5*\length+0.3) {\textcolor{blue}{$\bm{s}$}};
            %
            \node [] at (-\branchScale*\length*\sinval-0.1,\branchScale*\length*\sinval+\length+0.1) {$\varepsilon'$};
            \node [] at (-\branchScale*\length*\sinval-0.1,-\branchScale*\length*\sinval-\length-0.1) {$d$};
            \node [] at (\branchScale*\length*\sinval+0.1,-\branchScale*\length*\sinval-\length-0.1) {$\alpha$};
            \node [] at (\branchScale*\length*\sinval+0.1,\branchScale*\length*\sinval+\length+0.1) {$b$};
            \node [] at (-0.2,0) {$c$};
            \node [] at (-\branchScale*\length*\sinval/2+0.25,\branchScale*\length*\sinval/2+\length) {$a$};
            \node [] at (\branchScale*\length*\sinval/2-0.25,-\branchScale*\length*\sinval/2-\length) {$e$};
        \end{tikzpicture}
        \Bigg \rangle
        =
        \sum_{\eta', \beta} F^{b^* c a}_{s' \varepsilon'^* \eta'} F^{d c^* e^*}_{s \alpha^* \beta}
        \Bigg \vert
        \begin{tikzpicture}[baseline={([yshift=-.5ex]current bounding box.center)}, every node/.style={scale=0.9}]
            \pgfmathsetmacro{\length}{0.5}
            \pgfmathsetmacro{\Axesscaling}{1.68}
            \pgfmathsetmacro{\OffAxesscaling}{1.18}
            \pgfmathsetmacro{\sinval}{sin(45)}
            \pgfmathsetmacro{\lineW}{0.4}
            \pgfmathsetmacro{\markingPos}{0.65}
            \pgfmathsetmacro{\branchScale}{1.7}
            %
            \begin{scope}[decoration={markings, mark=at position \markingPos with {\arrow{Latex}}}]
                \draw[line width = \lineW pt, postaction={decorate}] (0,\length/3) -- (0,\length);
                \draw[line width = \lineW pt, postaction={decorate}] (0,-\length/3) -- (0,\length/3);
                \draw[line width = \lineW pt, postaction={decorate}] (0,-\length/3) -- (0,-\length);
                %
                \draw[line width = \lineW pt, postaction={decorate}] (-\branchScale*\length*\sinval+0,\branchScale*\length*\sinval+\length) -- (0,\length); 
                \draw[line width = \lineW pt, postaction={decorate}] (0,\length) -- (\branchScale*\length*\sinval+0,\branchScale*\length*\sinval+\length); 
                %
                \draw[line width = \lineW pt, postaction={decorate}] (-\branchScale*\length*\sinval+0,-\branchScale*\length*\sinval-\length) -- (0,-\length); 
                \draw[line width = \lineW pt, postaction={decorate}] (0,-\length) --  (\branchScale*\length*\sinval+0,-\branchScale*\length*\sinval-\length); 
            \end{scope}
            %
            \draw[line width = 0.7 pt, draw = blue,  -latex] (-\branchScale*\length*\sinval-0.3,\branchScale*\length*\sinval+0.5*\length) to[out = 270, in = 180] (0,\length/3);
            %
            \node [] at (-\branchScale*\length*\sinval-0.2,\length/3) {\textcolor{blue}{$\bm{s'}$}};
            %
            \draw[line width = 0.7 pt, draw = blue,  -latex] (\branchScale*\length*\sinval+0.3,-\branchScale*\length*\sinval-0.5*\length) to[out = 90, in = 0] (0,-\length/3);
            %
            \node [] at (\branchScale*\length*\sinval+0.1,-\length/3) {\textcolor{blue}{$\bm{s}$}};
            %
            \node [] at (-\branchScale*\length*\sinval-0.1,\branchScale*\length*\sinval+\length+0.1) {$\varepsilon'$};
            \node [] at (-\branchScale*\length*\sinval-0.1,-\branchScale*\length*\sinval-\length-0.1) {$d$};
            \node [] at (-\branchScale*\length*\sinval-0.1,-\branchScale*\length*\sinval-\length-0.1) {$d$};
            \node [] at (\branchScale*\length*\sinval+0.1,-\branchScale*\length*\sinval-\length-0.1) {$\alpha$};
            \node [] at (\branchScale*\length*\sinval+0.1,\branchScale*\length*\sinval+\length+0.1) {$b$};
            \node [] at (-0.2,-\length/3+0.15) {$c$};
            \node [] at (0.2,\length/3+0.2) {$\eta'$};
            \node [] at (-0.2,-\length/3-0.3) {$\beta$};
        \end{tikzpicture}
        \Bigg \rangle \\
        &=
        \sum_{\eta', \beta, q} F^{b^* c a}_{s' \varepsilon'^* \eta'} F^{d c^* e^*}_{s \alpha^* \beta} F^{s \beta^* c^*}_{s' \eta'^* q}
        \Bigg \vert
        \begin{tikzpicture}[baseline={([yshift=-.5ex]current bounding box.center)}, every node/.style={scale=0.9}]
            \pgfmathsetmacro{\length}{0.5}
            \pgfmathsetmacro{\Axesscaling}{1.68}
            \pgfmathsetmacro{\OffAxesscaling}{1.18}
            \pgfmathsetmacro{\sinval}{sin(45)}
            \pgfmathsetmacro{\lineW}{0.4}
            \pgfmathsetmacro{\markingPos}{0.65}
            \pgfmathsetmacro{\branchScale}{1.7}
            %
            \begin{scope}[decoration={markings, mark=at position \markingPos with {\arrow{Latex}}}]
                \draw[line width = \lineW pt, postaction={decorate}] (0,\length/3) -- (0,\length);
                \draw[line width = \lineW pt, postaction={decorate}] (0,-\length/3) -- (0,\length/3);
                \draw[line width = \lineW pt, postaction={decorate}] (0,-\length/3) -- (0,-\length);
                %
                \draw[line width = \lineW pt, postaction={decorate}] (-\branchScale*\length*\sinval+0,\branchScale*\length*\sinval+\length) -- (0,\length); 
                \draw[line width = \lineW pt, postaction={decorate}] (0,\length) -- (\branchScale*\length*\sinval+0,\branchScale*\length*\sinval+\length); 
                %
                \draw[line width = \lineW pt, postaction={decorate}] (-\branchScale*\length*\sinval+0,-\branchScale*\length*\sinval-\length) -- (0,-\length); 
                \draw[line width = \lineW pt, postaction={decorate}] (0,-\length) --  (\branchScale*\length*\sinval+0,-\branchScale*\length*\sinval-\length); 
            \end{scope}
            %
            \draw[line width = 0.7 pt, draw = blue,  -latex] (-\branchScale*\length*\sinval-0.3,\branchScale*\length*\sinval+0.5*\length) to[out = 270, in = 180] (0,-\length/3);
            %
            \node [] at (-\branchScale*\length*\sinval-0.2,-\length/3) {\textcolor{blue}{$\bm{s'}$}};
            %
            \draw[line width = 0.7 pt, draw = blue,  -latex] (\branchScale*\length*\sinval+0.3,-\branchScale*\length*\sinval-0.5*\length) to[out = 90, in = 0] (0,\length/3);
            %
            \node [] at (\branchScale*\length*\sinval+0.1,\length/3) {\textcolor{blue}{$\bm{s}$}};
            %
            \node [] at (-\branchScale*\length*\sinval-0.1,\branchScale*\length*\sinval+\length+0.1) {$\varepsilon'$};
            \node [] at (-\branchScale*\length*\sinval-0.1,-\branchScale*\length*\sinval-\length-0.1) {$d$};
            \node [] at (-\branchScale*\length*\sinval-0.1,-\branchScale*\length*\sinval-\length-0.1) {$d$};
            \node [] at (\branchScale*\length*\sinval+0.1,-\branchScale*\length*\sinval-\length-0.1) {$\alpha$};
            \node [] at (\branchScale*\length*\sinval+0.1,\branchScale*\length*\sinval+\length+0.1) {$b$};
            \node [] at (0.15,-\length/3+0.15) {$q$};
            \node [] at (-0.2,\length/3+0.2) {$\eta'$};
            \node [] at (-0.2,-\length/3-0.3) {$\beta$};
        \end{tikzpicture}
        \Bigg \rangle 
        = \cdots \\
        &=
        \sum_{q} C_\text{coef.}(q)
        \Bigg \vert
        \begin{tikzpicture}[baseline={([yshift=-.5ex]current bounding box.center)}, every node/.style={scale=0.9}]
            \pgfmathsetmacro{\length}{0.5}
            \pgfmathsetmacro{\Axesscaling}{1.68}
            \pgfmathsetmacro{\OffAxesscaling}{1.18}
            \pgfmathsetmacro{\sinval}{sin(45)}
            \pgfmathsetmacro{\lineW}{0.4}
            \pgfmathsetmacro{\markingPos}{0.65}
            \pgfmathsetmacro{\branchScale}{1.7}
            %
            \begin{scope}[decoration={markings, mark=at position \markingPos with {\arrow{Latex}}}]
            %
                \draw[line width = \lineW pt, postaction={decorate}] (0,-\length) -- (0,\length);
                %
                \draw[line width = \lineW pt, postaction={decorate}] (0,\length) -- (-\branchScale*\length*\sinval+0,\branchScale*\length*\sinval+\length); 
                \draw[line width = \lineW pt, postaction={decorate}]
                (\branchScale*\length*\sinval/2+0,\branchScale*\length*\sinval/2+\length) -- (\branchScale*\length*\sinval+0,\branchScale*\length*\sinval+\length);
                \draw[line width = \lineW pt, postaction={decorate}] (\branchScale*\length*\sinval/2+0,\branchScale*\length*\sinval/2+\length) -- (0,\length); 
                %
                \draw[line width = \lineW pt, postaction={decorate}] (-\branchScale*\length*\sinval/2+0,-\branchScale*\length*\sinval/2-\length) -- (0,-\length);
                \draw[line width = \lineW pt, postaction={decorate}] (-\branchScale*\length*\sinval+0,-\branchScale*\length*\sinval-\length) -- (-\branchScale*\length*\sinval/2+0,-\branchScale*\length*\sinval/2-\length); 
                \draw[line width = \lineW pt, postaction={decorate}] (0,-\length) -- (\branchScale*\length*\sinval+0,-\branchScale*\length*\sinval-\length); 
            \end{scope}
            %
            \draw[line width = 0.7 pt, draw = blue,  -latex] (-\branchScale*\length*\sinval-0.3,-\branchScale*\length*\sinval-0.5*\length) to[out = 45, in = 135] (-\branchScale*\length*\sinval/2+0,-\branchScale*\length*\sinval/2-\length);
            %
            \node [] at (-\branchScale*\length*\sinval-0.1,-\branchScale*\length*\sinval-0.5*\length+0.3) {\textcolor{blue}{$\bm{s'}$}};
            %
            \draw[line width = 0.7 pt, draw = blue,  -latex] (\branchScale*\length*\sinval+0.3,\branchScale*\length*\sinval+0.5*\length) to[out = 225, in = 315] (\branchScale*\length*\sinval/2+0,\branchScale*\length*\sinval/2+\length);
            %
            \node [] at (\branchScale*\length*\sinval+0.1,\branchScale*\length*\sinval+0.5*\length-0.3) {\textcolor{blue}{$\bm{s}$}};
            %
            \node [] at (-\branchScale*\length*\sinval-0.1,\branchScale*\length*\sinval+\length+0.1) {$\varepsilon'$};
            \node [] at (-\branchScale*\length*\sinval-0.1,-\branchScale*\length*\sinval-\length-0.1) {$d$};
            \node [] at (\branchScale*\length*\sinval+0.1,-\branchScale*\length*\sinval-\length-0.1) {$\alpha$};
            \node [] at (\branchScale*\length*\sinval+0.1,\branchScale*\length*\sinval+\length+0.1) {$b$};
            \node [] at (-0.2,0) {$q$};
            \node [] at (\branchScale*\length*\sinval/2-0.25,\branchScale*\length*\sinval/2+\length) {$\gamma$};
            \node [] at (-\branchScale*\length*\sinval/2+0.25,-\branchScale*\length*\sinval/2-\length) {$\tau'$};
        \end{tikzpicture}
        \Bigg \rangle,
\end{align*}
\endgroup
where, for each $q$, the coefficient of the alternative expression is
\begin{equation}
    C_\text{coef.}(q) = \sum_{\eta', \beta} F^{b^* c a}_{s' \varepsilon'^* \eta'} F^{d c^* e^*}_{s \alpha^* \beta} F^{s \beta^* c^*}_{s' \eta'^* q} F^{\alpha^* d \beta}_{s' q^* \tau'} F^{\varepsilon'^* b^* \eta'}_{s q \gamma}.
\end{equation}
Manipulating the $F$-symbols via the tetrahedral symmetry Eq.~\eqref{eq:tetrahedral_symmetry} and performing the above sum over either $\eta'$ or $\beta$ via the pentagon equation Eq.~\eqref{eq:pentagon_equation}, we obtain
\begin{equation}
    \left \{ 
    \begin{aligned}
        &\sum_{\rho} F^{b^* c a}_{s' \varepsilon'^* \rho} F^{e^* d c^*}_{s' \rho^* \tau'} F^{\tau' \rho^* e^*}_{s \alpha^* q^*} F^{\varepsilon'^* b^* \rho}_{s q \gamma} &&\text{if sum over $\beta$}, \\
        &\sum_{\rho} F^{d c^* e^*}_{s \alpha^* \rho} F^{a b^* c}_{s \rho^* \gamma} F^{\gamma \rho^* a}_{s' \varepsilon'^* q} F^{\alpha^* d \rho}_{s' q^* \tau'} &&\text{if sum over $\eta'$},
    \end{aligned} 
    \right.
\end{equation}
which are exactly the coefficients in Eq.~\eqref{eq:CommutationSPrimedFirst} and Eq.~\eqref{eq:CommutationSFirst} for fixed $q$. Hence, the order, in which  the strings $s$ and $s'$ are fused into the bordering edges does not matter. That is, $[G^s_p,G^{s'}_{p'}] = 0$, as long as they do not act on each other's controlled edge. The proofs for the remaining commutation relations in Eq.~\eqref{eq:Gsp_commute1} and Eq.~\eqref{eq:Gsp_commute2} are similar.

\subsection{The central equation}
\label{sec:Proof_centralequation}

Let us prove the central equation on a triangular plaquette. The proof on any polygon-shaped plaquette is similar. With the graphical definitions, we find 
\begingroup
\allowdisplaybreaks
\begin{align}
        & G^{\gamma}_p \left(\ket{\gamma} \bra{c}\right)_\text{ct} {G^{c}_p}^{\dagger} 
        \Bigg \vert
 
        \Bigg \rangle ,
        \label{eq:CentralEqonTriangularPlaquette_details}
\end{align}
\endgroup
 Now, it remains to show that the coefficient for every basis ket-vector (i.e. fixing $\alpha$ and $\beta$) is the same as that of $\sum_k P^\gamma_\text{ct} \left(\frac{d_k}{d_\gamma d_c} B^k_p \right) P^c_\text{ct}$. To do this, first, let us write the pentagon equation Eq.~\eqref{eq:pentagon_equation} in a different form via the tetrahedral symmetry Eq.~\eqref{eq:tetrahedral_symmetry}
\begin{equation}
    F^{i j p}_{k q^* r} F^{i q^* r}_{\ell^* s m} = \sum_{n} F^{r j^* k^*}_{n \ell s^*} F^{i p j}_{n s m} F^{q^* k p^*}_{n m^* \ell^*} \frac{d_n d_r}{\sqrt{d_k d_s d_j d_\ell}}.
    \label{eq:Alternative_pentagon_equation}
\end{equation}
Focusing on the $F$-symbols of Eq.~\eqref{eq:CentralEqonTriangularPlaquette_details}, we find
\begin{equation}
    \begin{aligned}
    F^{\ell_2 a^* b}_{c \ell_3 \ell_1} F^{\ell_1 c^* a}_{c \ell_1^* 0} & F^{\ell_1^* \ell_1 0}_{\gamma^* \gamma \alpha^*} F^{\ell_2 \ell_3 \ell_1}_{\gamma \alpha^* \beta} \\
    &= \sqrt{\frac{d_\alpha}{d_{\ell_1} d_\gamma}} \sqrt{\frac{d_a}{d_{\ell_1} d_c}}  F^{\ell_2 a^* b}_{c \ell_3 \ell_1} F^{\ell_2 \ell_3 \ell_1}_{\gamma \alpha^* \beta} \\ 
    &= \sum_k \frac{d_k}{d_\gamma d_c} F^{\ell_1 a c^*}_{k \gamma^* \alpha} F^{\ell_2 b a^*}_{k \alpha^* \beta} F^{\ell_3 c b^*}_{k \beta^* \gamma},
    \end{aligned}
\end{equation}
The first equality follows from the normalization of $F$-symbols in Eq.~\eqref{eq:F_normalization}, and we have used Eq.~\eqref{eq:Alternative_pentagon_equation} in the second equality to get the last line. The last line is exactly the coefficient of $\sum_k P^\gamma_\text{ct} \left(\frac{d_k}{d_\gamma d_c} B^k_p \right) P^c_\text{ct}$ on the same ket-vector. Hence, we have proven the central equation on the triangular plaquette.

\section{Anyon condensation on a lattice}
\label{sec:Anyon_condensation_review}

In this appendix, we briefly review the lattice realization of anyon condensation as discussed in Ref.~\onlinecite{Burnell_2012,PhysRevB.84.125434,PhysRevB.79.033109}. Specifically, we review condensation of abelian bosonic anyons through two examples: condensation of $\psi\bar{\psi}$ in the doubled-Ising string-net (Appendix~\ref{sec:Lattice_condensation_DI}) and condensation of $e$ or $m$ in the toric code model (Appendix~\ref{sec:Lattice_condensation_TC}). For a comprehensive review, we refer the reader to Ref.~\onlinecite{Burnell2018} and references therein.

\subsection{Condensation of $\psi\bar{\psi}$ in doubled-Ising}
\label{sec:Lattice_condensation_DI}

After condensing the abelian boson $\psi\bar{\psi}$ in the doubled-Ising string-net, the resultant system has a topological order that is the same as the toric code.\cite{Bais2009} For the lattice model, the condensation is achieved as follows.\cite{Burnell_2012,PhysRevB.84.125434}

First, we couple the doubled-Ising Hamiltonian with the shortest open-ended string-operator of $\psi\bar{\psi}$. That is, we add the operator $W^{\psi\bar{\psi}}_l$ for every edge $l$ as
\begin{equation}
    H_\text{D.I.} - J \sum_{l} W^{\psi\bar{\psi}}_l,
\end{equation}
where $H_\text{D.I.}$ is given by Eq.~\eqref{eq:Doubled_Ising_Hamiltonian}, $J>0$ is a real parameter controlling the strength of the coupling, and $W^{\psi\bar{\psi}}_l = (-1)^{n_1(l)}$ with $n_1(l) = 1$ if $l$ is in the state $\ket{1}$ and $n_1(l) = 0$ otherwise (see Appendix~\ref{sec:DI_fluxon_condensation} for a derivation). $W^{\psi\bar{\psi}}_l$ creates a pair of $\psi\bar{\psi}$ excitations, each at a plaquette bordering $l$ (see Fig.~\ref{fig:Square_octagon_lattice}). 

Next, we energetically favor the creation of $\psi\bar{\psi}$ excitations by increasing $J$. The system is then driven across a phase transition. The ground state of the resultant phase is a condensate of $\psi\bar{\psi}$ excitations. 

To see the $\psi\bar{\psi}$-condensed phase has the topological order of the toric code, let us take the $J \to +\infty$ limit. The coupling term $W^{\psi\bar{\psi}}_l$ imposes energy costs for every edge in the state $\ket{1}$. Taking the limit essentially removes every string-net configuration that contains a $1$-string. Treating $H_\text{D.I.}$ as a perturbation, we find a commuting projector Hamiltonian
\begin{equation}
    H = -\sum_v P Q_v P -\sum_p \frac{1}{2}\left(B^0_p + B^2_p\right),
\end{equation}
where $P = \prod_{l}\frac{1}{2}\left(\mathbbm{I}+W^{\psi\bar{\psi}}_l\right)$ is the projector onto the Hilbert space of the condensed phase. This Hamiltonian is exactly that of the toric code\cite{Stringnet}.

\subsection{Condensation of $e$ or $m$ in toric code}
\label{sec:Lattice_condensation_TC}

Condensing either the $e$ or the $m$ boson in the toric code model leads to the trivial phase without any topological order. On the lattice level, this is analyzed extensively in Ref.~\onlinecite{PhysRevB.79.033109}. Here, we quickly review the results of Ref.~\onlinecite{PhysRevB.79.033109} by following the discussion in Appendix~\ref{sec:Lattice_condensation_DI}.

Let us start by considering the condensation of the $e$ excitations. The condensation can be induced by coupling the toric code Hamiltonian (see Eq.~\eqref{eq:Toric_code_Hamiltonian}) with $W^e_l = X_l$, the shortest open-ended string-operator of $e$. That is, we consider the Hamiltonian
\begin{equation}
    H_\text{T.C.} - J \sum_l W^e_l.
\end{equation}
To see the condensed phase has the trivial topological order, we again take the $J \to +\infty$ limit. We see that the ground state is a product state given by $\otimes_l \ket{+}_l$, where $\ket{+} = \frac{1}{\sqrt{2}}\left(\ket{0}+\ket{1}\right)$ is the $+1$ eigenstate of $X$. Hence, no topological order.

Similarly, for the condensation of $m$ excitations, we consider the Hamiltonian
\begin{equation}
    H_\text{T.C.} - J \sum_l W^m_l,
\end{equation}
where $W^m_l = Z_l$. At the $J\to+\infty$ limit deep inside the condensed phase, we find the ground state is again a product state given by $\otimes_l \ket{0}_l$.

\clearpage

\end{document}